\documentclass[11pt]{article}

    \usepackage[dvipsnames]{xcolor}
    \usepackage[T1]{fontenc}
	\usepackage[utf8]{inputenc}
	\usepackage[english]{babel}
	\usepackage{amsfonts}
	\usepackage{amssymb}
        \usepackage{dsfont}
	    \usepackage{amsmath}
	
	\usepackage{enumerate}
	\usepackage{bbm}
	\usepackage{mathtools}
	\usepackage{subfig}
	\usepackage{booktabs}	
	\usepackage{setspace}
	\usepackage{graphicx}
	\usepackage{fullpage}
	
	\usepackage{soul}
	\usepackage{xspace}
	\usepackage{csquotes}
	\usepackage{placeins}
	\usepackage[footfont=normalsize,font=normalsize]{floatrow}
	\usepackage{floatrow}
	\usepackage{lscape}
	\usepackage{longtable}
	\usepackage{mathabx}
    \usepackage{accents}
    \usepackage{float}

    \usepackage{tikz}
    \usetikzlibrary{calc,intersections}

	\usepackage[]{natbib}
	\usepackage{bibunits}
	\defaultbibliography{CNA_Bibliography.bib}  
	\defaultbibliographystyle{aea}

\usepackage{geometry}
\usepackage{caption}
\usepackage{subcaption}

\usepackage{pgfplotstable}

\usepackage[explicit]{titlesec}

\titleformat{\section}
  {\normalfont  \fontsize{16}{14} \bf}
  {\thesection \,}
  {0.25em}
  {#1}
\titleformat{\subsection}
  {\normalfont \fontsize{14}{12} \bf}
  {\thesubsection}
  {1em}
  {#1}

\titleformat{\subsubsection}
  {\normalfont \fontsize{12}{10} \bf}
  {\thesubsubsection}
  {1em}
  {#1}

\usepackage{pgfplotstable}
	
	\usepackage{hyperref}
	\hypersetup{
    colorlinks=true,
    linkcolor=Red,
    filecolor=magenta,      
    urlcolor=blue,
    citecolor = Blue
}
	\usepackage{cleveref}

    \usepackage{rotating}
	\usepackage[framemethod=tikz]{mdframed}
	\usepackage{todonotes}
	\presetkeys{todonotes}{color=black!5,inline}{} 

    \usepackage{tcolorbox}
    \newtcbox{\feedback}{nobeforeafter,colframe=black,colback=white,boxrule=0.5pt,arc=2pt,
      boxsep=0pt,left=2pt,right=2pt,top=2pt,bottom=2pt,tcbox raise base}
      
    \usepackage{amsthm}

    \theoremstyle{definition}

\usepackage{array}
\newcolumntype{L}[1]{>{\raggedright\let\newline\\\arraybackslash}m{#1}}
\newcolumntype{C}[1]{>{\centering\let\newline\\\arraybackslash\hspace{0pt}}m{#1}}
\newcolumntype{R}[1]{>{\raggedleft\let\newline\\\arraybackslash\hspace{0pt}}m{#1}}

\usepackage{ragged2e}
\newlength\ubwidth

\def\eea{\end{eqnarray*}}
\def\bea{\begin{eqnarray*}}
\renewcommand{\[}{\begin{equation}}
\renewcommand{\]}{\end{equation}}
\newcommand{\bi}{\begin{itemize}}
\newcommand{\ei}{\end{itemize}}

\pgfplotsset{compat=1.18}
\hypersetup{
  pdfauthor={Raymond},
  pdftitle={The Market Effects of Algorithms},
  pdfdisplaydoctitle=true,
  pdfpagelayout=OneColumn,
  pdfstartview=FitH,
  pdfpagemode=UseOutlines,
  pdfpagemode=UseNone,
  bookmarksnumbered=false,
  verbose=false,
  colorlinks=true,
  citecolor=blue,
  linkcolor=blue,
  urlcolor=blue
}

\usepackage{pgfplotstable}

\usepackage{geometry}
\usepackage{caption}
\begin{document}
\thispagestyle{empty}

\vspace{20pt}
\begin{center}
\textcolor{white}{fill}\\
\vspace{20pt}

\LARGE{The Market Effects of Algorithms$^{*}$}\\\mbox{ }
\end{center}

\begin{singlespace}
\begin{center}
\begin{tabular}{ccccccc}
\large{Lindsey Raymond} \\
\large{MIT} \\
\end{tabular}
\end{center}

\end{singlespace}

\begin{center}
\vspace{.20in}
\normalsize{\today}

\vspace{10pt}
First Draft: December 2023\\
Please see \href{https://www.lindseyrraymond.com/assets/MarketEffectsofAlgorithms_LRaymond.pdf}{here} for latest version
\end{center}
\vspace{10pt}

\begin{abstract}
\vspace{-11pt}
\singlespacing
\noindent

While there is excitement about the potential for algorithms to optimize individual decision-making, changes in individual behavior will, almost inevitably, impact markets. Yet little is known about such effects. In this paper, I study how the availability of algorithmic prediction changes entry, allocation, and prices in the US single-family housing market, a key driver of household wealth. I identify a \textit{market-level} natural experiment that generates variation in the cost of using algorithms to value houses: digitization, the transition from physical to digital housing records. I show that digitization leads to entry by investors using algorithms, but does not push out investors using human judgment. Instead, human investors shift toward houses that are difficult to predict algorithmically. Algorithmic investors disproportionately purchase minority-owned homes, a segment of the market where humans may be biased. Digitization increases the average sale price of minority-owned homes by 5\% and reduces racial disparities in home prices by 45\%. Algorithmic investors, via competition, affect the prices paid by owner-occupiers and human investors for minority homes; such changes drive the majority of the reduction in racial disparities. The decrease in racial inequality underscores the potential for algorithms to mitigate human biases at the market level.
\end{abstract}


\bigskip

\hspace{-25pt}\rule[0.1ex]{0.33\textwidth}{0.2mm}\vspace{0.05in}\\\indent
\begin{singlespace}
\vspace{-25pt}
\noindent \scriptsize{\mbox{  }\mbox{  }\mbox{  }\mbox{  }\mbox{ }$^{*}$Correspondence to lraymond@mit.edu. 
I am especially grateful to Sendhil Mullainathan, Danielle Li, Erik Brynjolfsson, and Scott Stern for their invaluable feedback and advice. I thank Eric Budish, Mert Demirer, Brandon Enriquez, Bob Gibbons, Anh Nguyen, Manish Raghavan, Ashesh Rambachan, David Reshef, Daniel Rock, Frank Schilbach, Rob Seamans, Eric So, Evan Soltas, Bill Wheaton, Maggie Yellen, and various seminar participants for helpful comments and suggestions. I thank Max Feng for providing excellent research assistance and the Stanford Digital Economy Lab for financial support. The content is solely the responsibility of the author and does not necessarily represent the official views of Stanford University or the Massachusetts Institute of Technology.}
\end{singlespace}

\thispagestyle{empty}

 \setcounter{page}{0}

\clearpage

\newpage

Prediction plays a central role in many high-stakes decisions: Hiring depends on forecasting candidates' future performance; lending depends on risk of default; and investment relies on projections of returns. Recent advances in artificial intelligence (AI) and machine learning (ML) and the growing availability of digital data sparked interest in making such predictions algorithmically. A large and rapidly growing literature has begun to document when and how algorithms, formally specified, machine executable procedures, outperform human predictions. While algorithms have improved decision quality, their \textit{market-level} impacts remain less explored.

If algorithms change individual decisions, their use could also impact market-level outcomes, such as prices, or even change the nature of competition. These broader market dynamics mean that even if algorithms improve decision quality, individuals could be worse off. Yet, studies that focus solely on individual decision quality cannot capture impacts beyond the individual or firm level.

I provide early evidence on market effects by studying how digitization and algorithmic prediction by investors shape prices and allocations in the U.S. residential housing market. While the application of data and algorithms is growing across many industries, the single-family housing market offers an especially compelling setting for three reasons. First, single-family homes are economically significant: Housing is the largest asset market and important driver of household wealth \citep{corelogic2023, derenoncourtwealth2022}. Second, prediction is central to investor decisions, who buy houses to resell or rent out, and whose choices depend on forecasts of rental income, appreciation, and maintenance costs. Third, a natural experiment in the housing market overcomes the identification challenges that have constrained empirical work on market-level questions.

The central identification challenge is that the adoption of algorithmic prediction is not random. To address this, I develop a novel empirical strategy to identify market-wide effects by exploiting a simple yet fundamental insight: Machine learning algorithms require machine-readable data. Therefore, variation in the accessibility of machine-readable training data creates variation in the cost of accurate algorithmic prediction. 
 In the United States, useful training data for algorithms may come from public records on the housing stock and transactions collected by county governments for routine administrative tasks. Driven by the Open Government movement in the early 2000s, counties in the four states I study, Georgia, North Carolina, South Carolina, and Tennessee, began to transition from physical records to electronic database systems. The staggered rollout of digitization created variation in the cost of accessing housing market data, and thus algorithmic prediction, across counties and over time, enabling comparison of prices and allocations before and after digitization and between digitized and not-yet-digitized counties.

I have three sets of findings.

In my first set of results, I show that county digitization led to entry by algorithmic investors, investors augmented with algorithmic prediction methods. Single family houses are purchased by investors, who buy houses to rent out or resell, and owner-occupiers, who are individuals or households, purchasing houses to live in. Investors, which I identify in the data as corporate entities purchasing single-family homes to rent out, are manually classified into two groups based on their use of predictive technology. Algorithmic investors are those that either publicly disclose the use of algorithms or employ technical personnel capable of developing them. In contrast, human investors are firms with no evidence of algorithmic tools or relevant technical staff. County digitization leads to entry by algorithmic investors and a sharp and persistent increase in algorithmic investment, which is minimal before digitization. Post-digitization, algorithmic investor account for about 2 percent of all home sales and 20 percent of houses bought by investors, representing a sharp and substantial shift in market composition.


These initial findings could be misleading if unobserved county-level changes, such as policy changes, drive both digitization timing and housing market activity. To address potential violations of the parallel trends assumption, I exploit bureaucratic constraints that generate house-level variation in each county. Using not-yet-digitized houses in the same county or neighborhood as a control group, I conduct falsification tests and triple-difference analyses within counties and Census block groups. I show that county digitization leads to algorithmic investment in digitized houses, but has little effect on similar not-yet-digitized properties nearby. Additional robustness checks show that the predictive power of data from one county in another county is limited, and that digitization itself does not significantly affect human investor behavior. I also provide evidence that the classification procedure captures relevant differences across investor types.

To understand how digitization and the subsequent entry by algorithmic investors might impact prices and allocations, I develop a conceptual framework that contrasts human and algorithmic prediction, building on the existing evidence on human and algorithmic decisions. Algorithmic investors rely on machine-executable procedures that identify statistical patterns in data. Humans, by contrast, can incorporate unobservable information not captured in datasets, such as the quality of bathroom tile work, a yard's sunlight exposure, and ambient neighborhood noise. Yet this advantage in information richness comes with a tradeoff: Human predictions may be susceptible to cognitive limitations or biases.

The conceptual framework yields two empirical implications. First, if human biases lead to the systematic undervaluation of certain homes, this could create an arbitrage opportunity for algorithmic investors. If algorithmic investors disproportionately target those properties to exploit the mispricing, their entry could lead to disproportionately larger price increases for these homes as the mispricing is corrected. 

Second, the effectiveness of algorithmic prediction depends on the quality and comparability of administrative data collected. In houses where key features are missing, noisy, or hard to quantify, algorithmic models will perform poorly, giving human investors, who can rely on non-quantified information, a comparative advantage. As a result, human investors may shift toward these harder-to-predict segments, creating a market-wide reallocation driven by differences in predictive capabilities. This reallocation can influence prices even in areas where algorithmic investors are not active.

In my second set of results, I examine if there is algorithmic arbitrage of human biases by focusing on a particular bias that could lead to systematic undervaluation in the housing market: the role of race. I focus on race because it is one of the most extensively studied and well-documented potential biases in the housing market \citep[e.g.,][]{elsterMinoritiesPropertyValues2022, perryDEVALUATIONASSETSBLACK, freddiemac2021, cutlerRiseDeclineAmerican1999, box-couillard_racial_2024}. In support of the possibility that racial bias might lead to undervaluation of some houses, I document the existence of a 5\% \textit{race penalty}: a residual difference in sale price between similar minority and White-owned homes in the same Census block group. The key empirical challenge, for both algorithmic investors and the econometrician, is to understand if the race penalty is an arbitrage opportunity created by human errors or reflects unobserved differences in quality between minority and White-owned homes.

Consistent with the possibility that humans may undervalue minority-owned homes, algorithmic investors disproportionately buy such homes. Algorithmic investors’ share of transactions involving minority sellers is double their overall market share (4\% vs. 2\%). Moreover, house digitization is associated with a 50\% larger treatment effect on algorithmic investment for a minority-owned house than a comparable White-owned property in the same neighborhood. This disparity is not explained by differences in neighborhood composition. Together, these findings point to seller race as an important factor shaping algorithmic investment patterns.

With algorithmic entry and investment, particularly among minority homes, the average race penalty shrinks from 5.8\% before digitization to 3.2\% after, a 45 percent decline. In other words, after digitization observably similar houses sell closer to parity. This convergence occurs only in counties with algorithmic investment; digitization alone does not significantly change the race penalty in counties without such entry. One contributing factor is that algorithmic investors buy race-neutrally: on average, they do not exhibit a race penalty in the homes they purchase. Still, although algorithmic investors represent a disproportionate share of the market, their activity alone cannot fully account for the overall reduction in the race penalty.

Importantly, human investors and owner-occupiers drive much of the reduction in market-level racial disparities. With digitization, the race penalty among owner-occupiers shrinks from 5.5\% to 3.3\%, and from 8.9\% to 1.9\% among purchases by human investors. An Oaxaca-style decomposition shows that most of the market-wide decline stems from a falling race penalty among owner-occupiers. Two mechanisms could explain this effect: competition and the influence of comparable sales. Algorithmic investors may raise prices through competitive bidding, even in transactions they do not win, and their purchases may boost listing prices for nearby comparable homes, indirectly benefiting minority sellers. To test these channels, I use house-level variation in digitization as a quasi-random measure of exposure to algorithmic investor competition. Reductions in the race penalty are concentrated among digitized homes, with limited spillovers, suggesting that competition is the primary force narrowing racial disparities. 
This highlights a key market effect of algorithms: Through competition, algorithm activity can influence the behavior of home buyers not using algorithms. 

Although one explanation for the increasing prices of minority homes might be human mistakes, another possibility is that algorithmic investors are simply overpaying for unobservably bad
minority-owned houses. As emphasized in the conceptual framework, algorithms do not see all aspects of house quality available to humans, potentially leading to adverse selection. I assess these explanations using two complementary approaches. First, I examine whether unobserved differences in property appearance, such as interior condition or curb appeal, can explain the race penalty. I find that it persists even after controlling for deep learning–based image embeddings of house exteriors and interiors. Second, I test whether algorithmic investors earn lower returns on homes previously owned by minorities, which would be consistent with overpaying. I find no evidence that algorithmic investors, on average, earn lower margins on formerly minority-owned homes relative to comparable homes previously owned by White sellers.

So far, my price analysis has focused on the race penalty, the within-neighborhood price gap between similar homes. I also estimate the effect of county digitization on average sale prices. County digitization raises prices by 5\% for minority-owned homes and 2\% for White-owned homes, with an overall increase of 3\%. Prices dip slightly in the first year, consistent with early algorithmic entrants exploiting limited competition, but increases become apparent after about three years. This 5\% price increase among incumbent minority homeowners is economically meaningful, equivalent to 38\% of median Black household wealth and 25\% of median Hispanic household wealth \citep{bhuttaDisparitiesWealthRace2020}.

In my third set of results, I examine whether human investors reallocate toward parts of the market where data quality constrains algorithmic prediction, and the resulting impacts on prices. I build an extreme gradient boosted tree model that generates an ex-ante measure of property predictability that captures how well availability and informativeness of structured data predict sales price. While digitization does not change total human investment, human investors increase their activity in less predictable properties, consistent with specialization in the investor market.  Prices of these hard-to-predict houses also rises by 4 percent, even though algorithmic investment in is relatively limited.

Together, these findings illustrate how markets can amplify the effects of algorithmic adoption. The results echo the spirit of \citet{becker1957}, where market competition disciplines and ultimately displaces discriminatory actors. The magnitude and patterns of these effects raise questions about how even low levels of algorithmic adoption could be reshaping other markets.

This paper contributes to a growing empirical literature on the impacts of access to algorithmic recommendations, with a distinct focus on market-level effects. Comparing human decision-makers' choices with predictive models has a long history \citep{dawes1971, dawes1989, dawesbook2001}. Modern advances in ML, increased computing power, and data availability have renewed interest in these questions. I build on prior work that shows that algorithmic recommendations can lead to improvements ranging from better heart attack diagnosis to more accurately determining which defendants should get bail and or interview screening decisions.\footnote{For example, see \citet{autor2008, li_hiring_2025, Raghavan_2020, frankelSelectingApplicants2021, whitehouse2022, oecd2023} for applications in the labor market, \citet{einav2013, fuster2022, gillis_2019, dobbieetal, blattnerHowCostlyNoise2021} for consumer finance, \citet{mullainathanobermeyer2022, obermeyerPredictingFutureBig2016, kleinberg_inherent_2016, chouldechovaCaseStudyAlgorithmassisted2018, abaluckFixingMisallocationGuidelines2020, kleinberg2017, mullainathanPredictiveAlgorithmsAutomatic2023} for examples in the criminal justice system, health care, among other areas. See \citet{rambachanIdentifyingPredictionMistakes, mullainathanbail2017, kleinbergPredictionPolicyProblems2015b} for issues comparing human and machine predictions.} Other work shows that access to algorithms translates into improved productivity or efficiency.\footnote{See \citet{brynjolfsson_generative_2025} for the impacts of generative AI on productivity in customer service, \citet{harrisyellen2023} for the impact of the adoption of predictive maintenance on repair costs in a trucking company. See \citet{bubeck2023sparks, choiAIAssistanceLegal2023, peng2023check, noyExperimentalEvidenceProductivity2023} for additional effects of AI access on productivity, writing, and test taking capabilities.} However, not all studies find positive effects.\footnote{For instance, \citet{acemoglu2022} finds no detectable relationship between AI investments and firm performance, while \citet{babina2022} finds a positive relationship.}

While most research on algorithmic prediction focuses on individual decisions, a smaller body of work examines how ML-powered algorithms affect outcomes at the market level. \citet{calvanoAI, assad2023, calderwang2023, mackaybrownpricing} focus on the impact of ML-powered pricing algorithms on pricing and collusive behavior. Other studies focus on the impacts of automated algorithmic trading on the liquidity and pricing efficiency in financial markets \citep{hendershottDoesAlgorithmicTrading2011, chaboudRiseMachinesAlgorithmic2014, upsonMultipleMarketsAlgorithmic2017}. I extend this literature by examining how algorithmic prediction affects entry, prices, and allocation in the single-family housing market, an asset market with high transaction costs and longstanding equity concerns.

This paper also contributes the interdisciplinary literature on algorithmic bias. Initially, researchers and policy makers hoped that the use of algorithms could help mitigate human biases. For example, \citet{kleinberg2018discrimination} show that reliance on algorithms to grant bail could simultaneously reduce crime, jail populations, and racial disparities. However, many examples show algorithms directing disproportionately fewer opportunities or resources to minorities.\footnote{See \citet{smithAlgorithmsBias2021} for a summary of empirical work on algorithmic bias. See \citet{rambachan2019bias}, \citet{Rambachanetal}, \citet{bakalar2021fairness}, \citet{kleinberg_inherent_2016} and \citet{cowgilltucker2019} for theoretical work.} This paper is early evidence of the indirect effects of algorithms on racial bias that work via market competition.

Lastly, the paper contributes to research on the effects of investors in the single-family housing market. Growing concern over rising housing prices has spurred proposed policies to restrict single-family home ownership in the US and Europe. For example, in December 2023, Democrats introduced legislation in the House and Senate to ban hedge fund ownership of single-family homes \citep{kaysenNewLegislationProposes2023, merkleyEndHedgeFund}. A growing interdisciplinary literature has examined investor impacts on housing markets.\footnote{\citet{fields2018a, fields2022a} examine how technology-driven calculative agency enabled the financialization of the single-family housing market. \citet{raymondCorporateLandlordsInstitutional2016, raymondForeclosureEvictionHousing2018, raymondGentrifyingAtlantaInvestor2021} analyze institutional investors in Georgia and their effects on housing insecurity. \citet{millsLargeScaleBuy2019} provides early empirical evidence on institutional investor activity; \citet{gurun2023} study the rise in institutional ownership and the effects of investor mergers on rent and neighborhood safety; \citet{buchak2022} examine iBuyer firms (e.g., Zillow, Offerpad, Redfin, and Opendoor) and their effects on market liquidity; and \citet{franckeBuytoLiveVsBuytoLet2023} evaluate a ban on large institutional housing buyers in the Netherlands.} This paper extends this work by providing new empirical evidence on how algorithm-augmented investors shape market outcomes and interact with public data, with implications for both housing access and public data governance.

\section{The Economic Impacts of Algorithms}

One of the most robust findings in social science is that people routinely rely on mental shortcuts, or heuristics, to navigate complex and cognitively demanding decisions. While these strategies help manage complexity, they often lead to systematic errors in judgment \citep{tverskykahneman1974, halford_how_2005, shleifernoise}. Examples include confirmation bias, in which people selectively favor evidence that supports their existing beliefs; anchoring effects, where initial information disproportionately shapes later judgments; recency bias, where people over-weight recent events; and default effects, where people stick with pre-set options rather than actively choosing alternatives \citep{wason_failure_1960, ebbinghaus_memory_1913, tverskykahneman1974, Tversky_Kahneman_1982, optimaldefaults2003, thalerSMART2004}. People may also under-respond to new information, especially when it involves negative feedback \citep{mobiusupdating}. Beyond these general cognitive limitations, implicit biases based on race, gender, age, appearance, and language can consciously and unconsciously shape influence judgments \citep{Wistrich2017ImplicitBI, biddlebeauty, mobiusbeauty, DeprezSims2010AccentsIT, goldin2000orchestrating, greenwald_implicit_1995}. Decision quality may deteriorate due to fatigue, stress, and resource scarcity \citep{mani_poverty_2013, Levi2017DecisionFA, kaur_financial_2025}. Other work explores how human cognitive biases can shape markets \citep[e.g.][]{hirshleifer2015survey, shleifernoise, Salzman01012017, mullainathan2012}).

Motivated by evidence on human mistakes, researchers began investigating whether algorithms could improve decision quality. Early work, summarized in \citet{dawesbook2001}, compared human experts to simple statistical models and concluded: ``expert judgments are rarely impressively accurate and virtually never better than a mechanical judgment rule.'' \citet{tetlock_expert_2005} studied political forecasters over several decades, found that human forecasting ability was low, and showed that even simple statistical models outperformed both expert and non-expert forecasters. However, the adoption of algorithms remained limited, constrained by both the modest capabilities of early statistical models and the scarcity of digitized data on human decisions.

The rapid adoption of algorithms and the renewed focus on their impacts were driven by two key forces: advances in AI and ML, and the increasing digitization of economic activity. As more economic activity moved online, organizations began generating large datasets on decision making as a byproduct of routine operations.\footnote{For example, the adoption of electronic medical records in healthcare and Applicant Tracking Systems in hiring led to the automatic collection of rich data on hiring and medical professionals decisions \citep{li_hiring_2025, donnelly_systematic_2022}. } Concurrently, breakthroughs in ML, a subfield of AI, improved model architectures and training methods, enabling algorithms to learn complex, high-dimensional patterns from data without relying on task-specific, rule-based programming \citep{peter2021modern}. ML-based algorithms now outperform human experts in cognitively demanding tasks that historically resisted automation, including medical diagnosis, strategic gameplay such as Go, and financial forecasting \citep{mckinneybreastcancer2020, silver_mastering_2016, barboza_machine_2017}. These modeling advances were coupled with improvement in training models on large datasets without overfitting, enabling substantial gains in accuracy simply by increasing data volume \citep{halevy2009}. Collectively, these performance gains have contributed to rapidly growing adoption, with 62 percent of IT professionals reporting significant or moderate increases in AI or ML investments \citep{comptia}.

Researchers have begun to rigorously evaluate human and algorithmic performance in high-stakes decisions, revealing both substantial improvements over human judgment and challenges in human-AI collaboration. In a landmark study of bail decisions, \citet{kleinberg2017} demonstrate that machine learning algorithms can simultaneously improve accuracy and equity in the decision to grant bail. Their econometric approach grapples with two key challenges in the human-algorithm comparison: Humans may optimize different objectives than the algorithm (omitted-payoffs), and they may have access to additional information during in-person bail hearings (private information). 
In some contexts, algorithms can improve efficiency and equity of the decision process \citep[e.g.][]{li_hiring_2025, mullainathan_diagnosing_2022, rambachanIdentifyingPredictionMistakes, brynjolfsson_generative_2025, pmlr-v81-chouldechova18a, dobbiealgorecsbails2023, gruber_managing_2020}. In others, humans are overly skeptical of algorithmic recommendations (algorithm aversion) or too willing to adopt the AI recommendations \citep{vaccaro2024combinationshumansaiuseful, helanderHandbookHumanComputerInteraction2014, agarwalCombiningHumanExpertise, dietvorstAlgorithmAversionPeople2014, steyversThreeChallengesAIAssisted2024}. Other work has raised concerns about the potential for algorithmic bias to perpetuate or exacerbate discrimination in society \citep{rambachanalgobias2020, barocas-hardt-narayanan, chouldechovaroth2020}. 

Research on human and algorithmic decision quality has largely focused on settings where a single decision maker operates independently, such as a judge deciding whether to grant bail. Yet many prediction tasks occur in markets or other multi-agent settings, where one actor’s decisions influence others through prices, competition, or strategic interactions. For example, a firm might adopt an algorithm that identifies an overlooked talent pool, such as philosophy majors from Wesleyan University, as strong data analysts. If other firms follow, rising demand could drive up wages, affecting all firms seeking to hire these candidates. If algorithms change decisions, adoption could alter equilibrium outcomes.  On the other hand, if algorithm simply replicate human decisions, prices and allocations may remain unchanged. Yet there has been limited work on these questions. While single-agent studies offer valuable insights into individual decision quality, they are not designed to capture the equilibrium effects that can arise in markets.

Evidence on the market effects of algorithms has largely focused on two areas: the risks of algorithmic homogeneity and strategic interactions in financial markets. In pricing, algorithms may inadvertently or intentionally learn to collude. \citet{calvanoAI} raise this possibility theoretically, and empirical evidence suggests it is more than hypothetical: in the German retail gasoline market, adoption of algorithmic pricing by duopolist gas stations increased average margins by 28 percent, with no comparable change among monopolists or in highly competitive areas \citep{assad2023}. Related work examines other risks when many firms rely on similar algorithms \citep{Kleinberg2021AlgorithmicMA, fish2024algorithmiccollusionlargelanguage, calderwang2023, mackaybrownpricing}. From financial markets, research shows that technological innovations can also reduce welfare. \citet{budishHighFrequencyTradingArms2015} find that faster trading algorithms triggered an arms race for speed, raising entry costs without improving efficiency, and other work suggests high-frequency trading can disadvantage non-algorithmic investors and impair market functioning, though some studies find positive effects \citep{jones_what_2013, menkveld_economics_2016, hendershottDoesAlgorithmicTrading2011}. This paper provides early evidence on digitization and algorithmic decision-making in the housing market, a large and economically significant asset class where firms employ distinct proprietary algorithms.

\section{Setting: Algorithms in the Housing Market}
\subsection{The Single-Family Housing Market}
I study the market effects of algorithms in Georgia, Tennessee, North Carolina and South Carolina for three reasons: housing is the largest asset market in the U.S., housing investment requires prediction, and a natural experiment enables causal inference. I study single family houses, detached residences intended for a single household, comprise 66 percent of all U.S. housing stock and 86 percent of the \$43 trillion in total residential real estate value \citep{corelogic2023, neal2020}.\footnote{Single-family homes are detached dwellings built to be occupied by one household on their own plot of land \citep{neal2020, freddiemac2018}.} They also represent 53 percent of the rental market nationally \citep{freddiemac2018}. In the four states I study, single-family homes make up 66 percent of occupied housing in urban areas and 72 percent in rural areas, and are the largest single segment of the rental market \citep{u.s.censusbureauS2504PhysicalHousing2021, censusB25, neal2020, freddiemac2018}. These homes play a central role in household wealth accumulation and intergenerational wealth transmission \citep{derenoncourtwealth2022}.

\subsection{Housing Investment is a Prediction Problem}
Two groups of buyers participate in the housing market: owner-occupiers and investors. Owner-occupiers purchase homes to live in, making a consumption decision. Investors, by contrast, buy properties to rent out or resell, and their decisions depend on complex forecasts of future income, property values, and associated costs such as repairs, upgrades, and maintenance. Although owner-occupiers account for most single-family home purchases, investors are influential marginal buyers whose behavior can significantly shape prices and market dynamics \citep{bayer_investors_2021}.

Traditionally, investment in single-family houses was dominated by local investors due to the importance of proximity and deep knowledge of neighborhood-level housing markets. Often referred to as ``mom-and-pop'' investors, these firms typically owned a small number of properties while often working in adjacent fields such as construction or as real estate agents, where they have particular insight into local housing market dynamics \citep{fields2022a}.\footnote{In the rental market, \citet{freddiemac2018} estimates that 15.5 million investors  own 1 to 10 properties, which make up 98\% of all single-family investors who rent property.} The informational advantage of these local players led to the widespread belief that such mom-and-pop investors would maintain permanent dominance in the single-family home sector \citep{fields2018a, americanhomes4rentFormS11}.

While local mom-and-pop investors gain an informational edge from physical proximity, other aspects of housing investment favor machine learning. Houses are complex, idiosyncratic assets with thousands of potentially relevant features that often interact in nonlinear ways. Investors must assess both the property’s physical condition, including its structural soundness, features, and layout, and broader market conditions such as supply and demand dynamics, economic trends, neighborhood characteristics, and indicators of future development or change. The volume and complexity of this information make it difficult for humans to process but well suited to machine learning \citep{kahneman_thinking_2011, athey_impact_2018}. For example, the value of an extra bedroom may depend on total square footage, while the effect of school quality may vary with neighborhood demographics. These challenges are amplified by the thinness and heterogeneity of housing markets. Whereas human investors see only a limited set of transactions, algorithms can learn from large datasets covering diverse homes and neighborhoods.

\subsection{Human and Algorithmic Investors}
I classify investors into two categories based on their prediction technology: algorithmic and human. Algorithmic investors incorporate machine learning algorithms into their decision-making processes to determine which houses to buy and at what price. These algorithms are formally specified, machine-executable procedures that search across a wide space of candidate models, using historical data to optimize predictive performance \citep{mitchell2015}. By contrast, human investors rely on personal judgment and experience, often employing an acquisitions specialist to identify and evaluate properties. For example, Myers House Buyers, a prominent human investor in Georgia, staffs a team with diverse, non-technical backgrounds who emphasize their local roots and community ties. These acquisition specialists will attend open houses, speak with local business owners, and consider zoning or regulatory changes when evaluating properties.

Algorithmic investors, on the other hand, invest in data science teams that develop predictive models to screen and value properties. Their acquisitions teams, often located in cities like New York, San Francisco, or Austin, build these models and work with the generated recommendations. For instance, Amherst Residential reports that its proprietary platform, Amherst Explorer, automatically evaluates roughly 500 new listings per day across target markets, estimating rent, renovation costs, taxes, insurance, and other expenses to compute net operating income. Each morning, the acquisitions team receives a prioritized list of properties with projected returns, pre-screened by the algorithm \citep{amherstsfr2016}.\footnote{As Amherst describes: “Our data puts the dart on the dartboard, then our team puts it in the bullseye... As we evaluate and complete each transaction, machine learning enables us to incorporate live observations of our own portfolio into underwriting the next opportunity” \citep{kapoor_ai_2024}.}

Although building these technology platforms is expensive and requires specialized teams of data scientists and software engineers, algorithmic investors emphasize their necessity: ``[w]ithout using technology to filter and deliver automated valuations... it would be extremely time-consuming and inefficient to review and bid on these properties.... The entire process uses a vast amount of data that is impossible to distill into actionable information without the use of technology'' \citep{amherstsfr2016, christopher2023}.

\section{Digitization, Data, and the Empirical Strategy}
The main identification challenge is separating outcomes such as price increases that occur because algorithmic investors target markets with rising prices from price increases caused by algorithmic entry. Methods used to study algorithmic effects on individual decisions do not directly translate to measuring market-level impacts, which require a different empirical strategy.

The ideal experiment would randomize counties into  treatment and control groups. In treatment counties, machine learning algorithms \textit{could} be used to assess the value of houses, while in control counties, valuations would rely solely on human judgment. I would then compare market structure, prices, and allocations across the two groups. Such a comparison would identify the causal effect of algorithmic investors on market-level outcomes.

Designing an ideal experiment to measure market effects raises three challenges. First, causal identification must occur at the market level because the entry of algorithmic investors can change the behavior of owner-occupiers and human investors. These second-order effects are potentially central to market settings and especially important to capture. However, since randomizing counties is logistically infeasible, the empirical strategy must rely on a source of quasi-experimental variation in algorithm use. Second, this variation must also occur at the market level rather than the firm or individual level to avoid confounding from firm-specific factors such as differences in scale, capital, or strategy. Third, unlike domains where a single centralized algorithm guides decision-making, housing market firms use distinct proprietary models. This makes the treatment decentralized, so there is no single system to observe or ``switch on'', so the treatment must plausibly affect the ability to use any machine learning algorithm. The decentralized nature of algorithms mean the strategy requires a way to identify which firms actually rely on algorithmic valuation.

\subsection{Empirical Strategy: Digitization Creates Variation in the Cost of Accurate Algorithmic Prediction}
My empirical strategy rests on a simple observation: Machine learning algorithms depend on machine-readable training data. Differences in the cost of accessing relevant training data translate into differences in the cost of producing accurate algorithmic predictions.

In the housing market, potentially useful training data for predictive models are maintained by county governments, which by law collect detailed property records for administrative use. Counties collect two key types of public records: legal documents related to property transactions and titles, often managed by the recorder’s office, and detailed property characteristics and assessments for tax purposes, often the responsibility of the assessor’s office.\footnote{These records often date back to the county’s founding, in some cases as early as the 1600s.} Originally created to support core government functions such as tax collection, land-use planning, and property rights enforcement, these datasets are also widely used by engineers, lawyers, and urban planners. While similar records exist nationwide, their content and coverage vary widely, and the four states in my sample have some of the best quality records in the United States \citep{nolte_studying_2021}.\footnote{For instance, in ``non-disclosure'' states such as Alaska, Texas, and Utah, sale prices are not publicly recorded \citep{berrens2004}. Even in states that do record prices, data quality can be inconsistent and incomplete.}

By digitizing physical documents into structured online databases, counties reduced the cost of accessing public records.  The Open Government movement of the early 2000s accelerated digitization. This effort was part of a broader agenda to promote transparency and accessibility in government, assisted by falling costs of digital storage and formalized by federal initiatives such as the 2009 Open Government Directive \citep{gray2011beyondao, whitehouse2009}.\footnote{These efforts culminated in the 2007 Open Government Data Principles and were formally codified in the 2009 Open Government Directive \citep{schudson2020, gray2011beyondao, whitehouse2009, kavanaugh2010}} Besides making government more transparent, digitization saved the increasing cost of maintaining physical records.

Although county records appear to be a valuable data source, algorithmic investors may rely on alternative datasets and find these records redundant. To test this, the research design is inspired by a regression discontinuity around digitization: digitization represents a sharp drop in the cost of accessing training data, while other factors should smoothly. This structure allows me to examine, in the first set of results, whether algorithmic investment increases in response to the availability of digitized records.

\subsubsection{Digitization Data and Rollout}\label{sec:data:digitrollout}
\begin{figure}[ht!]
\begin{center}
\captionsetup{justification=centering}
\caption{\textsc{Figure \ref{fig:countydigit_balance}: County Digitization}}
\makebox[\linewidth]{
\begin{tabular}{c}
\textsc{A.  Share of Counties with Digitized Records}\\
\includegraphics[scale=0.35]{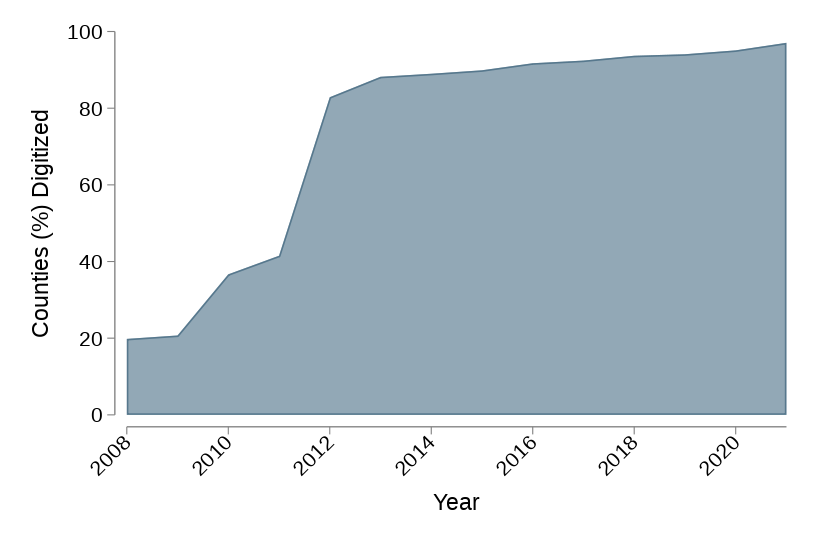} \\
\\
\textsc{B.  Balance Coefficient Plot of Early County Digitization}\\
\includegraphics[scale=0.35]{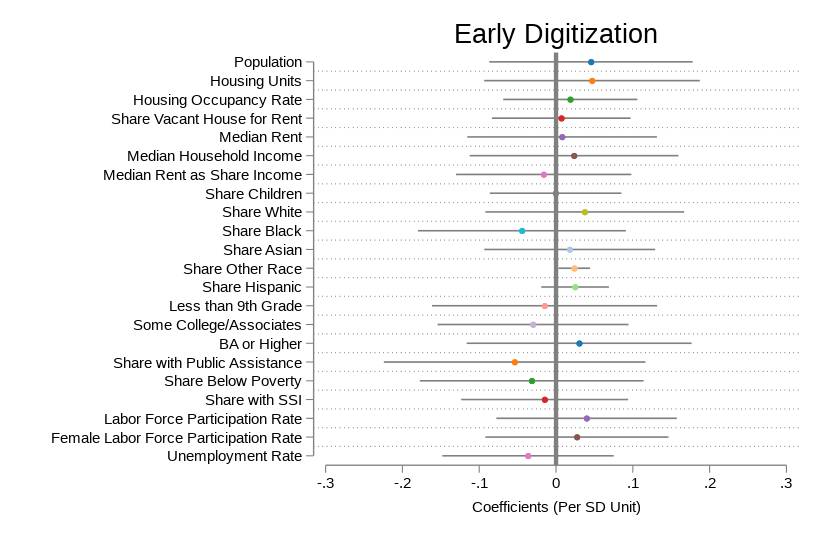}  \\
\end{tabular}
}
\label{fig:countydigit_balance}
\end{center}
\end{figure}

Panel A of Figure \ref{fig:countydigit_balance} shows the timing of digitization for 400 counties in North Carolina, South Carolina, Tennessee, and Georgia. The year of digitization is defined as the first year county recorder records became electronically accessible. I manually collected these dates from two complementary sources: county recorders’ offices and the Internet Archive.\footnote{The highly manual nature of this data collection is one reason the sample is limited to four states.} The primary source was direct interviews with county officials. To validate and supplement these interviews, I used historical snapshots of county websites from the Internet Archive to identify when records were first accessible remotely. While I measure digitization timing using transaction records, counties typically digitized and published property attribute data at the same time.

While state-level coordination generally led to sharp increases in digitization within each state, legal, technical, and financial barriers shaped specific completion times at the county level. Counties first had to secure legal authorization to maintain digital records, develop compatible software systems for hosting and managing these records online, and scan thousands of physical documents.

Panel B of Figure \ref{fig:countydigit_balance} examines whether baseline county characteristics predict early county digitization, controlling for state-fixed effects. Across a range of characteristics, including county size, educational attainment, housing occupancy and vacancy rates, and socioeconomic indicators, the coefficients are small and statistically insignificant. Even for county population, which shows one of the largest correlations with early digitization, the magnitude is modest: a one standard deviation increase in population is associated with only a 4.5 percentage point (15\%) higher likelihood of digitizing early. 
These results show that, conditional on state characteristics, early digitization was largely uncorrelated with observable county attributes, consistent with idiosyncratic bureaucratic and legal processing driving the rollout. 

\subsubsection{Transaction Data and Summary Statistics}\label{sec:data:summarystats}
My sample includes arm’s-length, single-family house transactions from 400 counties in Georgia, North Carolina, South Carolina, and Tennessee between 2009 and 2021.\footnote{Arm’s-length transactions are conducted at market prices between unrelated parties. Transactions between related parties, such as a wife transferring a home to her husband, are excluded because they may not reflect market value.} The primary data source is county recorder and assessor records, which provide the sale price, transaction date, buyer and seller identities, legal structure of each party, and detailed property characteristics, including geographic coordinates, year built, architectural style, number of bedrooms and bathrooms, roof construction, lot size, number of buildings, parking availability, and historical assessments of market, land, and improvement values.\footnote{As of 2023, records for these counties have been digitized back to the early 2000s, enabling historical analysis.} I supplement these data with demographic and socioeconomic information from the American Community Survey, the 2010 and 2020 Decennial Censuses, and Redfin and Zillow data on local housing market activity. I exclude multi-parcel sales, foreclosure sales, and non-arm’s-length transactions.

Table \ref{tab:sumstat_houses} provides details on the 6.7 million transactions in the sample, grouped into four categories: all sales, owner-occupiers, human investors, and algorithmic investors. With a sale price of \$207,248, the average home is 31 years old and has two bedrooms, two bathrooms, a garage, a parking space, and a fireplace. Reflecting regional construction norms in the Southeastern United States, most houses are single-story and lack basements. The median home stays on the market for about three months. Owner occupiers make up 89 percent of the overall transaction volume.

I also geocode each property to Census-defined tracts, block groups, and blocks using latitude and longitude coordinates. In my sample, Census tracts average 2,000 housing units; block groups, 700 units; and blocks, 30 units.\footnote{All geographic unit averages are based on the 2010 Decennial Census.} Because not all properties have coordinates precise enough for block-level geocoding, I use the block group as the primary unit of geographic analysis. Neighborhood demographic and housing characteristics come from the 2010 Decennial Census at the block group level, while socioeconomic variables, such as median income, educational attainment, labor force participation, and unemployment, are drawn from five-year ACS estimates at the tract level for each property.\footnote{Many counties in the sample fall below the population threshold for one-year ACS estimates (65,000 people). Due to suppression of block group-level ACS data for confidentiality, I use tract-level estimates for socioeconomic characteristics.}

Column 1 of Appendix Table \ref{atab:balance_county_inv} describes the average characteristics of the surrounding neighborhood for the typical transaction. On average, the block group has 978 housing units and an 87\% housing occupancy rate; 74\% of residents are White; and 51\% of adults aged 25 and older hold a bachelor’s degree or higher. In the broader Census tract, the unemployment rate averages 4.5\%, and the median rent-to-income ratio is 30\%.

\subsubsection{Dynamic Differences-in-Differences}\label{sec:diffindiff}
To estimate the causal impact of digitization in county $c$ and year $t$, I use a dynamic difference-in-differences specification with staggered timing:

\begin{equation} \label{eq:es_main}
y_{ct} = \delta_{t} + \alpha_{c} + \sum_{j \ne -1} \beta_{j} \mathds{1}[\tau_{ct} = j] + X_{ct}'\gamma + \epsilon_{ct}
\end{equation}
where $\tau_{ct} = t- E_c$ denotes event time relative the year of county digitization $E_c$. To assess whether digitization is associated with changes in algorithmic investment, the first outcome is the \textit{transaction market share} of algorithmic investors. The algorithmic investor share is defined as $y_{ct} =\frac{q^{algo}_{ct}}{q_{ct}}$, where $q^{algo}_{ct}$ is the number of houses purchased by algorithmic investors in county $c$ and year $t$, and $q_{ct}$ is the total number of home sales. The coefficients $\beta_j$ capture the average change in algorithmic investment in the $j$th year relative to the year before digitization. 

In my main specification, I include county-fixed effects ($\alpha_{c}$) to control for time-invariant county characteristics, such as proximity to major cities or natural amenities, and year-fixed effects ($\delta_{t}$) to account for time-varying shocks common across counties, such as changes in interest rates or broader macroeconomic conditions.  All regressions are weighted by the number of transactions in each county-year, with standard errors clustered at the county level. The comparison group consists of counties that had not yet digitized by year $t$, with counties that digitize after 2017 serving as the last-treated units. All counties eventually digitize, and digitization is an absorbing state; once a county digitizes, it remains treated in all subsequent years.

I also estimate these regression specifications using a range of robust difference-in-difference estimators. I use \citet{dechaisemartin2024} as my primary estimator, and examine robustness to alternative estimators from \citet{sun_estimating_2020}, \citet{borusyak2022revisiting}, \citet{callaway_2021}, and traditional two-way fixed effects ordinary least squares (OLS). These robust methods avoid problematic comparisons between already-treated and just-treated counties. In general, the robust estimators produce qualitatively similar, and often slightly larger, estimates than OLS.

I also incorporate a vector of pre-treatment time-varying socioeconomic and demographic county-level covariates, ($X_{ct}$), from the American Community Survey that may influence both digitization timing and investment patterns. For example, counties with faster-growing populations could tend to digitize earlier and be more attractive to investors. However, as shown in Table~\ref{tab:es_main}, adding in additional controls does not materially change the estimated treatment effects.

I also estimate a pooled difference-in-difference specification: 

\begin{equation} \label{eq:didpooled}
y_{ct} = \delta_{t} + \alpha_{c} + \beta D_{ct} + X_{ct}'\gamma + \epsilon_{ct}
\end{equation}
where $D_{ct}=\mathds{1}[t \ge E_c]$ is an indicator equal to one if county $c$ has digitized by year $t$, and zero otherwise.

Identification relies on three key assumptions: no anticipation, no spillovers, and parallel trends. First, market participants should not change behavior in anticipation of digitization. Second, digitization in one county should not affect outcomes in not-yet-digitized counties. Third, in the absence of digitization, treated and control counties would have followed similar trajectories. I evaluate these assumptions in Section~\ref{sec:diffrobustness}.

\subsubsection{Classifying Investors}\label{sec:identifyinvestors}
To measure the transaction share of algorithmic investors, I identify real estate investors and classify them as algorithmic or human. Because corporate structures provide legal and tax advantages, corporate ownership is a reliable proxy for investment activity. Following prior work, I define investors as corporate entities that purchase residential properties for rental or resale purposes \citep{RedfinInvestorHomePurchases, millsLargeScaleBuy2019}. I consolidate related subsidiaries by fuzzy matching on name and using the geographic proximity of corporate mailing addresses. Then, I exclude known non-investor corporate entities, such as government agencies, financial institutions acquiring properties through foreclosure, religious or nonprofit organizations, timeshare operators, homeowner associations, corporate relocation services, farms, builders, hotels, and vacation rental companies.

I classify investors as algorithmic-augmented or human based on observable features of their acquisition strategies and workforce. Each corporate entity is manually classified using business registrations, public filings, media coverage, company websites, interviews, and LinkedIn profiles. Firms are classified as algorithmic if they explicitly disclose using ``automated acquisition engines'' or ``predictive valuation models,'' or if they employ a dedicated data science or machine learning team focused on acquisitions. Because the use of machine learning in real estate investing required specialized expertise in the early 2010s, workforce composition provides a valuable complement to acquisition strategy disclosures. Firms without evidence of such technologies or technical staff are classified as human-led. Classifications are made at the firm level and remain fixed over time, as no firms in the sample switch between human-led and algorithmic strategies during the study period.

Using this procedure, I identify 35 algorithmic investors responsible for around 122,000 home purchases in the sample. The largest include Amherst Residential, Invitation Homes, and Progress Residential. Table \ref{tab:sumstat_houses} shows that, relative to human investors, algorithmic investors tend to purchase slightly more expensive homes (\$229,000 vs. \$223,000), with more bedrooms (2.7 vs. 2.2) and newer construction (average age 22 vs. 42 years). Appendix Table \ref{atab:balance_county_inv} shows Algorithmic investors buy in areas with higher educational attainment (70\% vs. 51\% with a BA or higher), higher occupancy rates (92\% vs. 84\%), and lower poverty rates (8\% vs. 15\%). Appendix Figure~\ref{afig:mapsactivity} illustrates that algorithmic investors operate in 55\% of sample counties, while human investors are active in all counties.

\subsubsection{Assessing the Investor Classification with Geographic Proximity}
The investor classification strategy raises two concerns. First, could some investors classified as human actually be using machine learning technology? Second, might algorithmic investors overstate their reliance on algorithms? To assess classification validity, I draw on an observable difference between human and machine prediction: human investors depend on local knowledge and typically need to invest near their headquarters to physically evaluate properties, whereas algorithmic investors rely on structured data and predictive models, enabling them to assess properties across geographically dispersed markets without physical proximity \citep{fields2018a}. Geographic differences in acquisition patterns between the two groups therefore provide a test of whether the classification captures distinct investment strategies.

Appendix Table \ref{atab:investor_buying_chars} summarizes the geographic distribution of investment activity by classified investor type. Only 3\% of algorithmic purchases occur in the same zip code as the firm’s headquarters, compared to 35\% for human investors. Similarly, 80\% of human investor purchases are in-state, versus just 21\% for algorithmic investors. Geographic scope also differs markedly: Human investors are active in an average of 1.5 zip codes per firm, while algorithmic investors operate across 86. If algorithmic investors relied primarily on manual sourcing by local teams, their activity would likely be more geographically concentrated. Likewise, if many human-led firms were quietly using algorithms, the differences in geographic reach and scale would be smaller.

\section{Digitization Leads to Algorithmic Investment}

\subsection{County Digitization and Algorithmic Investor Entry}
\begin{figure}[ht!]
\begin{center}
\captionsetup{justification=centering}
\caption{\textsc{Figure \ref{fig:raw_nhouses}: Algorithmic Investment, by Time to County Digitization}}
\makebox[\linewidth]{
\begin{tabular}{c}
\textsc{A. Algorithmic Investment, Raw Data} \\
\includegraphics[scale=0.35]{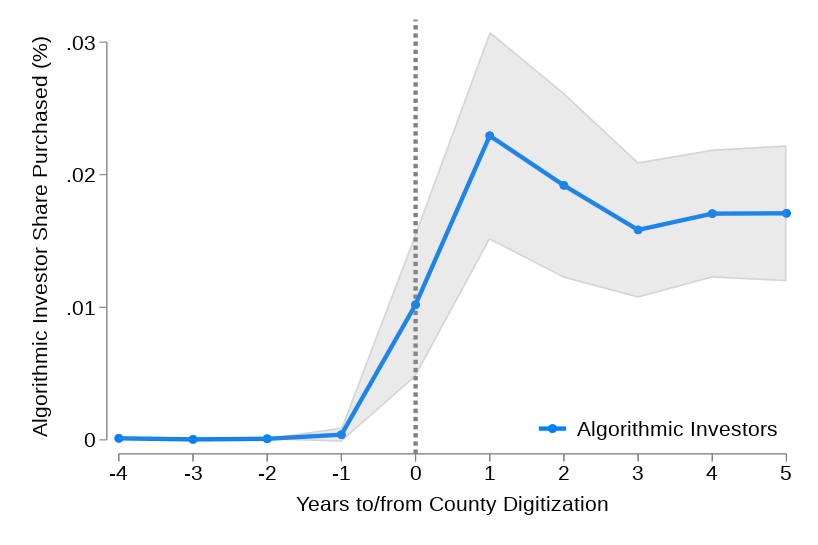} \\
\textsc{B. Effect of County Digitization on Algorithmic Investment, Event Study} \\
\includegraphics[scale=0.35]{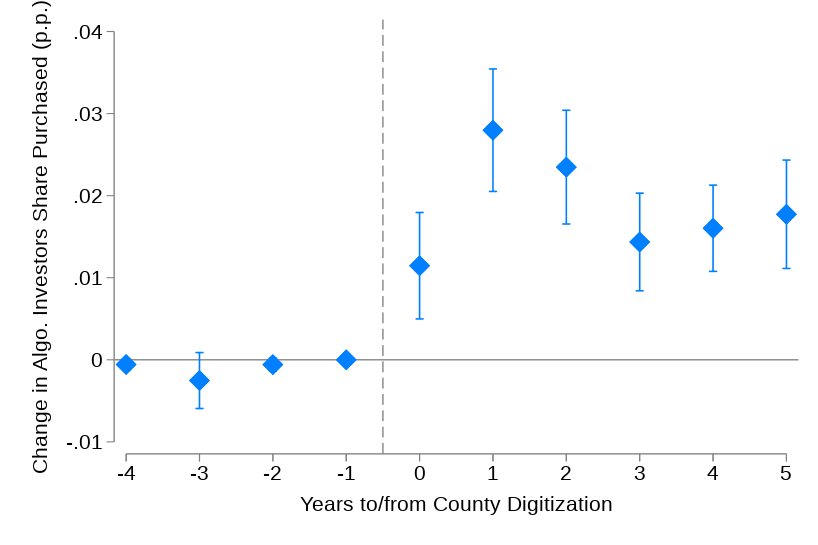} \\
\end{tabular}
}
\label{fig:raw_nhouses}
\end{center}
\end{figure}

Digitization is associated with a discrete and sustained increase in algorithmic investment. Panel A of Figure~\ref{fig:raw_nhouses} plots the county-year share of home purchases by algorithmic investors relative to the year of digitization. Algorithmic activity is negligible prior to digitization. In the year of digitization, their purchase share increases sharply to approximately 1\% of all homes sold and reaches about 2\% within three years. This increase is economically meaningful: a 2\% share of total transactions corresponds to 20\% of all investor purchases.\footnote{The average investor share post-digitization is 11.7\%, with algorithmic investors accounting for 2.3\%.} The sharp discontinuity in entry coinciding with digitization is consistent with algorithmic investors responding to reduced data access costs, rather than to unobserved factors that would be expected to evolve smoothly around the time of digitization.

Panel B of Figure~\ref{fig:raw_nhouses} presents the corresponding event study estimates, which closely mirror the patterns in the raw data. The algorithmic share of transactions increases by about one percentage point in the year of digitization and stabilizes at roughly two percentage points thereafter, compared with a pre-treatment share near zero. Table~\ref{tab:es_main} reports the full set of event-time coefficients, using \citet{dechaisemartin2024} to estimate Equation~\ref{eq:es_main}. Column (1) includes county and year fixed effects and county population. Column (2) adds demographic characteristics: racial composition (shares White, Black, and Hispanic), average family size, share of children, and share of adults with a bachelor’s degree or higher. Column (3) adds socioeconomic variables: unemployment rate, labor force participation, median household income, poverty rate, and share of residents receiving disability benefits. Column (4) further includes median rent, share of owner-occupied homes, and share of vacant homes listed for sale. Estimated treatment effects are quantitatively similar across all specifications.

Appendix Table \ref{atab:es_main_activity} examines the extensive margin, estimating the effect of digitization on algorithmic investor activity. Activity is defined as an indicator for whether any algorithmic investor purchases more than 10 houses in a county-year, which helps avoid capturing classification error. Column (1) shows that digitization is associated with a 30 percentage point increase in activity in the year of digitization, rising to about 53 percentage points three years later, from a baseline of near zero. Results in Columns (2)--(4), which include additional controls, are quantitatively similar.

The estimated effect of digitization on algorithmic investment is similar across alternative difference-in-differences estimators. Appendix Figure~\ref{afig:es_lnq_alts} presents event study estimates from  \citet{borusyak2022revisiting}, \citet{sun_estimating_2020}, and \citet{callaway_2021}, alongside the traditional two-way fixed effects model. Appendix Table~\ref{atab:dd_main_robust} reports the corresponding average treatment effects on the treated (ATT), aggregated across treatment cohorts. All methods show that digitization produces large, sustained increases in algorithmic investment, raising the share of homes purchased by algorithmic investors by about 2 percentage points. 

\subsection{Robustness}

\subsubsection{Exploiting Within-County, House-Level Digitization} \label{sec:diffrobustness}
Although county digitization timing is uncorrelated with observable characteristics, unobserved county-level shocks could influence both digitization and investment. For example, if a county digitizes as part of a broader modernization effort following a major manufacturing investment, and that investment also attracts algorithmic buyers, then increases in algorithmic activity may reflect underlying economic growth rather than a causal effect of digitization.  To address this, I exploit bureaucratic frictions in house record digitization to separate the causal effect of data availability from broader county-level trends or shocks. Because scanning thousands of records took time, counties typically digitized properties in batches, so some homes were added to the electronic database earlier than others. This process generated variation in data accessibility for otherwise similar homes within the same county: for digitized properties, information such as bedroom count and prior sale price was readily available online, whereas for not-yet-digitized properties, obtaining the same information often required an in-person visit.

\subsubsection{The Rollout of House Digitization} \label{sec:rollouthouse}
The timing of house-level digitization appears uncorrelated with observable house or neighborhood characteristics. Appendix Figure~\ref{afig:housedigit_balance} tests whether baseline characteristics predict early digitization, controlling for county fixed effects. Across a range of house-level variables such as sale price, age and indicators of financial distress and neighborhood characteristics, coefficients are small and statistically insignificant. The strongest predictor, house age, implies that a one standard deviation increase is associated with a 3 percentage point increase in the likelihood of early digitization within a county.

\subsubsection{Triple Difference Analysis and Falsification Tests}\label{sec:triplediff}
I estimate a triple-difference specification comparing digitized and not-yet-digitized houses within the same county, before and after county digitization:

\begin{align}
y_{dct} 
&= \alpha_c + \delta_t 
+ \sum_{j \ne -1}^{J} \left[ \beta^{\text{Digit}}_j \cdot \mathds{1}[d = 1] + \beta^{\text{NoDigit}}_j \cdot \mathds{1}[d = 0] \right] \cdot \mathds{1}[\tau_{ct} = j] 
+ X_{ct}'\gamma + \varepsilon_{dct}
\label{eq:triple_diff}
\end{align}

The dependent variable \( y_{dct} \) is the share of homes in digitization group \( d \in \{0,1\} \), county \( c \), and year \( t \) that were purchased by algorithmic investors. The coefficients \( \beta^{\text{Digit}}_j \) capture the changes in algorithmic investment for digitized homes in event time \( j \) relative to the year prior to digitization, while \( \beta^{\text{NoDigit}}_j \) measure effects for homes that have not yet been digitized, serving as a falsification test. 

\begin{figure}[ht!]
\begin{center}
\captionsetup{justification=centering}
\caption{\textsc{Figure \ref{fig:es_triplediff}: Algorithmic Investment, by County and House Digitization Status}}
\makebox[\linewidth]{
\begin{tabular}{c}
\textsc{A. Algorithmic Investment by House Digitization, Raw Data} \\
\includegraphics[scale=0.35]{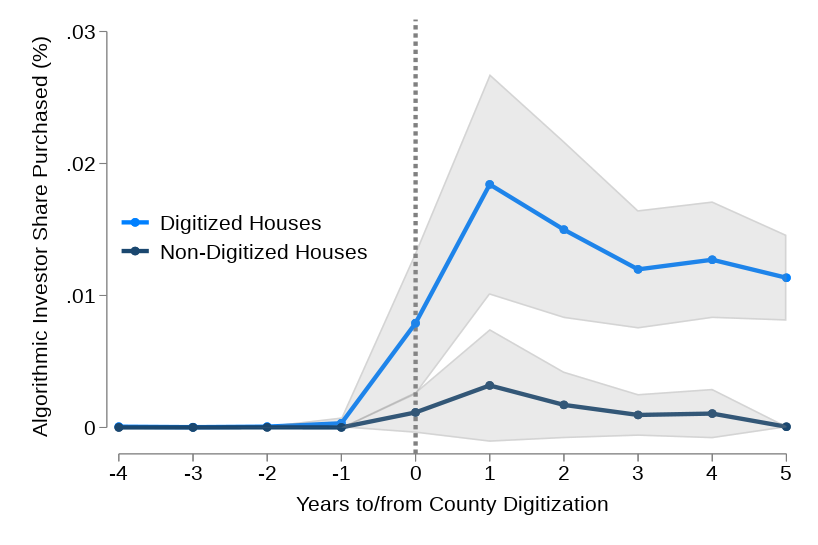} \\
\textsc{B. Effect of County Digitization on Algorithmic Investment, Event Study} \\
\includegraphics[scale=0.35]{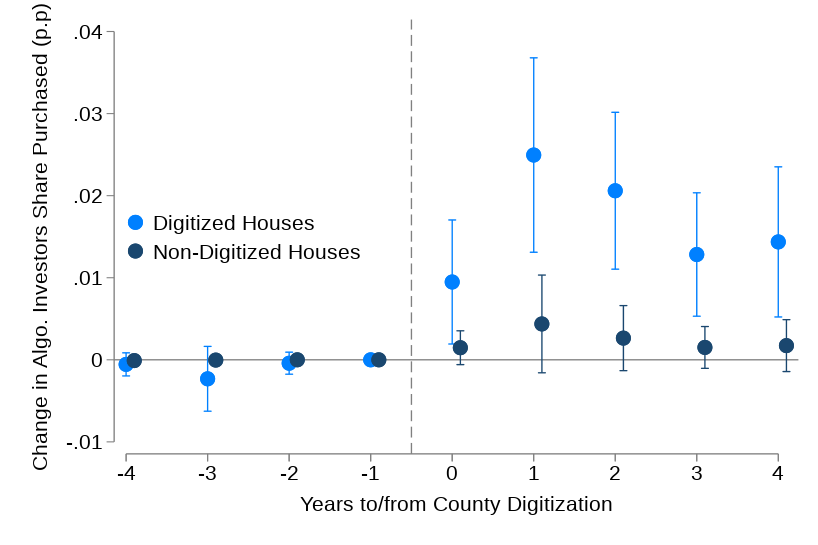} \\
\end{tabular}
}
\label{fig:es_triplediff}
\end{center}
\end{figure}

Figure~\ref{fig:es_triplediff} Panel A plots algorithmic investor share by event time, separately for digitized and not-yet-digitized homes. Panel B presents corresponding event-study estimates and 95\% confidence intervals. Algorithmic investment rises sharply for digitized homes following digitization, while investment in not-yet-digitized homes remains minimal. This divergence suggests that algorithmic investment is not driven by county-level shocks such as economic growth, which would affect all houses regardless of digitization status. 

To address the possibility of more localized unobserved shocks, I conduct a second falsification test that compares investment across digitized and not-yet-digitized homes within neighborhoods:

\begin{equation} \label{eq:fs_triple_diff_falsification}
y_{icbt} 
= \alpha_b + \delta_t  
+ \beta^{\text{Digit}}  D_{ct}  D^{\text{house}}_{it}
+ \beta^{\text{NoDigit}} D_{ct} \left(1 - D^{\text{house}}_{it} \right)
+ X_{ibt}'\gamma + \varepsilon_{icbt}
\end{equation}
$y_{icbt}$ is an indicator equal to $1$ if house $i$ in county $c$ in census block group $b$ and year $t$ was purchased by an algorithmic investor. $X_{ibct}$ is a vector of pre-determined house characteristics and includes neighborhood ($\alpha_b$) and year fixed effects ($\delta_t$). The regression is estimated with OLS, and standard errors are clustered at the neighborhood. 

Table~\ref{tab:triple_diff_falsification} reports the results. County digitization leads to algorithmic investment in digitized houses but not in nearby not-yet-digitized houses.  The coefficients on \textit{County Digitized} $\times$ \textit{Not-Yet-Digitized House} are statistically indistinguishable from zero in Columns 1--3. The 95\% confidence intervals rule out increases larger than 0.14 percentage points. In contrast, \textit{County Digitized} $\times$ \textit{Digitized House} implies county digitization leads to a 1.2 percentage point average increase in algorithmic investment in digitized houses.

\subsubsection{Additional Verification of Investor Classification Using House--Level Digitization}\label{sec:additionalclassification}
Table~\ref{tab:triple_diff_falsification} also supports the validity of the investor classification procedure. First, algorithmic investors respond strongly to the availability of digitized data, even when comparing nearby homes, consistent with reliance on data-driven models rather than manual human strategies. Columns 4–6 of Table~\ref{tab:triple_diff_falsification} present a falsification test: since county digitization should not increase human investment. Indeed, county digitization has no effect or small negative effects on human investment.

Additional support comes from the house-level analysis in Appendix Table~\ref{atab:housedigit_fs_byrace}. This specification replaces county-level digitization indicators with a measure of whether a given home is digitized, conditioning on block group-by-year fixed effects. Column (1) shows that digitized homes are 1.5 percentage points more likely to be purchased by an algorithmic investor than by a human investor or owner-occupier, a statistically significant 83\% increase. Column (2) confirms that this effect is specific to algorithmic investors by restricting the sample to investor transactions. Column (4) presents a falsification test, again finding no effect for human investor purchases.

\subsubsection{Simulation Evidence on Cross-Country Prediction Accuracy}\label{sec:robustness:sutva}
A key assumption is the Stable Unit Treatment Value Assumption (SUTVA), which requires that digitization in one county does not affect outcomes in counties that have not yet digitized. This assumption could be violated if models trained on digitized records in County A accurately predict in County B, allowing algorithmic investors to enter County B before it digitizes.

To assess the plausibility of such cross-county spillovers, I simulate the transferability of house price prediction models across counties. For each of 400 counties, I use pre-digitization transactions to train a gradient-boosted tree model on the natural log of sale price, then evaluate its performance both within the training county and in a randomly selected different county. I run this simulation repeated 100 times per county.

Appendix Figure~\ref{afig:outofsample_countydatasim} shows that out-of-county prediction performance deteriorates substantially. In dollar terms, within-county models yield an average prediction error of about \$20,000 (10\% of sale price), while out-of-county errors rise to nearly \$80,000 (39\% of sale price), with a right-skewed distribution where some out-of-county accuracy is substantially worse. These simulation results suggests that the institutional differences across counties mean that digitized data in one country does not generalize well.

\subsubsection{Effects of Digitization on Human Investors}\label{sec:robustness:humaninvs}

Digitization could impact more than algorithmic investment by lowering search costs, improving transparency, or shaping decisions by human investors. To see if there is evidence of this, I examine the impact of county digitization on human investor activity, average sale price, transaction volume, median days on market, and the sale-to-list price ratio. Panel A of Appendix Table~\ref{atab:county_falsification} reports OLS difference-in-differences estimates for the 45\% of counties where algorithmic investors never enter. 

However, algorithmic investors may avoid weaker housing markets, potentially biasing these OLS estimates. To address this, Panel B of Appendix Table~\ref{atab:county_falsification} presents 2SLS estimates that instrument for algorithmic entry using pre-digitization data quality: the share of assessor records with complete information on bedrooms, bathrooms, stories, and building count. This variable proxies for the usefulness of digitized data to algorithmic models and reflects long-standing differences in local record-keeping that are plausibly exogenous to market quality. Appendix Section~\ref{asec:instrumentingentry} provides additional discussion.

Results from both OLS and 2SLS show no detectable effect of digitization on market outcomes in counties without algorithmic entry. This interpretation is consistent with qualitative evidence from interviews and disclosures, which indicate that human investors do not rely heavily on digitized records when making decisions.

\section{Prices and Allocation Results} \label{sec:mainresults}
\subsection{Conceptual Framework} \label{sec:conceptualframework}

This section outlines a conceptual framework, informed by prior work on human and algorithmic decision making \citep[e.g.][]{mullainathan_diagnosing_2022, kleinberg2017, rambachanIdentifyingPredictionMistakes, li_hiring_2025}, to frame how the shift from human to algorithmic prediction may affect prices and allocations in the housing market.

\subsubsection{Setup}
Suppose each house can be represented by a vector of characteristics $(x,z)$ , which include observable characteristics $x \in \mathbb{R}^d$ (e.g., square footage, bedrooms, neighborhood demographics) and $z \in \mathbb{R}^k$ captures unobservable or hard-to-quantify attributes (e.g., interior finishes, ambient noise, odor from a nearby farm). Each house has some underlying common value, denoted by $Y$, where $ Y = f(x, z) + \varepsilon\ $, $f(x, z)$ is the underlying true optimal prediction function based on $(x,z)$ that relates house characteristics to value, and $\varepsilon$ is a mean-zero error term independent of $(x,z)$ capturing idiosyncratic variation. \footnote{Optimal prediction does not imply perfect accuracy; some prediction error is due to irreducible error \citep{hastie2009elements}.}

\subsubsection{Human Prediction}
Humans often have access to a rich set of information about a house, including the observable characteristics \(x\) (e.g., number of bedrooms) as well as hard-to-quantify or unobservable attributes captured in \(z\), such as noise level in the yard. However, using all this information requires cognitive effort: When valuing a property, individuals must weigh and compare the many differences across houses. For example, how should recent kitchen renovations be weighed against the need for a new roof? Prior research suggests that humans frequently rely on heuristics when making such complex decisions. For example, \citet{mullainathanobermeyer2022} show that physicians often default to simplified diagnostic rules that overweight salient features, such as chest pain, when predicting heart attacks.

To capture both the richness of human-perceived information and the possibility of human cognitive limitations, I model the human prediction function $h(x, z)$ as:
\[
h(x, z) = f(x, z) + \Delta(x, z),
\]
 where \(\Delta(x, z)\) is a statistical bias term that captures systematic deviations in human predictions from the optimal benchmark. These may arise from cognitive limitations, reliance on heuristics, or behavioral biases, and may vary with with property characteristics or other contextual factors. For instance, buyers may overvalue amenities like pools in warm weather but undervalue them in colder climates \citep{busse2012}.\footnote{ While these human errors may vary across houses, I will write $ \Delta(x, z)$ as $\Delta$ for notational simplicity.}

\subsubsection{Algorithmic Prediction}
In contrast, algorithmic prediction relies on supervised learning models that are trained on historical data to minimize a predefined loss function, such as mean squared error. The algorithmic prediction function  $m(x)$ can be written as:
\[
m(x) = \mathbb{E}[f(x, z) \mid x],
\]
which represents the best approximation of the true value function conditional on the observable covariates \(x\). Unlike humans, algorithms are not subject to behavioral biases, such as assigning different values to a house based on the weather at the time of viewing. However, algorithms are limited by the data on which they are trained and cannot access the unobserved characteristics in \(z\).

\subsubsection{Comparing Human and Algorithmic Predictions}
For any house $(x,z)$, the difference between human and algorithmic prediction can be written as:
\begin{equation} \label{eq:tradeoff}
    \underbrace{m(x) - h_i(x,z)}_{\text{algo-human difference}} = \underbrace{E[f(x,z)|x] - f(x,z)}_{\text{ information gap}} - \underbrace{\Delta}_{\text{human error}}
\end{equation}

Equation~\eqref{eq:tradeoff} shows that disagreement between human and algorithmic predictions arises from two sources: 1) an information gap, reflecting that algorithms cannot observe all the house characteristics available to humans and 2) the human error term that captures how human cognitive limitations  may cause human predictions to diverge from the optimal benchmark. The decomposition highlights a tradeoff. If the unobserved information $z$ is important to estimating $Y$, humans may outperform. However, if there are houses where human judgment introduces systematic and substantial mistakes, algorithms may have an advantage.  

Modeling this as a tradeoff is motivated by the empirical finding that digitization affects algorithmic investor entry. If humans had an information advantage and made no systematic errors, human prediction would always outperform algorithms, and algorithmic investors would not enter the market. Conversely, if human prediction were cognitively constrained and lacked any informational advantage, algorithms would always outperform humans, and digitization would drive human investors out entirely. Instead, digitization leads algorithmic investors to account for about 20 percent of the investor market while human investors remain active, suggesting that the relative advantage of human and algorithmic prediction varies.

\subsubsection{Empirical Implications}
This conceptual framework highlights two ways the shift from human to algorithmic prediction might impact prices and allocations. The first implication concerns houses subject to systematic human errors. If a set of houses were consistently undervalued by humans mistakes these properties could offer an arbitrage opportunity for algorithmic investors, whose predictions are not subject to the same distortions.\footnote{This implication is consistent in spirit with the \citet{becker1957} model of taste-based discrimination.} Algorithmic entry should lead to a narrowing of the relative price gap between systematically undervalued houses and comparable properties. Also, algorithmic investors should be disproportionately more likely to purchase these undervalued homes. These predictions differ from the effects of a standard demand shock, which would raise the price of all homes uniformly, rather than narrowing the relative price gap. I explore these implications in the context of racial bias in Section \ref{sec:racepenalty}.

The second implication for market prices and allocations concerns the informational gap between humans and algorithms. In the U.S. housing market, the features in $x$ are shaped by county-level decisions about data collection and reporting, as well as the characteristics of the existing housing stock. These baseline differences create heterogeneity in the relative advantage of algorithmic versus human prediction across properties. Consequently, the effects of digitization on investment should depend on how well the available data capture the key determinants of housing value. Houses with high-quality, informative data should see larger increases in algorithmic investment after digitization. In contrast, when available data are sparse or uninformative, digitization may have limited effects or even shift human investment toward properties where their informational advantage is greater.\footnote{This pattern is consistent with the predictions of a Roy model, in which individuals sort into sectors where their comparative advantage yields the highest returns \citep{roy1951}.} I examine the price and quantity implications of this specialization in Section~\ref{sec:specialization}.

\subsection{The Race Penalty}\label{sec:racepenalty}
\subsubsection{Background on the Race Penalty}
Persistent racial and ethnic disparities have been well documented in the U.S. housing market. Homes in majority-Black neighborhoods are valued significantly lower than comparable homes in predominantly White areas \citep{harris1999, perryDEVALUATIONASSETSBLACK, elsterMinoritiesPropertyValues2022}.\footnote{Proposed mechanisms include preferences for living near similar people, differences in local amenities, risk premiums, and structural barriers such as redlining and discriminatory lending \citep{lewisracialcomposition, caseyRaceEthnicitySocioeconomic, tessumairpollution2021, cutlerRiseDeclineAmerican1999}.} These gaps persist even when comparing two similar houses in the same neighborhood. Repeat-sales evidence shows that Black sellers earn lower annualized returns, receive less in distressed sales, and obtain lower appraisals, even after adjusting for home and neighborhood characteristics \citep{drukker2024racial, kermaniRacialDisparitiesHousing2021, freddiemac2021, perryBiasedAppraisalsDevaluation2021}.\footnote{On the buyer side, \citet{box-couillard_racial_2024} and \citet{bayerRacialEthnicPrice2017} find that minority buyers often pay a premium when purchasing from White sellers, even for the same property.} Audit studies also suggest that race can influence valuation, even absent overt discrimination. For instance, appraisers seem to select different sets of comparable homes depending on the actual or perceived race of the homeowner \citep{lilienjakeFaultyFoundationsMysteryShopper, howell_appraised_2022}.\footnote{Research on implicit bias in housing valuation has coincided with media coverage and legal cases, including reports of minority homeowners “whitewashing” their homes to obtain higher appraisals \citep{kamindebraHomeAppraisedBlack2023, howell2018NeighborhoodsRA}.} This body of evidence suggests that race shapes how information is used in human valuation, making it a natural lens for examining how the shift from human to algorithmic prediction may alter pricing dynamics.

\subsubsection{Evidence on the Race Penalty}
Motivated by this prior work, I document the existence of a race penalty prior to county digitization. To account for unobservable differences in amenities or risk premiums across neighborhoods, I restrict the analysis to \textit{within-neighborhood} comparisons, comparing two observably similar houses in the same census block group.

I estimate the \textbf{race penalty}  using a hedonic regression in Equation \ref{eq:racepenalty}: 

\begin{equation} \label{eq:racepenalty}
\ln(p_{ibt}) = 
\beta_M \text{Minority}_{it} + \mathbf{X}_{ibt}'\boldsymbol{\gamma} + \lambda_{bt} + \varepsilon_{ibt}
\end{equation}

where $\ln(p_{ibt})$ is the natural logarithm of sales price of house $i$ in block group $b$ and year $t$. $\text{Minority}_{it}$ equals 1 if the seller for house $i$ is Black or Hispanic in year $t$, and 0 otherwise. The vector $\mathbf{X}_{ibt}$ includes observable housing characteristics, such as square footage, number of bedrooms, age, number of bathrooms, lot size, number of stories, and building type, and block group-by-year fixed effects absorb time-varying local housing market conditions.\footnote{The full list of observable characteristics in $\mathbf{X}_{ibt}$ is available in Appendix Section \ref{asec:housechars}.} The race penalty, $\beta_M$, captures the average sales price gap between minority and White homeowners, conditional on observed home characteristics and local price trends. 

Seller race is inferred from first and last names combined with geographic location, using two approaches: Bayesian Improved Surname Geocoding (BISG) and a machine learning method (\texttt{Ethnicolr}), both described in Appendix Section~\ref{asec:inferringrace}. The two approaches yield similar results; I use Ethnicolr in the main specification, as it produces slightly more conservative estimates of the race penalty.

Table~\ref{tab:racepenalty_by_geo_ethnicolr} shows that, prior to digitization, minority-owned homes sell for less than observably similar homes owned by White sellers. Within the same Census tract (about 2,000 households), the estimated race penalty is 7 log points. The gap narrows to 5.8 log points at the Census block group level (about 700 households). Although the full sample cannot be geocoded to the block level, I calculate the penalty for the subsample with block-level matches (about 30 households), finding a gap of 2.9 log points. The gap is statistically significant at all geographic levels and economically meaningful; A 5.8 log point gap corresponds to \$11,958 and a 2.9 log point gap corresponds to \$6,010.  Appendix Table~\ref{atab:racepenalty_by_geo_bisg} reports similar magnitudes and patterns when seller race is inferred using BISG.

The conceptual framework suggests two possible explanations for the observed price disparity. First, the gap may reflect unobserved housing characteristics such as interior condition, noise, or yard maintenance that are visible to human buyers but not captured in the data and are correlated with minority ownership. Second, the disparity could result from biases correlated with the seller’s race. Because the homeowner’s race changes upon sale, any race-related price discount implies a temporary mispricing rather than a persistent difference in house quality. To distinguish between these explanations, I analyze how digitization affects the magnitude of the race penalty.

\subsection{The Impact of Digitization on Racial Disparities} \label{sec:digitonracepenalty}
County digitization reduces the race penalty by nearly half. In the preferred specification (Column 2 of Table~\ref{tab:racepenalty_by_geo_ethnicolr}), which includes Census block group fixed effects, the gap narrows by 2.6 log points, from 5.8 to 3.2 log points, equivalent to a reduction of 45\%. The coefficient on ``Minority Seller $\times$ County Digitization'' measures this differential effect on sale prices received by minority homeowners.\footnote{Because geography-by-year fixed effects absorb the average effect of digitization, this interaction term identifies how digitization changes the relative outcomes for minority sellers.} Across all specifications, the positive and statistically significant coefficients imply a narrowing of roughly 40 percent. Panel A of Figure~\ref{fig:racepenalty_subsampleanalysis} plots the average race penalty before and after digitization. 

Appendix Figure~\ref{afig:residprice} provides a nonparametric illustration of the convergence in residualized sale prices. Panel A plots the distributions for White and minority sellers prior to digitization. A Kolmogorov–Smirnov (K–S) test rejects the null hypothesis that the distributions are the same, with a D-statistic of 0.048. Panel B shows the distributions after digitization, where the K–S D-statistic falls to 0.031, indicating greater similarity between minority and White sellers. 

One possible explanation for the observed price convergence is that digitization reduces informational frictions, such as by improving transparency around home values, which in turn narrows the race penalty. Panel B of Figure~\ref{fig:racepenalty_subsampleanalysis} shows that the race penalty declines significantly only in counties where algorithmic investors enter after digitization ($p < 0.001$), while the change is smaller and statistically insignificant in counties without such entry ($p = 0.328$).\footnote{Appendix Table~\ref{atab:racepenalty_coeff_tests} reports coefficient test results.} This pattern suggests that digitization alone is not sufficient to reduce racial disparities; rather, the reduction appears linked to algorithmic investment.

\subsection{Mechanism for Reduction in Racial Disparities}
\subsubsection{Direct Effects of Algorithmic Investors}
Algorithmic investors reduce race-based pricing disparities through both their pricing behavior and their selection of properties. Panel C of Figure~\ref{fig:racepenalty_subsampleanalysis} shows that, among homes purchased by algorithmic investors, who enter only after county digitization, the race penalty is statistically indistinguishable from zero. In other words, algorithmic investors pay similar prices for observably similar homes regardless of the seller’s race. 

Second, algorithmic investors purchase disproportionately from minority homeowners. In counties where they are active, algorithmic investors account for about 4\% of sales by minority homeowners, roughly twice their overall transaction share. These purchases are not limited to racially segregated areas; the average algorithmic investment occurs in a majority White Census tract, indicating that the pattern is not simply driven by neighborhood racial composition. The effects of digitization on algorithmic investment are also stronger for minority-owned properties. Column 3 of Appendix Table~\ref{atab:housedigit_fs_byrace} shows that digitization leads to a 53\% larger increase in algorithmic purchases for homes owned by minority sellers relative to observably similar homes owned by White sellers in the same Census block group. These quantity patterns are consistent with the conceptual prediction that algorithmic investors, facing an arbitrage opportunity, would target undervalued homes. However, algorithmic investors represent a small share of overall transactions, so these price and quantity effects alone cannot account for the 45\% reduction in the race penalty.

\subsubsection{Indirect Effects of Algorithmic Investors on Racial Disparities}
The housing market is thin, highly intermediated, and subject to binding short-run supply constraints, making it susceptible to spillovers from demand shocks. Prior research shows that even indirect shocks can propagate through local markets via buyer behavior, visual appeal, or nearby foreclosures.\footnote{For example, \citet{aiello_when_2022} show that wealth shocks to homebuyers affect the price of their next purchase and spread through markets. \citet{glaeser_computer_2018} find that a one–log-point increase in the visual appeal of neighboring homes raises a property’s value by 0.4 log points. \citet{cohen2016sales} document that foreclosures reduce nearby home values by about 2\%.} If algorithmic investors influence the behavior of other buyers, they could generate market-wide effects that extend beyond their own transactions.

The race penalty shrinks among human investors and owner occupiers in markets where algorithmic investors are active. Panel C of Figure~\ref{fig:racepenalty_subsampleanalysis} shows that, following county digitization, the race penalty declines by about 50 percent for owner-occupiers ($p < 0.0001$) and by 80 percent for human investors ($p = 0.044$).\footnote{See Appendix Table~\ref{atab:racepenalty_coeff_tests}.} Panel D confirms that this pattern is limited to counties with algorithmic investor entry. In counties that digitize but do not attract algorithmic investors, the racial price gap shows no statistically significant change. These results suggest that algorithmic entry shifts market dynamics in ways that raise the prices paid by other buyers for minority-owned homes.

Two mechanisms could explain how algorithmic activity reduces the racial penalty among other buyers. First, competition: Even when algorithmic investors do not win a transaction, their bids, or the expectation of their bids, may pressure others to raise their offers. This competitive effect would directly raise sale prices of digitized, minority-owned homes exposed to algorithmic investor competition. Second, spillovers through price anchoring: Algorithmic purchases may affect the reference points used to set prices. Buyers, agents, and appraisers rely heavily on comparable recent transactions, so higher prices paid by algorithms could lift valuations for similar nearby homes, even if those homes are not themselves targeted by algorithmic investors.\footnote{This could occur through behavioral anchoring or institutional mechanisms such as appraisal standards.}

To separate these channels, I use the staggered rollout of digitization at both the house and county level to estimate a triple-difference regression among human investors and owner occupiers purchases. Algorithmic investors concentrate their activity in digitized homes within digitized counties and rarely buy not-yet-digitized homes. Asa result, house digitization creates quasi-random variation in exposure to algorithmic bidding across houses within counties and years.\footnote{This interpretation assumes that the absence of algorithmic investment in non-digitized homes reflects lower exposure to algorithmic bids, conditional on year, house type, and neighborhood.} My regressions specification interacts minority seller status with house-level and county-level digitization, comparing changes in the racial price gap for digitized versus non-digitized homes before and after county digitization. A larger decline in the race penalty for digitized homes would indicate competition as the primary channel, while a stronger effect for non-digitized homes would point to price anchoring. Appendix~\ref{asec:triplediff_racepenalty} provides additional details on the specification.

Direct exposure to algorithmic bidding narrows the racial price gap by nearly half in human investors and owner occupiers transactions. Column (1) of Appendix Table~\ref{atab:racepenalty_mechanism} shows that the coefficient on \textit{County Digitized} $\times$ \textit{Digitized House} $\times$ \textit{Minority Seller} is large and statistically significant among owner-occupier and human investor purchases. In contrast, the coefficient on \textit{County Digitized} $\times$ \textit{Digitized House}, which captures spillover effects for non-digitized homes that are harder for algorithmic investors to target, is small and statistically insignificant. This indicates that algorithmic presence alone does not improve outcomes for minority sellers unless their properties are directly targeted.

Columns (2) and (3) show similar patterns in subsamples of owner-occupiers and human investors. For owner-occupiers, the narrowing of the race penalty is concentrated in digitized homes, consistent with competition as the primary channel. For human investors, the point estimate on the spillover coefficient is larger than the competition effect, although statistically insignificant, suggesting that price anchoring may play a greater role for this group. Overall, these results show that algorithmic investors influence not only the prices they pay but also the broader competitive environment, meaning their market impact cannot be inferred from transaction share alone.

\subsubsection{Decomposition of Race Penalty}
\begin{figure}[ht!]
\begin{center}
\captionsetup{justification=centering}
\caption{\textsc{Figure \ref{fig:racepenalty_decomp}: Decomposition of the Aggregate Change in the Race Penalty}}
\makebox[\linewidth]{
\begin{tabular}{c}
\includegraphics[scale=0.75]{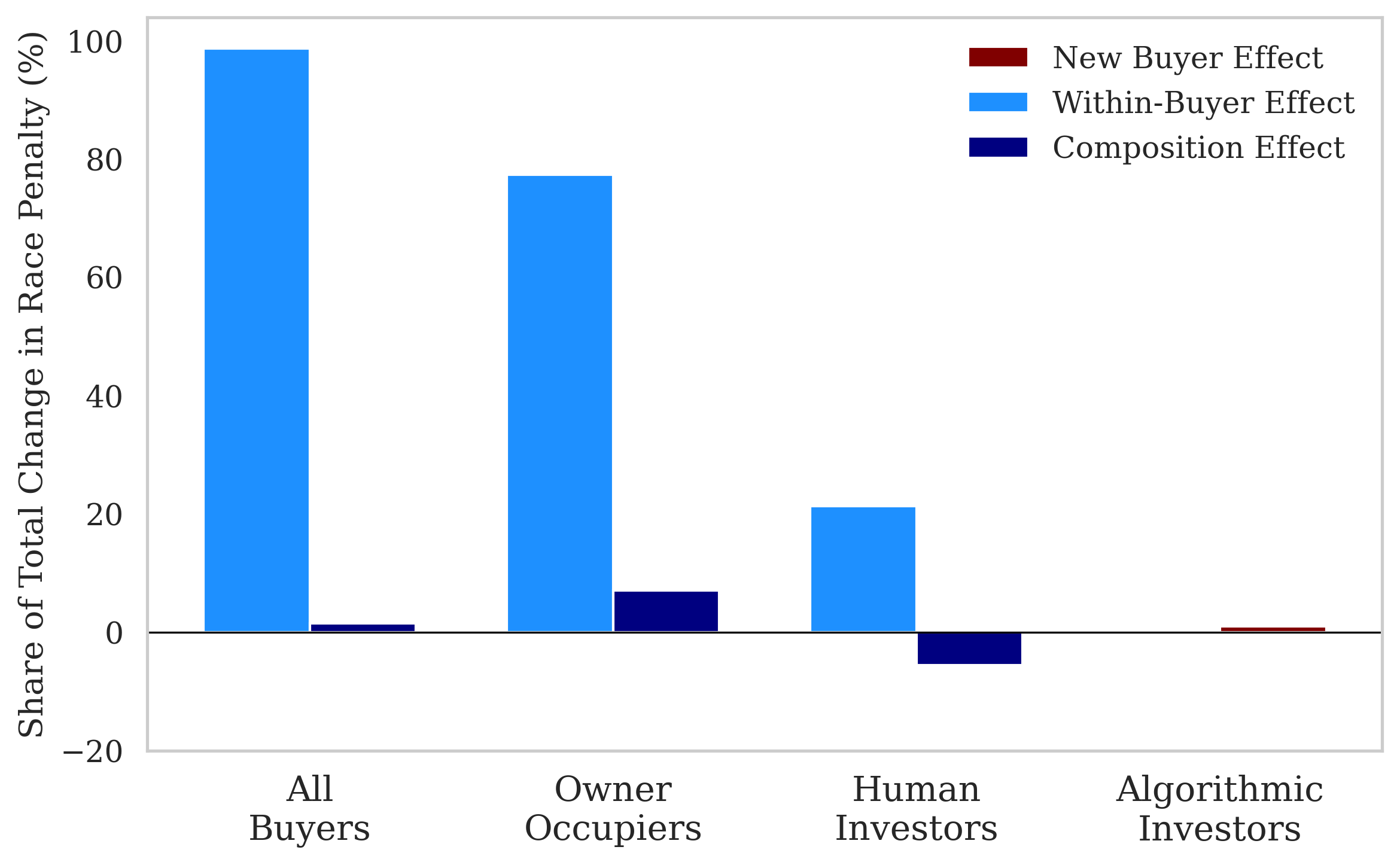}   \\
\end{tabular}
} 
\label{fig:racepenalty_decomp}
\end{center}
\end{figure}

Most of the decline in the race penalty is explained by changes within buyer types rather than compositional changes. Figure~\ref{fig:racepenalty_decomp} shows that 98\% of the reduction is attributable to within-buyer effects, based on an Oaxaca-style decomposition that separates changes in buyer composition from changes in the average race penalty within buyer types.\footnote{Appendix Section~\ref{asec:oaxaca_decom} provides the full derivation.} Algorithmic buyers, who enter only after digitization and exhibit no racial price gap, contribute little to the overall reduction (``the new buyer effect''). Human investors show the largest decline in their own race penalty, but owner-occupiers account for about 80\% of the total drop because they are the bulk of overall transaction volume. Taken together, these results suggest that competitive effects among owner-occupiers explain most of the overall reduction in the race penalty.

\subsection{Adverse Selection or Human Error?}

\subsubsection{Deep Learning Image Analysis} \label{sec:deeplearn}
To assess whether the decline in the race penalty reflects overpayment by algorithmic investors or improved market efficiency, I first test whether controlling for additional information on home quality from interior and exterior images can explain the gap. If homes sold by minority owners are on average less well-maintained due to wealth or liquidity constraints \citep{harris1999}, and this is not captured in structured data, these omitted factors could partly account for the observed price gap.

I match home listings to scraped Zillow images for properties currently listed for sale, yielding 35,000 transactions with matched photos. Appendix Figure~\ref{afig:scrapedimages} provides examples, and Appendix Table~\ref{atab:summary_stats_houseswithimages} compares characteristics of the full and image-matched samples. Visual features are extracted using a Vision Transformer model pre-trained on ImageNet-21k \citep{dosovitskiy2020, rw2019timm}, which generates embedding vectors capturing latent aspects of house appearance. These embeddings are included as controls in the race gap regressions to test whether differences in interior or exterior condition account for the race penalty. Appendix Section~\ref{asec:images} details the image processing pipeline.

Appendix Table~\ref{atab:racepenalty_images} reports the results. Columns 1–2 show baseline estimates of the race penalty for the image-matched sample under different geographic fixed effects. Adding exterior image embeddings (Columns 3--4) modestly reduces the gap from 5.9 to 5.1 log points in the block-group-by-year specification, but a substantial disparity remains. Adding interior embeddings (Columns 5--6) produces little further change. The persistence of the race gap after controlling for both exterior and interior visual features suggests that differences in appearance and condition cannot fully account for the observed disparities in sale price.

\subsubsection{Margin Analysis} \label{sec:margin_analysis}

Visual features capture only one dimension of unobserved home quality and are available for only a subset of the sample. To complement this approach, I examine gross margins to test whether algorithmic investors systematically overpay for homes sold by minority owners. If so, their average returns on these purchases should be lower.

I use two proxies for gross margin. The first is the resale margin, defined as the difference between the natural log of resale price and original purchase price. This measure captures realized returns but is available only for properties that were resold. The second is the assessment margin, defined as the difference between the natural log of the next available tax assessor valuation and the purchase price. This measure covers all properties, since assessments occur regardless of resale, but reflects estimated rather than realized market values.\footnote{For small changes, the log difference approximates the rate of return.} I estimate regressions including block group-by-year, resale or assessment year fixed effects, and observable housing characteristics. Table~\ref{tab:assessresalemargin} shows no statistically significant difference in gross margins by seller race across both measures, different geographic specifications and alternative definitions of minority status. On average, algorithmic investors are not systematically earning lower returns on purchases from minority homeowners.

A potential concern is that algorithmic investors may invest more in post-purchase improvements for homes acquired from minority sellers, which could mask differences in initial quality. To investigate this, Appendix Table~\ref{atab:assessimp} analyzes reported improvement expenditures from tax assessor records. Consistent with prior research, Columns 2, 4 and 6, show that minority owner-occupiers spend less on improvements than White owners, on average \citep{bayer2016_borrowing}. In Columns 1, 3 and 5, however, I find no statistically significant difference in improvement spending among algorithmic investors. This suggests that differences in post-purchase investment cannot explain the similar gross margins observed for homes purchased from minority and White sellers.

\subsection{County Digitization Increases House Sale Prices}
Up to this point, the analysis has focused on relative prices, specifically the racial price gap, by comparing observably similar homes within neighborhoods. I now examine the effect of digitization on average sale prices. 

\begin{figure}[ht!]
\begin{center}
\captionsetup{justification=centering}
\caption{\textsc{Figure \ref{fig:es_logprices}: Effect of County Digitization on Average Sales Prices}}
\makebox[\linewidth]{
\begin{tabular}{c}
\textsc{A. Effect of Digitization on Ln(Sales Price), All Homeowners} \\
\includegraphics[scale=0.35]{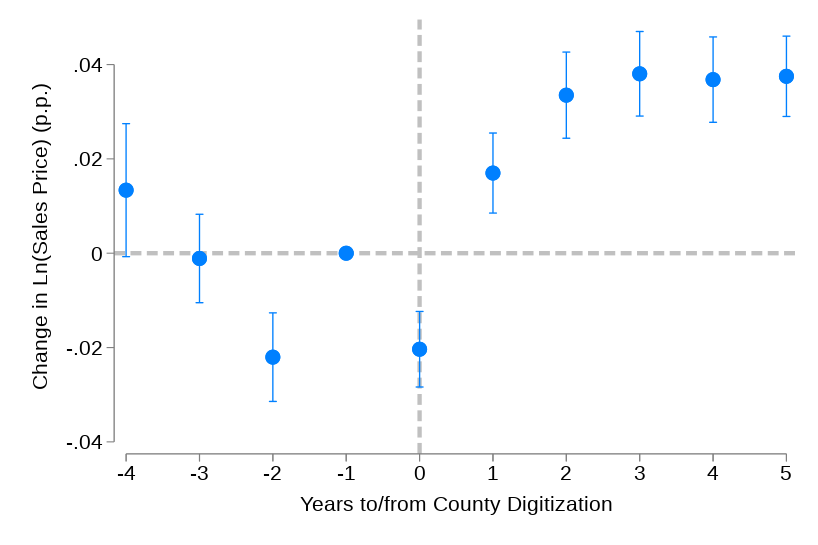} \\
\textsc{B. Effect of Digitization on Ln(Sales Price), by Homeowner Race} \\
\includegraphics[scale=0.35]{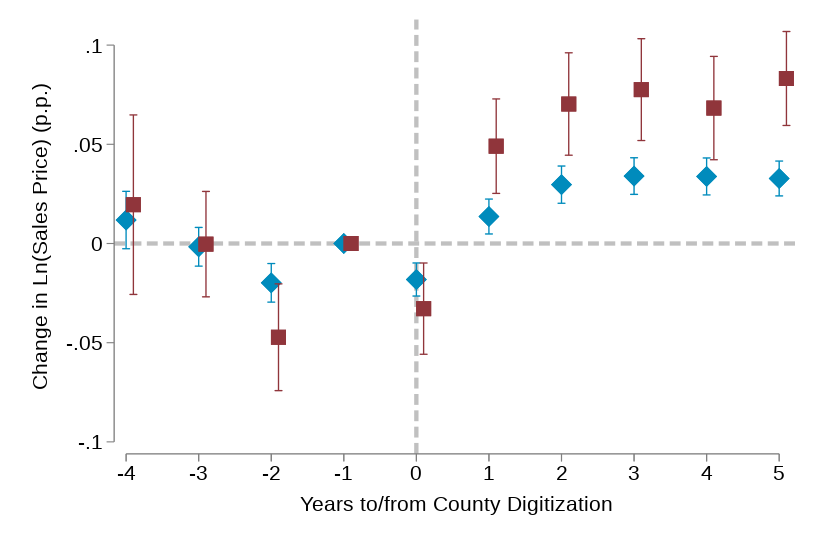} \\
\end{tabular}
}
\label{fig:es_logprices}
\end{center}
\end{figure}

Panel A of Figure~\ref{fig:es_logprices} plots event study estimates of the impact of county digitization on house sale prices, controlling for house characteristics and census block group and year fixed effects. Prices are stable in the pre-digitization period, then begin to rise in the year after digitization. By year three, average sale prices are about 3 log points higher, and the increase persists through year five.

Panel B separates results by seller race. Prices for homes sold by minority owners rise by roughly 5 log points within three years of digitization, compared to an increase of about 2 log points for White-owned homes. These effects emerge in the year following digitization and grow over time. The results indicate that the narrowing of the racial price gap is driven primarily by faster price growth for homes sold by minority owners, rather than price declines for White-owned properties.

Although the percentage gains in sale prices are modest, they translate into large wealth increases for minority sellers. Based on 2019 Survey of Consumer Finances data from \citet{bhuttaDisparitiesWealthRace2020}, the estimated price effects imply a 38\% increase in median wealth for Black households and a 25\% increase for Hispanic households, whose median wealth was \$24,100 and \$36,100, respectively. For the median White household, the same price effect represents only a 2\% gain. These calculations highlight the disproportionate impact of housing appreciation could have on wealth for the typical minority households.\footnote{Figures are based on median household wealth by race overall, not conditional on homeownership.}


\subsection{Digitization and Specialization of Human Investors}\label{sec:specialization}
I now turn to the second empirical implication outlined in Section \ref{sec:conceptualframework}, which concerns how digitization may reshape the behavior of human investors. Appendix Table \ref{atab:county_falsification} shows that digitization does not affect the overall level of human investment, suggesting that algorithmic entry does not crowd out human investors in aggregate.  However, this average effect may mask important heterogeneity. Digitization could lead to changes in the composition of properties purchased by human investors.

\subsubsection{Measuring House Predictability}
To systematically assess changes in composition, I construct a measure of how well a property’s sale price can be predicted from available structured data, which I call baseline \textit{predictability}. This measure serves as a proxy for where human judgment may have an advantage, or where the information gap $E[f(x,z)|x] - f(x,z)$ is large. Simply counting the number of recorded characteristics for a house would fail to capture the varying predictive value of different variables. For example, a single recorded feature such as the number of bedrooms conveys far more relevant information than the number of sheds.

To account for this, I calculate predictability using out-of-sample errors from a supervised machine learning model trained on pre-digitization transactions. To ensure the measure reflects only pre-existing information unaffected by post-digitization price dynamics or algorithmic entry, I restrict the training sample to the pre-digitization periods. The pre-digitization sample is randomly split into training and test sets, and I estimate prices using Extreme Gradient Boosting (XGBoost) \citep{Chen_2016}, a non-parametric, tree-based method well suited to the nonlinear and context-dependent returns to housing characteristics. Appendix Section~\ref{asec:buildingpredictability} describes model training and tuning.

The model predicts prices within 10 percent of the observed value for roughly 90 percent of homes in the held-out test set. Appendix Figure~\ref{afig:xgboost_resids} Panel A shows the distribution of residuals. I define baseline prediction error as the difference between the predicted and actual price in the test set. To capture heterogeneity in both direction and magnitude, potentially reflecting asymmetries in mispricing or arbitrage opportunities, I divide the residual distribution into sextiles. I compute to sets of terciles separately for overpredicted and underpredicted homes. Appendix Figure~\ref{afig:xgboost_resids} Panel B shows residual distributions by sextile.

\subsubsection{Empirical Strategy: Heterogeneity by Predictability}
I use this predictability measure to estimate how digitization affects human investment differentially across properties. The empirical strategy exploits within-county, quasi-random variation in the timing of house-level digitization, which shifts exposure algorithmic investors. I include census block group–by–year fixed effects to absorb both local investment trends and time-varying neighborhood-level shocks.
I restrict the sample to properties that appear in the pre-digitization hold-out test set used to compute predictability. This restriction reduces sample size but ensures that the predictability measure avoids any contamination from changes in market dynamics post-digitization.

\subsubsection{Human Investor Specialization}

\begin{figure}[ht!]
\begin{center}
\captionsetup{justification=centering}
\caption{\textsc{Figure \ref{fig:total_specialization}: Heterogeneous Effects of Digitization by House Predictability}}
\makebox[\linewidth]{
\begin{tabular}{c}
\textsc{A. Total Effect of Digitization on Human Investment} \\
\includegraphics[scale=0.25]{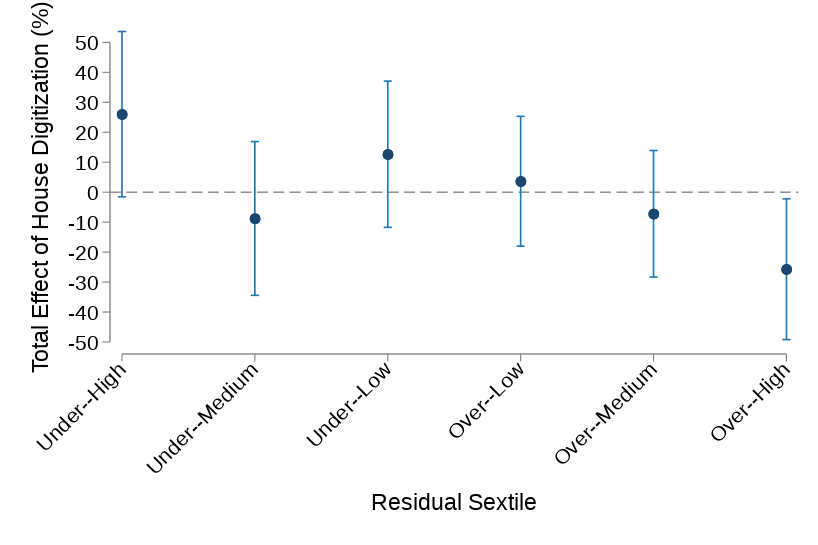} \\
\textsc{B. Total Effect of Digitization on Owner-Occupier Purchase} \\
\includegraphics[scale=0.25]{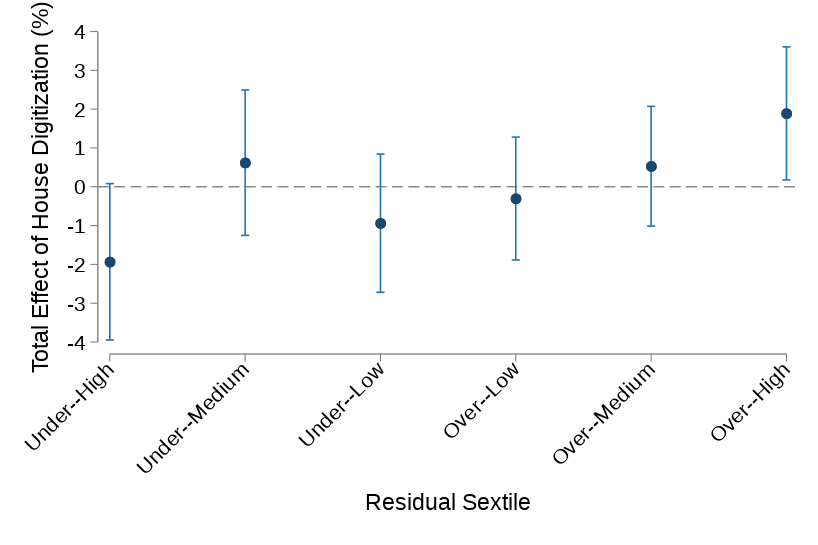} \\
\textsc{C. Total Effect of Digitization on Ln(Sales Price)} \\
\includegraphics[scale=0.25]{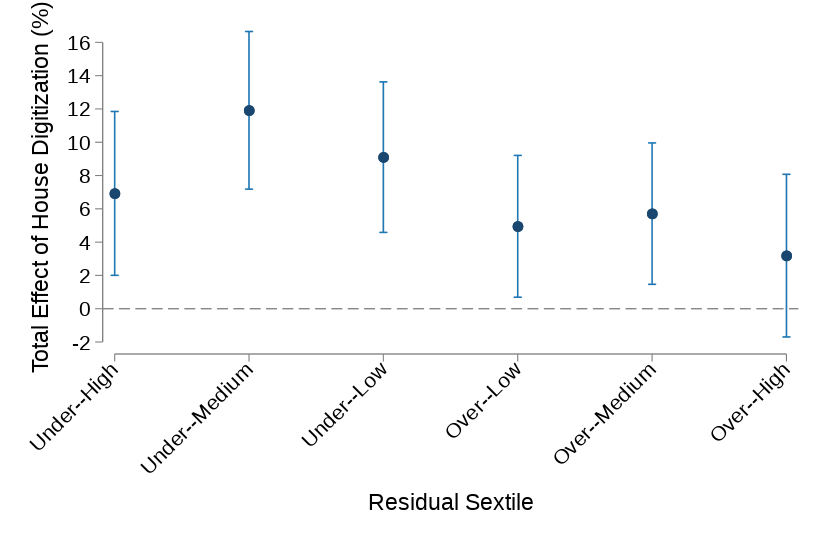} \\
\end{tabular}
}
\label{fig:total_specialization}
\end{center}
\end{figure}

Figure~\ref{fig:total_specialization} and Appendix Table~\ref{atab:specialization} show that digitization’s impact on human investor activity varies systematically with a property’s baseline predictability. Human investment rises by roughly 25 percent in the sextile where the machine learning model most underpredicts prices (``Under High''), meaning observed prices are much higher than predicted. These properties likely have value not captured in structured data, perhaps due to unobserved or difficult-to-codify features. Algorithmic activity in this segment is negligible, with average market share near zero, consistent with digitization enabling human investors to specialize in segments where they hold a comparative advantage.

By contrast, in the sextile where the model most overpredicts prices (``Over High''), human investment falls by about 25 percent after digitization. This segment attracts more algorithmic activity than the ``Under High'' segment, although still less than in the middle sextiles, because many of these properties appear undervalued to predictive models. Panel~B shows that owner-occupier purchases also rise here, potentially because algorithms push owner occupiers away from highly predictable segments where algorithmic activity is concentrated. Together, these patterns suggest that investor responses depend not only on the magnitude of prediction error but also on its direction.

Panel~C shows that prices in the ``Under High'' segment rise by about 7 log points following digitization, despite negligible algorithmic activity. This illustrates how algorithmic entry can affect prices indirectly, through spillovers into untargeted segments. In thin or segmented markets, such indirect effects can be economically meaningful. More broadly, these results underscore that measuring the market impact of algorithms requires accounting for both direct and indirect effects.

\section{Conclusion}
Progress in machine learning and the widespread availability of digitized data are transforming decision-making across a range of economic domains. This paper provides early empirical evidence of how algorithms reshape market outcomes, focusing on the U.S. housing market. I study the staggered digitization of property records across four states, which lowered the cost of accurate algorithmic prediction and facilitated the entry of algorithmic investors. Following digitization, average house prices rise and the racial price penalty narrows by 45 percent. This convergence reflects not only algorithmic firms disproportionately purchasing homes from minority owners, but also shifts in the behavior of owner-occupiers and human investors. Human investors adapt by reallocating toward segments where available data make algorithmic valuation less effective. Together, the findings show that algorithmic prediction influences individual transactions while also triggering market-wide changes in entry, allocation, and prices.

This work is subject to some qualifications and raises avenues for future research. First, the analysis focuses on four Southeastern housing markets in the early 2010s, a period and setting with unusually high-quality administrative data and relatively thin, inelastic markets. Effects may differ in more liquid or standardized markets, or in regions with less comprehensive data. The period studied also captures an early stage of private-sector machine learning adoption; since then, large language models and alternative data sources have greatly expanded the scope and accessibility of predictive tools, potentially further blurring the line between human and algorithmic decision-making.

Second, this work does not address how algorithmic investment affects housing supply. While the study period shows no change in the quantity of new homes, algorithmic buyers could influence the types of homes built, encouraging more uniform, algorithm-friendly construction. Such general equilibrium effects on supply may be particularly important in markets with more elastic production, such as labor markets, and merit further study.

Finally, the results underscore that even modest algorithmic adoption can produce sizable market-level effects, especially in settings where human decisions are biased. A growing body of evidence suggests that systematic human errors can outweigh informational advantages in decision quality, but whether and how this pattern holds across markets remains an open question \citep{mullainathanbail2017, mullainathanobermeyer2019, rambachanIdentifyingPredictionMistakes, kahnemanNoiseFlawHuman2021}. As prediction technologies advance, it will become increasingly important to understand their spillover effects and how these vary across institutional contexts.

\clearpage
\begin{spacing}{0.8}
\bibliography{CNA_Bibliography.bib}
\end{spacing}



\clearpage

\begin{table}[ht!]
\begin{center}
                \caption{\textsc{Table \ref{tab:sumstat_houses}: House Transaction Summary Statistics}}
                  \vspace{20pt}        
\scalebox{.8}{\makebox[\linewidth]{{
\def\sym#1{\ifmmode^{#1}\else\(^{#1}\)\fi}
\begin{tabular}{l*{4}{c}}
\toprule
                    &\multicolumn{1}{c}{(1)}&\multicolumn{1}{c}{(2)}&\multicolumn{1}{c}{(3)}&\multicolumn{1}{c}{(4)}\\
                    &\multicolumn{1}{c}{All}&\multicolumn{1}{c}{Owner Occupiers}&\multicolumn{1}{c}{Human Investors}&\multicolumn{1}{c}{Algo. Investors}\\
\midrule
Sale Price          &  207,248.27         &  205,177.59         &  223,101.26         &  229,012.25         \\
                    &(754,033.05)         &(547,549.42)         &(1,811,861.00)         &(283,873.25)         \\
\addlinespace
Bedrooms            &        2.11         &        2.09         &        2.21         &        2.71         \\
                    &      (3.22)         &      (3.14)         &      (4.10)         &      (1.47)         \\
\addlinespace
Bathrooms           &        2.14         &        2.14         &        2.02         &        2.42         \\
                    &      (2.37)         &      (1.43)         &      (6.43)         &      (0.99)         \\
\addlinespace
Partial Baths       &        0.27         &        0.27         &        0.22         &        0.41         \\
                    &      (0.48)         &      (0.48)         &      (0.48)         &      (0.50)         \\
\addlinespace
Stories             &        1.25         &        1.25         &        1.13         &        1.55         \\
                    &      (0.75)         &      (0.73)         &      (0.93)         &      (0.63)         \\
\addlinespace
Other Buildings     &        0.06         &        0.06         &        0.15         &        0.04         \\
                    &      (0.58)         &      (0.41)         &      (1.43)         &      (0.23)         \\
\addlinespace
Garage              &        0.56         &        0.57         &        0.42         &        0.81         \\
                    &      (0.50)         &      (0.49)         &      (0.49)         &      (0.40)         \\
\addlinespace
Fireplace           &        0.59         &        0.59         &        0.53         &        0.79         \\
                    &      (0.49)         &      (0.49)         &      (0.50)         &      (0.41)         \\
\addlinespace
Basement            &        0.17         &        0.17         &        0.13         &        0.15         \\
                    &      (0.37)         &      (0.38)         &      (0.33)         &      (0.36)         \\
\addlinespace
Parking Spaces      &        0.75         &        0.77         &        0.54         &        0.83         \\
                    &      (8.83)         &      (8.93)         &      (8.67)         &      (0.98)         \\
\addlinespace
House Age           &       31.13         &       30.28         &       42.07         &       22.49         \\
                    &     (25.76)         &     (25.34)         &     (28.89)         &     (16.16)         \\
\addlinespace
Age Since Remodel   &       24.44         &       23.70         &       33.34         &       19.94         \\
                    &     (21.08)         &     (20.54)         &     (25.39)         &     (14.86)         \\
\midrule
N                   &   6,727,758         &   5,991,240         &     614,694         &     121,824         \\
\bottomrule
\multicolumn{5}{l}{\footnotesize mean coefficients; sd in parentheses}\\
\multicolumn{5}{l}{\footnotesize \sym{*} \(p<0.05\), \sym{**} \(p<0.01\), \sym{***} \(p<0.001\)}\\
\end{tabular}
}
}}
  \label{tab:sumstat_houses}
\end{center}
\end{table}

\begin{singlespace}
\footnotesize
\noindent \textsc{Notes}: This table shows characteristics from the sample, defined as arm's-length transactions from Georgia, North Carolina, South Carolina, and Tennessee over 2009--2021. Column 1 includes all transactions; Column 2 shows houses purchased by owner-occupiers (i.e., those buying houses to live in); Column 3, houses purchased by human investors; Column 4, houses purchased by algorithmic investors. All data are at the transaction level and come from Attom Data. 
\end{singlespace}
\normalsize


\clearpage

\begin{table}[ht!]
\begin{center}
                \caption{\textsc{Table \ref{tab:es_main}: Impact of County Digitization on Algorithmic Investors’ Transaction Share}}
                  \vspace{20pt}        
\scalebox{.8}{\makebox[\linewidth]{{
\def\sym#1{\ifmmode^{#1}\else\(^{#1}\)\fi}
\begin{tabular}{l*{4}{l}}
\toprule
                &\multicolumn{4}{c}{Algorithmic Investor Share}     \\\cmidrule(lr){2-5}
                &\multicolumn{1}{c}{(1)}         &\multicolumn{1}{c}{(2)}         &\multicolumn{1}{c}{(3)}         &\multicolumn{1}{c}{(4)}         \\
\midrule
$\hat{\beta}_{-4}$&-0.003         &-0.001\sym{**} &-0.001\sym{*}  &-0.001\sym{*}  \\
                &(0.002)         &(0.000)         &(0.000)         &(0.000)         \\
\addlinespace
$\hat{\beta}_{-3}$&-0.001         &-0.003         &-0.003         &-0.003         \\
                &(0.000)         &(0.002)         &(0.002)         &(0.002)         \\
\addlinespace
$\hat{\beta}_{-2}$&-0.001         &-0.000         &-0.001         &-0.001         \\
                &(0.000)         &(0.000)         &(0.000)         &(0.000)         \\
\addlinespace
$\hat{\beta}_{0}$&0.009\sym{***}&0.011\sym{***}&0.011\sym{***}&0.011\sym{***}\\
                &(0.003)         &(0.003)         &(0.003)         &(0.003)         \\
\addlinespace
$\hat{\beta}_{1}$&0.024\sym{***}&0.028\sym{***}&0.028\sym{***}&0.028\sym{***}\\
                &(0.003)         &(0.004)         &(0.004)         &(0.004)         \\
\addlinespace
$\hat{\beta}_{2}$&0.019\sym{***}&0.023\sym{***}&0.024\sym{***}&0.023\sym{***}\\
                &(0.003)         &(0.004)         &(0.004)         &(0.004)         \\
\addlinespace
$\hat{\beta}_{3}$&0.016\sym{***}&0.014\sym{***}&0.014\sym{***}&0.014\sym{***}\\
                &(0.003)         &(0.003)         &(0.003)         &(0.003)         \\
\addlinespace
$\hat{\beta}_{4}$&0.018\sym{***}&0.016\sym{***}&0.016\sym{***}&0.016\sym{***}\\
                &(0.002)         &(0.003)         &(0.003)         &(0.003)         \\
\addlinespace
$\hat{\beta}_{5}$&0.017\sym{***}&0.018\sym{***}&0.018\sym{***}&0.018\sym{***}\\
                &(0.003)         &(0.003)         &(0.003)         &(0.003)         \\
\midrule
Dependent Variable Mean&0.003         &0.003         &0.003         &0.003         \\
Observations    &3127        &3127        &3127         &3127         \\
Average Total Effect&0.017         &0.018         &0.019         &0.019         \\
Joint Eq. Effects&0.000         &0.000         &0.000         &0.000         \\
Joint Sig. Placebo&0.007         &0.053         &0.075         &0.086         \\
County + Year FE&Yes         &Yes         &Yes         &Yes         \\
Demographics    & No         &Yes         &Yes         &Yes         \\
Economics       & No         & No         &Yes         &Yes         \\
Housing         & No         & No         & No         &Yes         \\
\bottomrule
\multicolumn{5}{l}{\footnotesize Standard errors in parentheses}\\
\multicolumn{5}{l}{\footnotesize \sym{*} \(p<0.10\), \sym{**} \(p<0.05\), \sym{***} \(p<0.01\)}\\
\end{tabular}
}
}}
  \label{tab:es_main}
\end{center}
\end{table}

\begin{singlespace}
\footnotesize
\noindent \textsc{Notes}: This table reports estimates of the impact of county digitization on the transaction share of algorithmic investors, defined as the share of home sales in a county-year purchased by algorithmic buyers. All regressions correspond to Equation~\ref{eq:es_main} and are estimated using the estimator of \citet{dechaisemartin2024}. Specifications include county and year fixed effects and are weighted by the number of housing transactions. Column 1 controls for county population. Column 2 adds more demographic characteristics: the share of the population that is White, Black, and Hispanic; average family size; and share of adults with a college degree or higher. Column 3 adds economic characteristics: median household income; poverty rate; share receiving Supplemental Security Income; share receiving public assistance; unemployment rate; and labor force participation rate. Column 4 additionally controls for housing market characteristics: the share of owner-occupied houses; the share of vacant houses listed for sale; and median rent. All data are measured at the county-year level and drawn from the American Community Survey (five-year estimates), Attom Data, and county administrative records.
\end{singlespace}
\normalsize



\clearpage

\begin{table}[ht!]
\begin{center}
                \caption{\textsc{Table \ref{tab:triple_diff_falsification}: Effect of County Digitization on Investment, by House Digitization}}
                  \vspace{20pt}        
\scalebox{.8}{\makebox[\linewidth]{{
\def\sym#1{\ifmmode^{#1}\else\(^{#1}\)\fi}
\begin{tabular}{l*{6}{c}}
\toprule
                    &\multicolumn{3}{c}{Algorithmic Investment}                       &\multicolumn{3}{c}{Human Investment}                             \\\cmidrule(lr){2-4}\cmidrule(lr){5-7}
                    &\multicolumn{1}{c}{(1)}         &\multicolumn{1}{c}{(2)}         &\multicolumn{1}{c}{(3)}         &\multicolumn{1}{c}{(4)}         &\multicolumn{1}{c}{(5)}         &\multicolumn{1}{c}{(6)}         \\
\midrule
County Digitized $\times$ Not-Yet-Digitized House&    -0.00163         &    -0.00148\sym{*}  &    0.000423         &    -0.00364\sym{*}  &    -0.00325\sym{*}  &    -0.00206         \\
                    &   (0.00120)         &  (0.000890)         &  (0.000485)         &   (0.00206)         &   (0.00174)         &   (0.00133)         \\
\addlinespace
County Digitized $\times$ Digitized House&      0.0116\sym{***}&      0.0116\sym{***}&     0.00990\sym{***}&     0.00115         &    0.000765         &   -0.000455         \\
                    &   (0.00126)         &  (0.000939)         &  (0.000524)         &   (0.00207)         &   (0.00175)         &   (0.00131)         \\
\midrule
Dependent Variable Mean&      0.0181         &      0.0181         &      0.0183         &    0.000135         &    0.000135         &    0.000137         \\
R-squared           &      0.0725         &      0.0792         &       0.134         &      0.0647         &      0.0733         &       0.162         \\
Adjusted R-squared  &      0.0715         &      0.0767         &      0.0790         &      0.0638         &      0.0707         &       0.108         \\
Observations        &     6727743         &     6727720         &     6654496         &     6727743         &     6727720         &     6654496         \\
Location FE         &       Tract         & Block Group         &       Block         &       Tract         & Block Group         &       Block         \\
Year FE             &         Yes         &         Yes         &         Yes         &         Yes         &         Yes         &         Yes         \\
Housing FE          &         Yes         &         Yes         &         Yes         &         Yes         &         Yes         &         Yes         \\
\bottomrule
\multicolumn{7}{l}{\footnotesize Standard errors in parentheses}\\
\multicolumn{7}{l}{\footnotesize \sym{*} \(p<0.10\), \sym{**} \(p<0.05\), \sym{***} \(p<0.01\)}\\
\end{tabular}
}
}}
  \label{tab:triple_diff_falsification}
\end{center}
\end{table}

\begin{singlespace} 
\footnotesize
\noindent \textsc{Notes}: This table reports estimates of the impact of county-level digitization on investment activity in digitized and non-digitized homes, based on Equation~\ref{eq:triple_diff}. Standard errors are clustered at the relevant geographic level. Column 1--3 shows the average treatment effect of county digitization on algorithmic investment (an indicator if a house is purchased by an algorithmic investor), separately by house digitization status. Columns 4--6 shows another falsification exercise to test if county digitization changes human investment (an indicator if a house is purchased by a human investor). All regressions include year and geographic fixed effects, and control for housing characteristics including the number of bedrooms, bathrooms, and stories; lot size; interior square footage; presence of a basement; foundation type; construction type; and roof material described in Appendix Section \ref{asec:housechars}. Not all properties can be geocoded to the Census block level, but all can be matched at the block group and tract levels. The data are at the transaction level and are drawn from Attom Data, county government records, and the American Community Survey (five-year estimates). 
\end{singlespace}
\normalsize

\clearpage
\begin{table}[ht!]
\begin{center}
                \caption{\textsc{Table \ref{tab:racepenalty_by_geo_ethnicolr}: Effect of County Digitization on the Race Penalty}} 
                  \vspace{20pt}        
\scalebox{.8}{\makebox[\linewidth]{{
\def\sym#1{\ifmmode^{#1}\else\(^{#1}\)\fi}
\begin{tabular}{l*{3}{c}}
\toprule
                    &\multicolumn{1}{c}{(1)}         &\multicolumn{1}{c}{(2)}         &\multicolumn{1}{c}{(3)}         \\
\midrule
Minority Seller     &     -0.0711\sym{***}&     -0.0577\sym{***}&     -0.0292\sym{***}\\
                    &   (0.00502)         &   (0.00458)         &   (0.00544)         \\
\addlinespace
Minority Seller $\times$ County Digitization&      0.0292\sym{***}&      0.0257\sym{***}&      0.0118\sym{**} \\
                    &   (0.00521)         &   (0.00478)         &   (0.00568)         \\
\midrule
Dependent Variable Mean&       11.90         &       11.91         &       12.05         \\
R-squared           &       0.612         &       0.657         &       0.791         \\
Adjusted R-squared  &       0.599         &       0.626         &       0.708         \\
Observations        &     2569946         &     2554547         &     1682143         \\
Location            &Tract x Year         &Block Group x Year         &Block x Year         \\
Housing             &         Yes         &         Yes         &         Yes         \\
\bottomrule
\multicolumn{4}{l}{\footnotesize Standard errors in parentheses}\\
\multicolumn{4}{l}{\footnotesize \sym{*} \(p<0.10\), \sym{**} \(p<0.05\), \sym{***} \(p<0.01\)}\\
\end{tabular}
}
}}
  \label{tab:racepenalty_by_geo_ethnicolr}
\end{center}
\end{table}

\begin{singlespace}
\footnotesize
\noindent \textsc{Notes}:  This table reports estimated race penalty from Equation~\ref{eq:racepenalty} (\textit{Minority Seller}) and the marginal effect of county digitization on the race penalty (\textit{Minority Seller $\times$ County Digitization}). All regressions control for house characteristics, detailed in Appendix Section \ref{asec:housechars}, and include year-by-geography fixed effects. Standard errors are clustered at the relevant geographic level. \textit{Minority Seller} is an indicator for whether the seller is identified as Black or Hispanic (vs. White). Race is inferred using the \texttt{Ethnicolr} procedure. The main effect of county digitization drops out due to the inclusion of geography-by-year fixed effects. Data are at the house transaction level and come from Attom Data, the U.S. Census, and county governments. 
\end{singlespace}
\normalsize

\clearpage
\begin{table}[ht!]
\begin{center}
                \caption{\textsc{Table \ref{tab:assessresalemargin}: Resale and Assessment Margin among Algorithmic Investor Purchases}}
                  \vspace{20pt}        
\scalebox{.6}{\makebox[\linewidth]{{
\def\sym#1{\ifmmode^{#1}\else\(^{#1}\)\fi}
\begin{tabular}{l*{8}{c}}
\toprule
                    &\multicolumn{2}{c}{Resale Margin}          &\multicolumn{2}{c}{Assessment Margin}      &\multicolumn{2}{c}{Resale Margin (BISG)}   &\multicolumn{2}{c}{Assessment Margin (BISG)}\\\cmidrule(lr){2-3}\cmidrule(lr){4-5}\cmidrule(lr){6-7}\cmidrule(lr){8-9}
                    &\multicolumn{1}{c}{(1)}         &\multicolumn{1}{c}{(2)}         &\multicolumn{1}{c}{(3)}         &\multicolumn{1}{c}{(4)}         &\multicolumn{1}{c}{(5)}         &\multicolumn{1}{c}{(6)}         &\multicolumn{1}{c}{(7)}         &\multicolumn{1}{c}{(8)}         \\
\midrule
Minority Seller     &     0.00677         &     0.00798\sym{**} &    0.000878         &     0.00251         &                     &                     &                     &                     \\
                    &   (0.00417)         &   (0.00380)         &   (0.00251)         &   (0.00252)         &                     &                     &                     &                     \\
\addlinespace
Minority Seller     &                     &                     &                     &                     &      0.0106\sym{**} &     0.00791\sym{*}  &     0.00757\sym{***}&     0.00598\sym{***}\\
                    &                     &                     &                     &                     &   (0.00474)         &   (0.00452)         &   (0.00209)         &   (0.00204)         \\
\midrule
Dependent Variable Mean&      0.0622         &      0.0601         &      0.0855         &      0.0761         &      0.0696         &      0.0672         &      0.0541         &      0.0407         \\
R-squared           &       0.343         &       0.379         &       0.737         &       0.766         &       0.297         &       0.337         &       0.763         &       0.781         \\
Adjusted R-squared  &       0.194         &       0.160         &       0.704         &       0.719         &       0.122         &      0.0842         &       0.726         &       0.730         \\
Observations        &       27710         &       24398         &       93186         &       87539         &       20838         &       18041         &       61039         &       56162         \\
Race Classification &   Ethnicolr         &   Ethnicolr         &   Ethnicolr         &   Ethnicolr         &        BISG         &        BISG         &        BISG         &        BISG         \\
Location            &Tract x Year         &Block Group x Year         &Tract x Year         &Block Group x Year         &Tract x Year         &Block Group x Year         &Tract x Year         &Block Group x Year         \\
Housing             &         Yes         &         Yes         &         Yes         &         Yes         &         Yes         &         Yes         &         Yes         &         Yes         \\
Sample              &Algo. Investors         &Algo. Investors         &Algo. Investors         &Algo. Investors         &Algo. Investors         &Algo. Investors         &Algo. Investors         &Algo. Investors         \\
\bottomrule
\multicolumn{9}{l}{\footnotesize Standard errors in parentheses}\\
\multicolumn{9}{l}{\footnotesize \sym{*} \(p<0.10\), \sym{**} \(p<0.05\), \sym{***} \(p<0.01\)}\\
\end{tabular}
}
}}
  \label{tab:assessresalemargin}
\end{center}
\end{table}

\begin{singlespace}
\footnotesize
\noindent \textsc{Notes}:  This table reports estimates of gross margin using two proxies: resale margin, defined as the difference between the natural log of the resale price and the original purchase price, and assessment margin, defined as the difference between the natural log of the next available assessor valuation and the original purchase price. \textit{Minority Seller} is an indicator for whether the home was sold by a Black or Hispanic homeowner (vs.\ White). Race is inferred using \texttt{Ethnicolr} in Columns 1--4 and Bayesian Improved Surname Geocoding (BISG) in Columns 5--8. All regressions include fixed effects for resale or assessment year, and for original sale year interacted with geographic unit, and housing characteristics. Standard errors are clustered at the relevant geographic level. The data are at the house transaction level and come from Attom Data. 
\end{singlespace}
\normalsize



\clearpage
\begin{footnotesize} 
\begin{singlespace}
\noindent \textsc{Notes}: Panel A shows the cumulative share of counties that adopted digitized housing records over time, weighted by the number of housing transactions. Panel B examines predictors of early digitization, defined as adoption prior to the mean county-year. Each coefficient comes from a separate regression of the early digitization indicator on a standardized baseline county characteristic, controlling for state fixed effects. Coefficients represent the change in the probability of early digitization associated with a one-standard-deviation increase in the independent variable. Horizontal bars denote 95\% confidence intervals, and standard errors are clustered at the county level. Demographic and housing data are from the 2010 Decennial Census; economic variables are from the 2005–2009 American Community Survey.
\end{singlespace}
\end{footnotesize}

\clearpage

\begin{footnotesize} 
\begin{singlespace}
\noindent \textsc{Notes}:  Panel A plots the transaction share of houses purchased by algorithmic investors in each county-year,  by time to digitization. Panel B presents event-study estimates and 95\% confidence intervals of the effect of county digitization on the algorithmic investors transaction share, using the robust estimator of \citet{dechaisemartin2020}. The regression specification follows Equation \ref{eq:es_main} and includes county and year fixed effects and county population. Standard errors are clustered at the county level, and observations are weighted by size of county house market. Data are at the county-year level and comes from five-year American Community Survey county-level extracts, Attom Data Solutions, and county governments.
\end{singlespace}
\end{footnotesize}


\clearpage

\begin{footnotesize} 
\begin{singlespace}
\noindent \textsc{Notes}: These figures show the impact of county digitization on the algorithmic investor transaction share, separately for \textit{digitized houses} and \textit{not-yet-digitized houses}. Panel A shows the transaction market share for digitized houses (the number of digitized houses purchased by algorithmic investors in each county-year divided by total house sales), and transaction share for non-digitized houses (the number of not-yet-digitized houses purchased by algorithmic investors in each county-year divided by total house sales). Panel B plots the coefficients and 95 percent confidence intervals for the corresponding event study estimates using the robust estimator of \citet{dechaisemartin2024}. All specifications follow the regression framework described in Equation~\ref{eq:triple_diff} and include county and year fixed effects, as well as controls for county population. Regressions are weighted by the number of transactions in each county-year, and standard errors are clustered at the county level. Data are at the county-year level and are drawn from Attom Data, the U.S. Census, and county government records.

\end{singlespace}
\end{footnotesize}


\clearpage
\begin{figure}[ht!]
\begin{center}
\captionsetup{justification=centering}
\caption{\textsc{Figure \ref{fig:racepenalty_subsampleanalysis}: The Race Penalty}}
\makebox[\linewidth]{
\begin{tabular}{cc}
\textsc{A. By County Digitization}  & \textsc{B. By County Digitization and Entry}  \\
\includegraphics[scale=0.3]{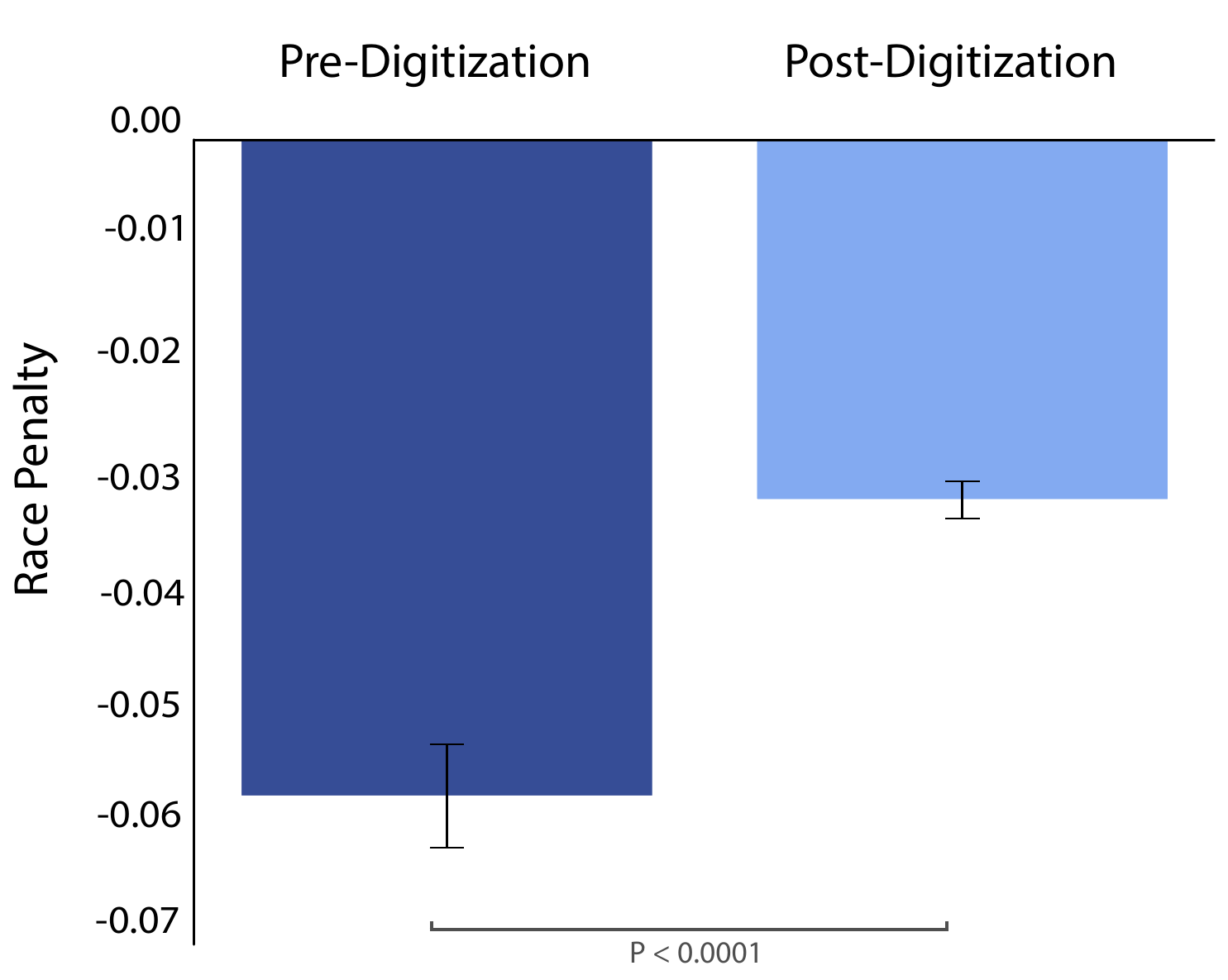}  & \includegraphics[scale=0.3]{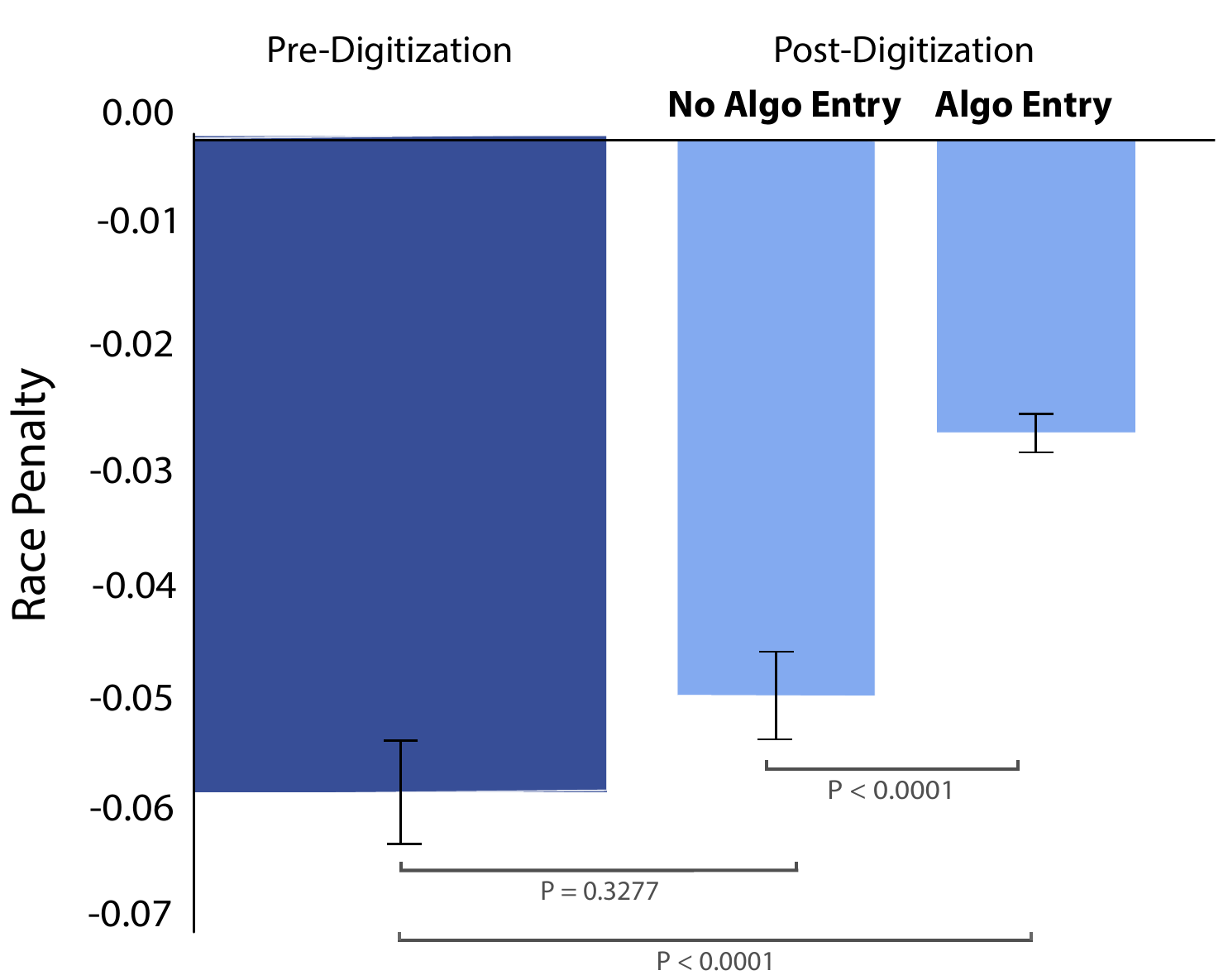} \\
\textsc{C. By County Digitization and Buyer} & \textsc{D. By County Digitization, By Buyer and Entry} \\
\includegraphics[scale=0.3]{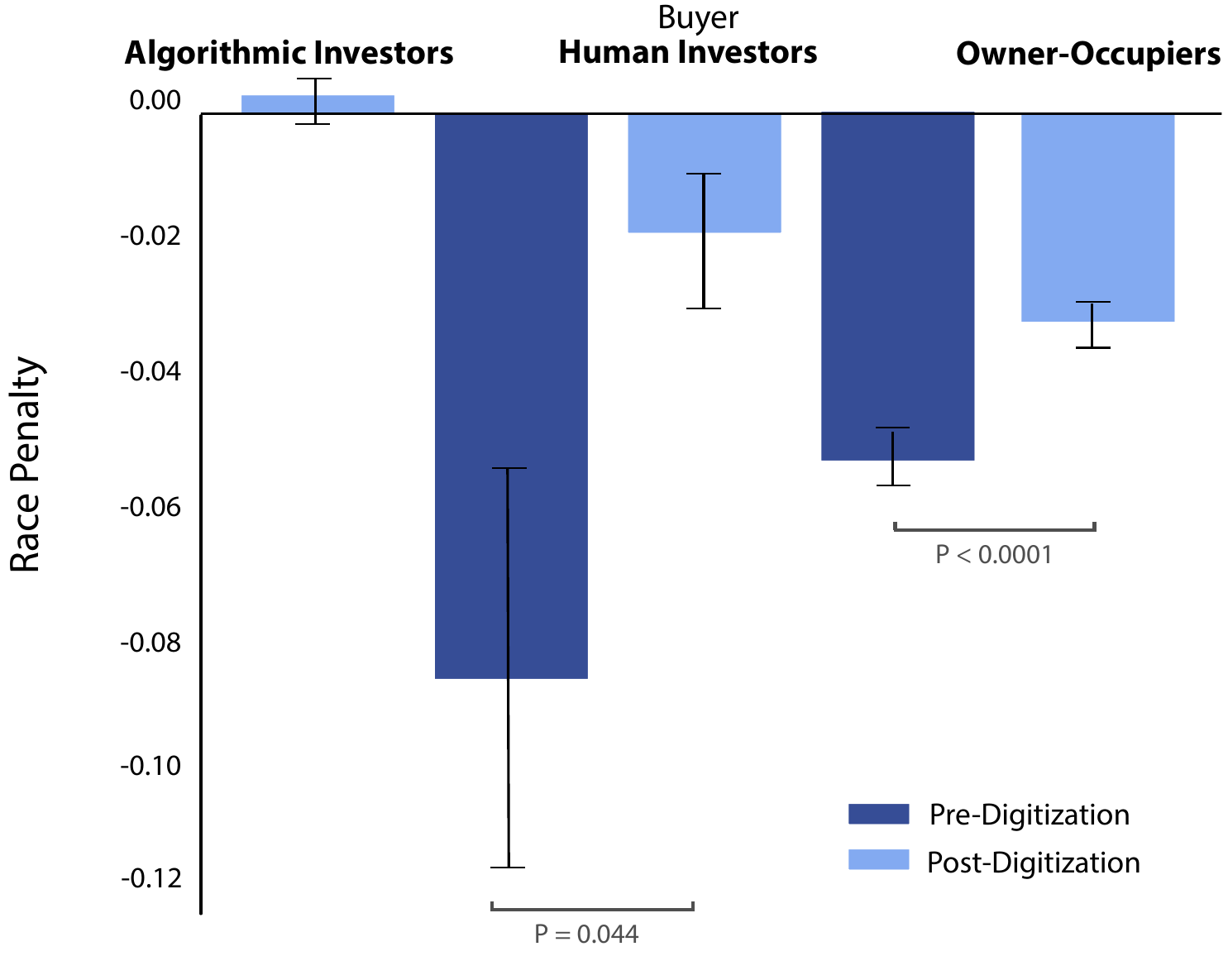} & \includegraphics[scale=0.3]{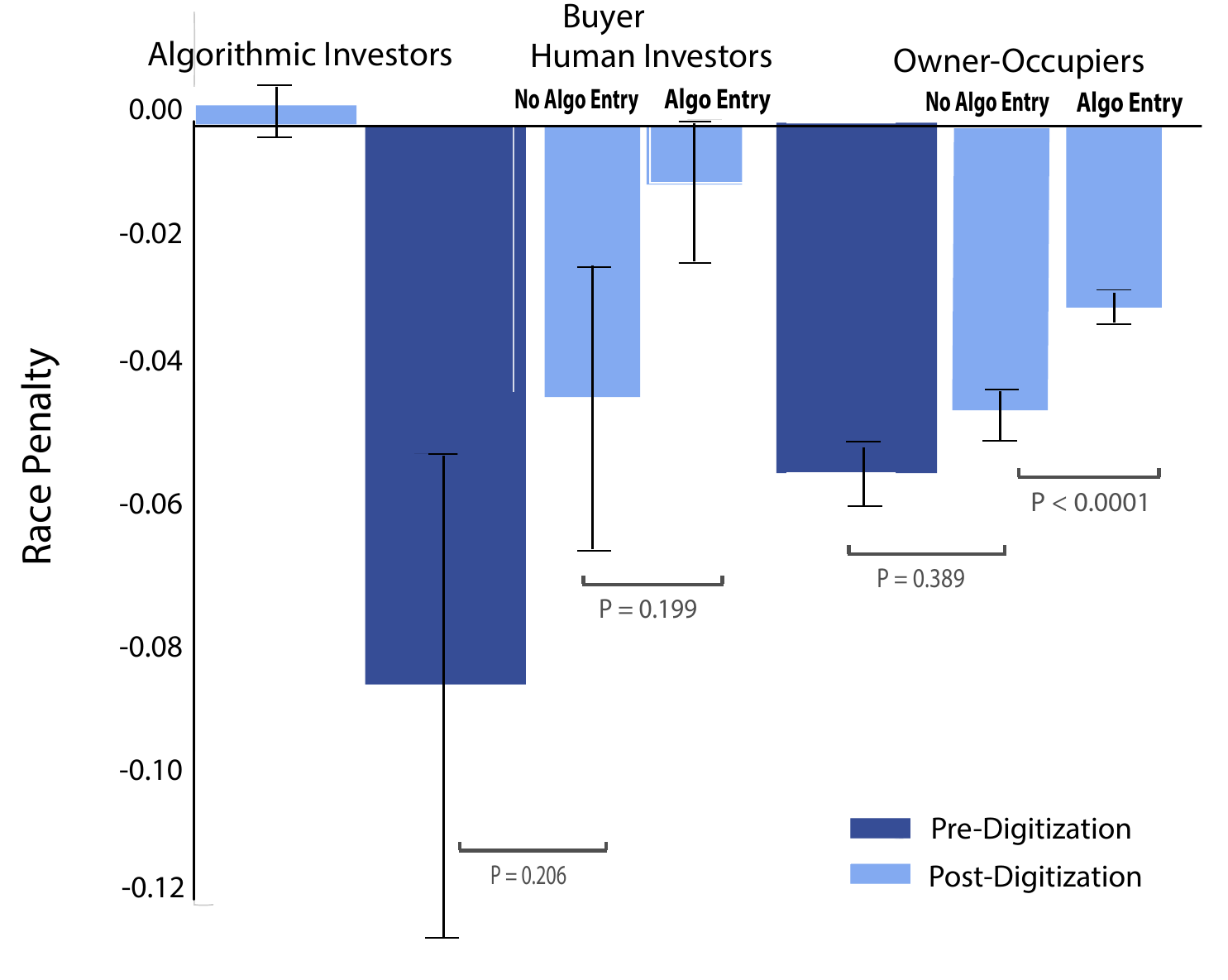}  \\
\end{tabular}
} 
\label{fig:racepenalty_subsampleanalysis}
\end{center}
\end{figure}

\begin{footnotesize} 
\begin{singlespace}
\noindent \textsc{Notes}:  This figure plots race penalty coefficients from separate regressions of Equation \ref{eq:racepenalty} across different subsamples defined by the timing of digitization, entry, and identity of the buyer. Panel A compares the race penalty before and after county digitization. ``Pre-Digitization'' includes all transactions in county-years prior to digitization; ``Post-Digitization'' includes those in county-years following digitization. Panel B further disaggregates the post-digitization period into counties with no algorithmic investor entry (``Post-Digitization / No Algo Entry'') and those with entry (``Post-Digitization / Algo Entry'').  Panel C reports coefficients separately for transactions involving algorithmic investors, human investors, and owner-occupiers, before and after digitization. Panel D breaks out the post-digitization period by buyer type and by whether algorithmic investors are active in the county. All regressions include controls for housing characteristics, and Census block group-by-year fixed effects. Appendix Table~\ref{atab:racepenalty_coeff_tests} reports tests of equality across subsamples. Data are at the house-transaction level and are drawn from Attom Data, the U.S. Census, and county government records.

\end{singlespace}
\end{footnotesize}

\clearpage

\begin{footnotesize} 
\begin{singlespace}
\noindent \textsc{Notes}: This figure shows the results from an Oaxaca-like decomposition of the change in the race penalty, described in Appendix Section \ref{asec:oaxaca_decom}. The within-buyer effect is the change due to the buyer's changing race penalty; the composition effect is the change driven by changes in transaction shares. The ``new buyer'' effect captures the contribution of algorithmic investors, who appear after county digitization and mechanically have a small contribution since their post-period race penalty is close to zero. Data are at the house transaction level and come from Attom Data, the U.S. Census, and county governments. 

\end{singlespace}
\end{footnotesize}

\clearpage

\begin{footnotesize} 
\begin{singlespace}
\noindent \textsc{Notes}: This figure presents event-study estimates of the effect of county digitization on the natural log of house sale prices. Panel A reports results for minority and White homeowners combined. Panel B disaggregates effects by seller race: Red diamonds indicate estimates for homes sold by minority owners; blue diamonds for those sold by White owners. All regressions include block group and year fixed effects, as well as controls for observable housing characteristics listed in Appendix Section \ref{asec:housechars}. Standard errors are clustered at the block group level. Due to the high dimensionality of fixed effects, regressions are estimated using OLS. The sample is limited to transactions for which seller race can be inferred using \texttt{Ethnicolr}. Data are at the house-transaction level and are drawn from Attom Data, the U.S. Census, and county government records.
\end{singlespace}
\end{footnotesize}

\clearpage

\begin{footnotesize} 
\begin{singlespace}
\noindent \textsc{Notes}:  These figures report coefficients and 95\% confidence intervals from regressions of: (i) human investor purchase (Panel A), (ii) owner-occupier purchase (Panel B), and (iii) log sale price (Panel C) on a house-level digitization indicator, estimated by sextile of baseline predictability. Predictability is defined as the out-of-sample prediction error, $y - \hat{y}$, from an XGBoost model predicting sale price, as described in Appendix~\ref{asec:buildingpredictability}. Coefficients capture the total effect of digitization within each sextile of the prediction error distribution: ``Under–High'' denotes the tercile with the largest positive residuals (actual prices exceed predicted prices), while ``Over–High'' denotes the tercile with the largest negative residuals. Sextiles are constructed separately for positive and negative residuals. Predictability scores are assigned using the pre-digitization validation sample, and the analysis is restricted to repeat-sale properties. All regressions include block group–by–year fixed effects; standard errors are clustered at the block group level. The unit of observation is a house transaction. Data are from Attom Data and county governments.

\end{singlespace}
\end{footnotesize}

\begin{appendix}
\renewcommand{\thefigure}{A.\arabic{figure}}
\setcounter{figure}{0}
\renewcommand{\thetable}{A.\arabic{table}}
\setcounter{table}{0}

\newpage
\clearpage
\noindent \textbf{\Large{Appendix Materials -- For Online Publication}}


\clearpage
\begin{table}[ht!]
\begin{center}
                \caption{\textsc{Table \ref{atab:balance_county_inv}: Neighborhood Characteristics of Transactions}}
                  \vspace{20pt}        
\scalebox{.8}{\makebox[\linewidth]{{
\def\sym#1{\ifmmode^{#1}\else\(^{#1}\)\fi}
\begin{tabular}{l*{5}{c}}
\hline\hline
                    &\multicolumn{1}{c}{(1)}&\multicolumn{1}{c}{(2)}&\multicolumn{1}{c}{(3)}&\multicolumn{1}{c}{(4)}&\multicolumn{1}{c}{(5)}\\
                    &\multicolumn{1}{c}{All}&\multicolumn{1}{c}{Owner Occupiers}&\multicolumn{1}{c}{Algo Investors}&\multicolumn{1}{c}{Human Investors}&\multicolumn{1}{c}{Difference (3)-(4)}\\
\hline
Housing Units       &     978.11 &     981.75 &    1168.14 &     904.95 &      263.19***\\
                    &   (590.48) &   (589.01) &   (705.13) &   (568.69) &      (2.15)   \\
Housing Occupancy Rate&      86.80 &      86.98 &      91.89 &      84.05 &        7.84***\\
                    &    (13.60) &    (13.44) &     (4.28) &    (15.76) &      (0.02)   \\
Share Children      &      23.84 &      23.78 &      27.99 &      23.59 &        4.40***\\
                    &     (6.31) &     (6.26) &     (4.56) &     (6.79) &      (0.02)   \\
Share White         &      74.08 &      75.58 &      59.04 &      62.49 &       -3.44***\\
                    &    (24.64) &    (23.36) &    (27.15) &    (31.27) &      (0.09)   \\
Share Black         &      20.89 &      19.44 &      34.26 &      32.36 &        1.89***\\
                    &    (23.60) &    (22.18) &    (27.15) &    (31.13) &      (0.09)   \\
Median Rent as Share Income&      30.01 &      29.84 &      30.04 &      31.62 &       -1.58***\\
                    &     (7.67) &     (7.62) &     (7.81) &     (7.92) &      (0.02)   \\
BA or Higher        &      50.97 &      50.56 &      70.31 &      50.84 &       19.47***\\
                    &    (31.87) &    (31.73) &    (28.89) &    (32.57) &      (0.09)   \\
Labor Force Participation Rate&      63.28 &      63.26 &      68.94 &      62.35 &        6.59***\\
                    &     (9.82) &     (9.80) &     (7.22) &    (10.13) &      (0.02)   \\
Unemployment Rate   &       4.49 &       4.43 &       4.29 &       5.16 &       -0.88***\\
                    &     (2.88) &     (2.81) &     (2.93) &     (3.47) &      (0.01)   \\
Share Below Poverty &      12.30 &      12.08 &       8.16 &      15.33 &       -7.17***\\
                    &     (9.11) &     (8.85) &     (6.35) &    (11.15) &      (0.02)   \\
\hline
N                   &   6,727,758&   5,991,240&     121,824&     614,694&     736,518   \\
\hline\hline
\end{tabular}
}
}}
  \label{atab:balance_county_inv}
\end{center}
\end{table}

\begin{singlespace}
\footnotesize
\noindent \textsc{Notes}: This table shows neighborhood characteristics of transactions in the sample, defined as arm's-length transactions from Georgia, North Carolina, South Carolina, and Tennessee over 2009--2021.  Column 1 includes all transactions in the sample. Columns 2-4 restrict to transactions by owner-occupiers, algorithmic investors, and human investors, respectively. Column 5 reports the difference in means between algorithmic and human investors (Column 3 and Column 4). Demographic and housing characteristics are measured at the Census block group level using data from the 2010 Decennial Census; socioeconomic characteristics are measured at the Census tract level using data from the 2005–2009 American Community Survey.
\end{singlespace}
\normalsize

\clearpage
\begin{table}[ht!]
\begin{center}
                \caption{\textsc{Table \ref{atab:investor_buying_chars}: Summary Statistics on Investor Buying Behavior}}
                  \vspace{20pt}        
\scalebox{.8}{\makebox[\linewidth]{{
\def\sym#1{\ifmmode^{#1}\else\(^{#1}\)\fi}
\begin{tabular}{l*{3}{c}}
\toprule
                    &\multicolumn{1}{c}{(1)}&\multicolumn{1}{c}{(2)}&\multicolumn{1}{c}{(3)}\\
                    &\multicolumn{1}{c}{All Investors}&\multicolumn{1}{c}{Algo Investors}&\multicolumn{1}{c}{Human Investors}\\
\midrule
Purchases per Year  &        1.57         &      453.69         &        1.44         \\
                    &     (20.16)         &  (1,107.78)         &      (2.70)         \\
\addlinespace
Distinct Active Zip Codes&        1.50         &       85.92         &        1.48         \\
                    &      (3.43)         &    (134.25)         &      (2.17)         \\
\addlinespace
Distinct Zips Active Each Year&        1.21         &       57.11         &        1.20         \\
                    &      (2.06)         &     (88.11)         &      (1.09)         \\
\addlinespace
Purchases in Firm Zip&        0.35         &        0.03         &        0.35         \\
                    &      (0.46)         &      (0.14)         &      (0.46)         \\
\addlinespace
Purchases in Firm State&        0.80         &        0.21         &        0.80         \\
                    &      (0.40)         &      (0.37)         &      (0.40)         \\
\bottomrule
\multicolumn{4}{l}{\footnotesize mean coefficients; sd in parentheses}\\
\multicolumn{4}{l}{\footnotesize \sym{*} \(p<0.05\), \sym{**} \(p<0.01\), \sym{***} \(p<0.001\)}\\
\end{tabular}
}
}}
  \label{atab:investor_buying_chars}
\end{center}
\end{table}

\begin{singlespace}
\footnotesize
\noindent \textsc{Notes}: This table reports firm-level summary statistics on the purchasing activity of classified investor firms. Column 1 presents averages across all investor firms in the sample. Columns 2 and 3 report averages separately for algorithmic and human investors, respectively. Means are reported with standard deviations in parentheses. All data are drawn from Attom Data.
\end{singlespace}
\normalsize

\clearpage
\begin{landscape}
\begin{figure}[ht!]
\begin{center}
\captionsetup{justification=centering}
\caption{\textsc{Figure \ref{afig:mapsactivity}: Algorithmic Investor Activity}}
\makebox[\linewidth]{
\begin{tabular}{c}
\includegraphics[scale=0.8]{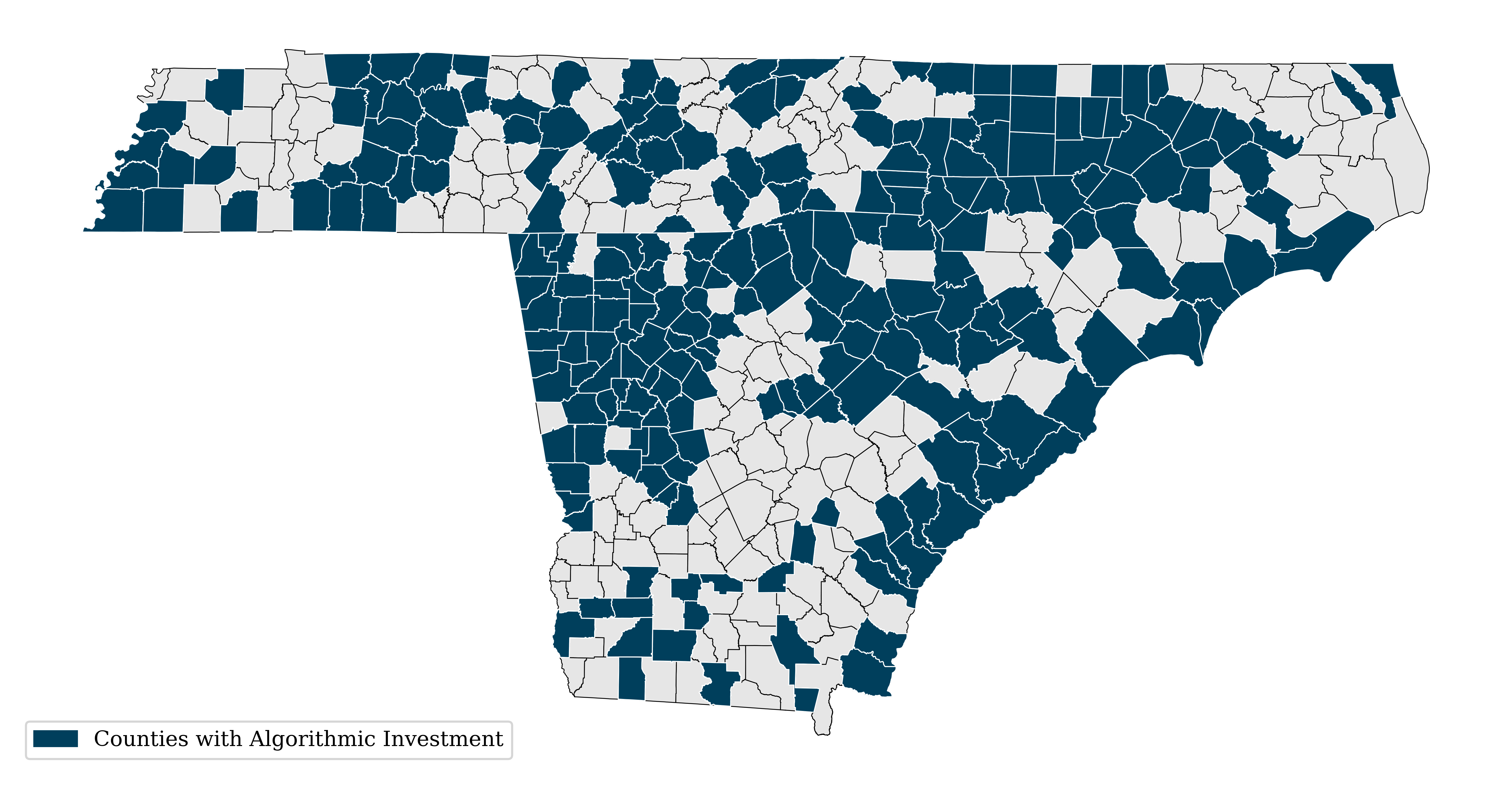} 
\end{tabular}
}
\label{afig:mapsactivity}
\end{center}
\end{figure}

\begin{footnotesize} 
\begin{singlespace}
\noindent \textsc{Notes}: This map displays the geographic distribution of investor activity across counties in Georgia, North Carolina, South Carolina, and Tennessee. Counties shaded in blue indicate the presence of algorithmic investment. All data are drawn from Attom Data Solutions. 
\end{singlespace}
\end{footnotesize}
\end{landscape}


\clearpage

\begin{table}[ht!]
\begin{center}
                \caption{\textsc{Table \ref{atab:es_main_activity}: The Impact of County Digitization on Algorithmic Investment}}
                  \vspace{20pt}        
\scalebox{.8}{\makebox[\linewidth]{{
\def\sym#1{\ifmmode^{#1}\else\(^{#1}\)\fi}
\begin{tabular}{l*{4}{l}}
\toprule
                &\multicolumn{4}{c}{Algorithmic Investor Activity}  \\\cmidrule(lr){2-5}
                &\multicolumn{1}{c}{(1)}         &\multicolumn{1}{c}{(2)}         &\multicolumn{1}{c}{(3)}         &\multicolumn{1}{c}{(4)}         \\
\midrule
$\hat{\beta}_{-4}$&-0.111         &0.002         &0.005         &0.004         \\
                &(0.097)         &(0.004)         &(0.010)         &(0.010)         \\
\addlinespace
$\hat{\beta}_{-3}$&-0.018         &-0.098         &-0.091         &-0.090         \\
                &(0.016)         &(0.085)         &(0.075)         &(0.074)         \\
\addlinespace
$\hat{\beta}_{-2}$&-0.018         &-0.015         &-0.019         &-0.020         \\
                &(0.016)         &(0.013)         &(0.017)         &(0.018)         \\
\addlinespace
$\hat{\beta}_{0}$&0.297\sym{***}&0.357\sym{***}&0.361\sym{***}&0.360\sym{***}\\
                &(0.062)         &(0.077)         &(0.074)         &(0.074)         \\
\addlinespace
$\hat{\beta}_{1}$&0.474\sym{***}&0.555\sym{***}&0.563\sym{***}&0.561\sym{***}\\
                &(0.042)         &(0.053)         &(0.051)         &(0.052)         \\
\addlinespace
$\hat{\beta}_{2}$&0.532\sym{***}&0.564\sym{***}&0.575\sym{***}&0.574\sym{***}\\
                &(0.056)         &(0.058)         &(0.056)         &(0.056)         \\
\addlinespace
$\hat{\beta}_{3}$&0.654\sym{***}&0.579\sym{***}&0.592\sym{***}&0.591\sym{***}\\
                &(0.050)         &(0.063)         &(0.060)         &(0.060)         \\
\addlinespace
$\hat{\beta}_{4}$&0.671\sym{***}&0.566\sym{***}&0.574\sym{***}&0.574\sym{***}\\
                &(0.051)         &(0.075)         &(0.072)         &(0.073)         \\
\addlinespace
$\hat{\beta}_{5}$&0.680\sym{***}&0.609\sym{***}&0.611\sym{***}&0.615\sym{***}\\
                &(0.052)         &(0.074)         &(0.073)         &(0.071)         \\
\midrule
Dependent Variable Mean&0.094         &0.094         &0.094         &0.094         \\
Observations    &3130        &3130        &3130         &3130         \\
Average Total Effect&0.557         &0.540         &0.548         &0.548         \\
Joint Eq. Effects&0.000         &0.000         &0.000         &0.000         \\
Joint Sig. Placebo&0.835         &0.852         &0.749         &0.728         \\
County + Year FE&Yes         &Yes         &Yes         &Yes         \\
Demographics    & No         &Yes         &Yes         &Yes         \\
Economics       & No         & No         &Yes         &Yes         \\
Housing         & No         & No         & No         &Yes         \\
\bottomrule
\multicolumn{5}{l}{\footnotesize Standard errors in parentheses}\\
\multicolumn{5}{l}{\footnotesize \sym{*} \(p<0.10\), \sym{**} \(p<0.05\), \sym{***} \(p<0.01\)}\\
\end{tabular}
}
}}
  \label{atab:es_main_activity}
\end{center}
\end{table}

\begin{singlespace} 
\footnotesize
\noindent \textsc{Notes}: This table reports the effect of county-level digitization on the extensive margin of algorithmic investor activity, defined as an indicator for whether algorithmic investors purchases more than 10 houses in a given county-year. All regressions correspond to Equation~\ref{eq:es_main} and are estimated using the estimator of \citet{dechaisemartin2024}. Specifications include county and year fixed effects and are weighted by the number of housing transactions. Column 1 controls for county population. Column 2 adds more demographic characteristics: the share of the population that is White, Black, and Hispanic; average family size; and share of adults with a college degree or higher. Column 3 further adds economic characteristics: median household income; poverty rate; share receiving Supplemental Security Income; share receiving public assistance; unemployment rate; and labor force participation rate. Column 4 additionally controls for housing market characteristics: the share of owner-occupied houses; the share of vacant houses listed for sale; and median rent. All data are measured at the county-year level and drawn from the American Community Survey (five-year estimates), Attom Data, and county administrative records.
\end{singlespace}
\normalsize

\clearpage
\begin{figure}[ht!]
\begin{center}
\captionsetup{justification=centering}
\caption{\textsc{Figure \ref{afig:es_lnq_alts}: Alternative Event Studies, Effect of County Digitization on Algorithmic Investors' Share Purchased}}
\makebox[\linewidth]{
\begin{tabular}{c}
\includegraphics[scale=0.4]{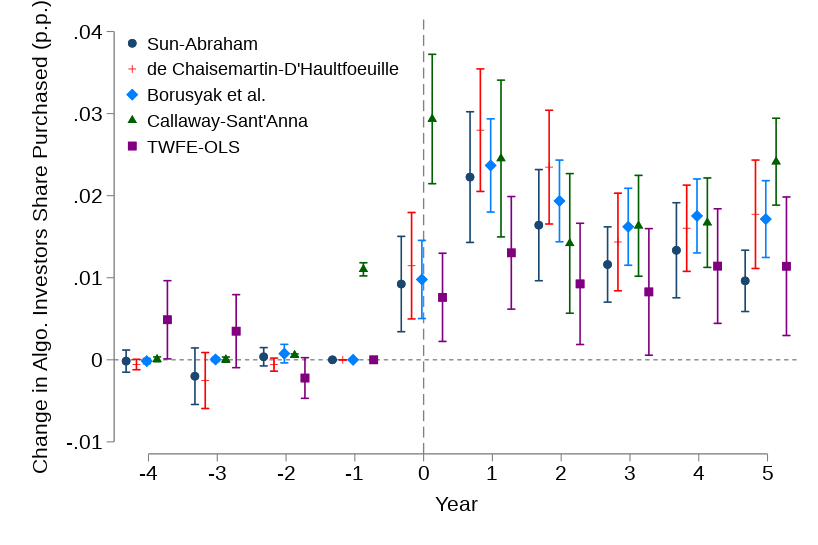} 
\end{tabular}
}
\label{afig:es_lnq_alts}
\end{center}
\end{figure}

\begin{footnotesize} 
\begin{singlespace}
\footnotesize
\noindent \textsc{Notes}: This figure plots event study coefficients and 95\% confidence intervals estimating the effect of county digitization on the transaction share of algorithmic investors. Estimates are constructed using several difference-in-differences approaches, including those of \citet{borusyak2022revisiting}, \citet{callaway_2021}, \citet{dechaisemartin2024}, and \citet{sun_estimating_2020}, as well as a standard two-way fixed effects estimator. All specifications follow Equation~\ref{eq:es_main}, include county and year fixed effects, and control for county population. Observations are weighted by the number of transactions, and standard errors are clustered at the county level. The data are at the county-year level and are drawn from Attom Data Solutions, the five-year American Community Survey, and county government records. 
\end{singlespace}
\end{footnotesize}

\clearpage
\begin{table}[ht!]
\begin{center}
                \caption{\textsc{Table \ref{atab:dd_main_robust}: Alternative Estimates of County Digitization on Algorithmic Investors' Share}}
                  \vspace{20pt}        
\scalebox{1}{\makebox[\linewidth]{{
\def\sym#1{\ifmmode^{#1}\else\(^{#1}\)\fi}
\begin{tabular}{l*{1}{cccc}}
\hline\hline
                &\shortstack{Point\\Estimate}&\shortstack{Standard\\Error}&\shortstack{Lower Bound\\ 95\% Confidence\\Interval}&\shortstack{Upper Bound\\ 95\% Confidence\\Interval}\\
\hline
Borusyak-Jaravel-Spiess&    0.018&    0.002&    0.013&    0.022\\
Callaway-Sant'Anna&    0.019&    0.003&    0.013&    0.025\\
DeChaisemartin-D'Haultfoeuille&    0.020&    0.003&    0.013&    0.026\\
Sun-Abraham     &    0.014&    0.002&    0.009&    0.018\\
TWFE-OLS        &    0.009&    0.003&    0.004&    0.015\\
\hline\hline
\end{tabular}
}
}}
  \label{atab:dd_main_robust}
\end{center}
\end{table}

\begin{singlespace}
\footnotesize
\noindent \textsc{Notes}: This table reports the effect of county data digitization on the share of houses purchased by algorithmic investors. Average treatment effects on the treated (ATT) are estimated using multiple robust difference-in-differences approaches, including those proposed by \citet{borusyak2022revisiting}, \citet{callaway_2021}, \citet{dechaisemartin2024}, and \citet{sun_estimating_2020}, as well as a traditional two-way fixed effects estimator. Regressions follow Equation \ref{eq:es_main}  include county and year fixed effects, control for county population, and weight observations by the number of housing transactions. Standard errors are clustered at the county level. The data are at the county-year level and are drawn from Attom Data Solutions, the five-year American Community Survey, and county government records.
\end{singlespace}
\normalsize

\clearpage
\begin{figure}[ht!]
\begin{center}
\captionsetup{justification=centering}
\caption{\textsc{Figure \ref{afig:housedigit_balance}: Balance Coefficient Plot of Early House Digitization}}
\makebox[\linewidth]{
\begin{tabular}{c}
\includegraphics[scale=0.4]{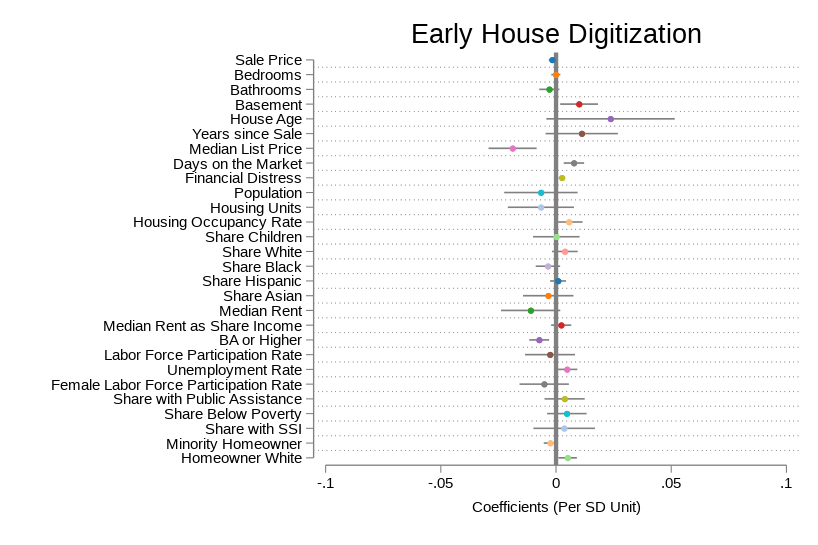}  \\
\end{tabular}
}
\label{afig:housedigit_balance}
\end{center}
\end{figure}
\begin{footnotesize} 
\begin{singlespace}
\noindent \textsc{Notes}: This figure presents tests of whether baseline neighborhood and house characteristics predict the timing of house digitization, defined as digitization prior to the mean house, controlling for county fixed effects. All independent variables are standardized. Coefficients represent the change in the probability of early digitization (in percentage points) associated with a one-standard-deviation increase in the independent variable. Horizontal bars denote 95\% confidence intervals, and standard errors are clustered at the county level. Data on demographics, population, and housing occupancy and rent come from the 2010 Decennial Census at the Census block group level; data on economic characteristics come from the 2005-2009 American Community Survey five-year survey at the census tract level. Additional data on homeowner financial status, house prices, and market activity come from Attom Data and Zillow. 
\end{singlespace}
\end{footnotesize}

\clearpage
\begin{table}[ht!]
\begin{center}
                \caption{\textsc{Table \ref{atab:housedigit_fs_byrace}: Effects of House Digitization on Investment}} 
                  \vspace{20pt}        
\scalebox{.8}{\makebox[\linewidth]{{
\def\sym#1{\ifmmode^{#1}\else\(^{#1}\)\fi}
\begin{tabular}{l*{4}{c}}
\toprule
                    &\multicolumn{3}{c}{Algorithmic Investment}                       &\multicolumn{1}{c}{Human Investment Falsification}\\\cmidrule(lr){2-4}\cmidrule(lr){5-5}
                    &\multicolumn{1}{c}{(1)}         &\multicolumn{1}{c}{(2)}         &\multicolumn{1}{c}{(3)}         &\multicolumn{1}{c}{(4)}         \\
\midrule
House Digitized     &      0.0154\sym{***}&      0.0647\sym{***}&      0.00447\sym{***}   &    -0.00383\sym{*}  \\
                    &   (0.00129)         &   (0.00714)         &    (0.000784)   &   (0.00205)         \\

\addlinespace
Minority Seller     &                     &                     &   -0.000245         &                     \\
                    &                     &                     &  (0.000347)         &                     \\
\addlinespace
House Digitized $\times$ Minority Seller&                     &                     &     0.00235\sym{***}&                     \\
                    &                     &                     &  (0.000495)         &                     \\
\midrule
Dependent Variable Mean&      0.0181         &       0.165         &      0.0181         &      0.0914         \\
R-squared           &       0.149         &       0.533         &       0.165         &       0.121         \\
Adjusted R-squared  &       0.119         &       0.434         &      0.0945         &      0.0897         \\
Observations        &     6725091         &      686247         &     2724654         &     6725091         \\
F-Statistic         &       35.14         &       57.32         &       26.25         &       187.4         \\
Location FE         &Block Group x Year         &Block Group x Year         &Block Group x Year         &Block Group x Year         \\
Housing FE          &         Yes         &         Yes         &         Yes         &         Yes         \\
Sample              &         All         &   Investors         &         All         &         All         \\
\bottomrule
\multicolumn{5}{l}{\footnotesize Standard errors in parentheses}\\
\multicolumn{5}{l}{\footnotesize \sym{*} \(p<0.10\), \sym{**} \(p<0.05\), \sym{***} \(p<0.01\)}\\
\end{tabular}
}
}}
  \label{atab:housedigit_fs_byrace}
\end{center}
\end{table}

\begin{singlespace}
\footnotesize
\noindent \textsc{Notes}: This table reports regression estimates of the relationship between house digitization and purchase by algorithmic or human investors. Column 1 estimates the effect of digitization on the probability that a property is purchased by an algorithmic investor rather than a human buyer (either an owner-occupier or a human investor). Column 2 restricts the sample to transactions with investor buyers and estimates the effect of digitization on the probability that the buyer is an algorithmic investor rather than a human investor. Column 3 examines heterogeneity by seller race by interacting the digitization variable with an indicator for a minority home seller. Column 4 estimates the effect of digitization on the probability that a property is purchased by a human investor. All specifications control for housing characteristics and include block group–by–year fixed effects. Standard errors are in parentheses. The data are at the transaction level and drawn from Attom Data and the Census.
\end{singlespace}
\normalsize

\clearpage
\begin{figure}[ht!]
\begin{center}
\captionsetup{justification=centering}
\caption{\textsc{Figure \ref{afig:outofsample_countydatasim}: Evaluating the predictive power of county data}}
\makebox[\linewidth]{
\begin{tabular}{c}
\includegraphics[scale=0.5]{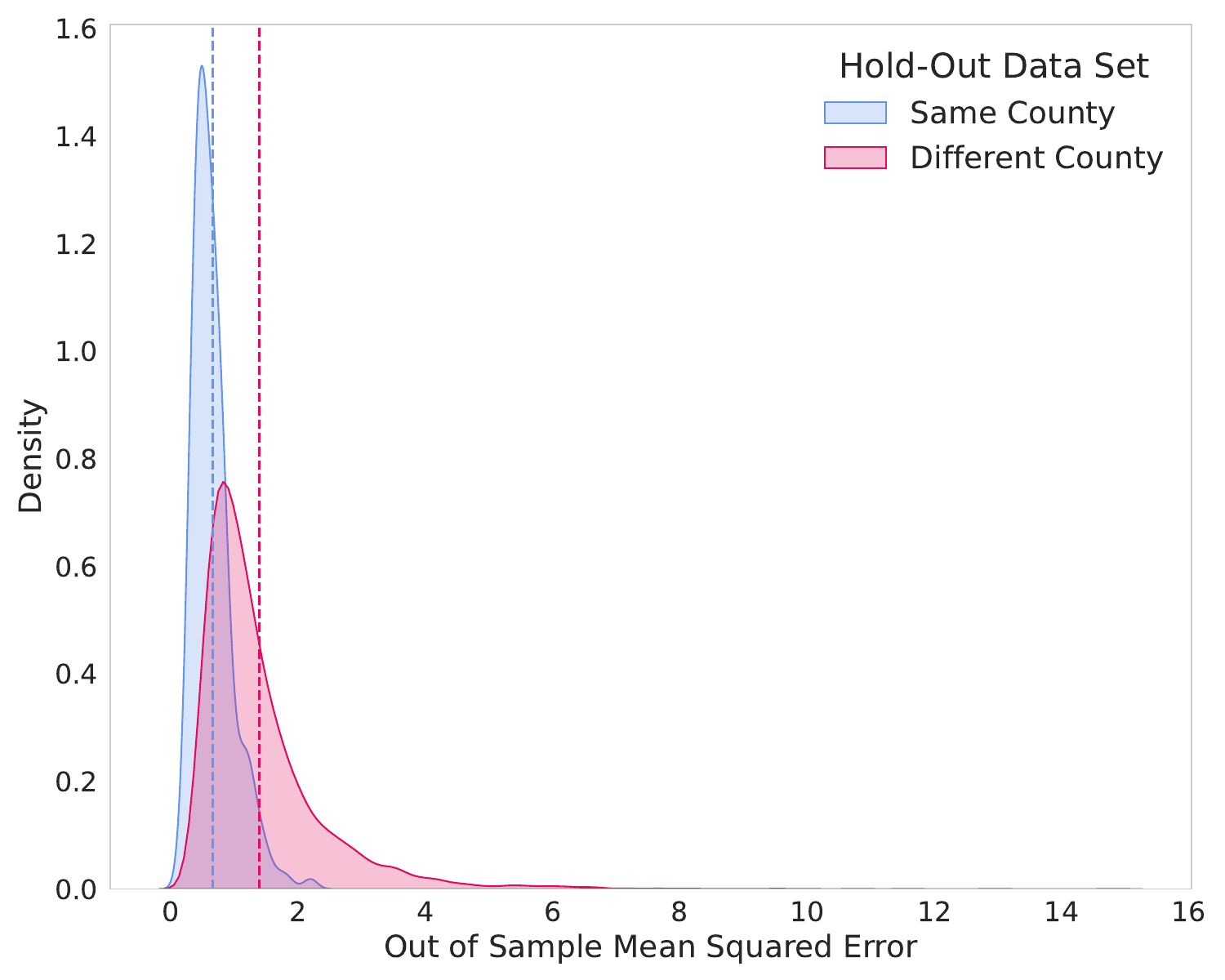} 
\end{tabular}
}
\label{afig:outofsample_countydatasim}
\end{center}
\end{figure}

\begin{footnotesize} 
\begin{singlespace}
\noindent \textsc{Notes}: This figure shows results from a simulation assessing how machine learning models trained in one county predict outcomes in another. For each of 400 counties, a gradient boosted machine is trained on pre-digitization transactions to predict the log of house sale prices. Model performance is then evaluated on two out-of-sample test sets: (1) a holdout sample from the same county (``same county'' in blue) and (2) a random sample from a different county (``different county'' in red), using only pre-digitization data in both cases. The simulation is repeated 100 times per county. The plot reports the average out-of-sample mean squared error for both evaluation types. Vertical dotted lines indicate the average mean squared error: 0.66 for the same-county test set (blue) and 1.394 for the different-county test set (red).

\end{singlespace}
\end{footnotesize}


\clearpage
\begin{table}[ht!]
\begin{center}
\caption{\textsc{Table \ref{atab:county_falsification}: Effects of County Digitization on Housing Market Dynamics}}
\vspace{20pt}        

\textbf{Panel A: OLS Results} \\
\vspace{10pt}
\scalebox{.6}{\makebox[\linewidth]{{
\def\sym#1{\ifmmode^{#1}\else\(^{#1}\)\fi}
\begin{tabular}{l*{6}{l}}
\toprule
                &\multicolumn{1}{c}{(1)}&\multicolumn{1}{c}{(2)}&\multicolumn{1}{c}{(3)}&\multicolumn{1}{c}{(4)}&\multicolumn{1}{c}{(5)}&\multicolumn{1}{c}{(6)}\\
                &\multicolumn{1}{c}{Total Human Investment}&\multicolumn{1}{c}{Human Investor Share}&\multicolumn{1}{c}{Mean Ln. House Price}&\multicolumn{1}{c}{Number of Sales}&\multicolumn{1}{c}{Median Days on the Market}&\multicolumn{1}{c}{Sales to List Ratio}\\
\midrule
County Digitization&1.959         &-0.006\sym{**} &-0.005         &-6.357         &-7.489         &-0.003         \\
                &(6.210)         &(0.002)         &(0.016)         &(40.577)         &(11.181)         &(0.004)         \\
\midrule
Dependent Variable Mean&24.930         &0.075         &10.851         &325.507         &156.537         &0.935         \\
R-squared       &0.870         &0.723         &0.934         &0.926         &0.449         &0.400         \\
Adjusted R-squared&0.856         &0.692         &0.927         &0.918         &0.366         &0.309         \\
Observations    &2056.000         &2056.000         &2052.000         &2056.000         &1467.000         &1466.000         \\
County + Year FE&Yes         &Yes         &Yes         &Yes         &Yes         &Yes         \\
Controls        &            &            &            &            &            &            \\
controls        &SocioEc + Housing         &SocioEc + Housing         &SocioEc + Housing         &SocioEc + Housing         &SocioEc + Housing         &SocioEc + Housing         \\
\bottomrule
\multicolumn{7}{l}{\footnotesize Standard errors in parentheses}\\
\multicolumn{7}{l}{\footnotesize \sym{*} \(p<0.10\), \sym{**} \(p<0.05\), \sym{***} \(p<0.01\)}\\
\end{tabular}
}
}}

\vspace{20pt}

\textbf{Panel B: 2SLS Results} \\
\vspace{10pt}
\scalebox{.5}{\makebox[\linewidth]{{
\def\sym#1{\ifmmode^{#1}\else\(^{#1}\)\fi}
\begin{tabular}{l*{7}{c}}
\toprule
            &\multicolumn{1}{c}{(1)}&\multicolumn{1}{c}{(2)}&\multicolumn{1}{c}{(3)}&\multicolumn{1}{c}{(4)}&\multicolumn{1}{c}{(5)}&\multicolumn{1}{c}{(6)}&\multicolumn{1}{c}{(7)}\\
            &\multicolumn{1}{c}{First Stage}&\multicolumn{1}{c}{Total Human Investment}&\multicolumn{1}{c}{Human Investor Share}&\multicolumn{1}{c}{Mean Ln. House Price}&\multicolumn{1}{c}{Number of Sales}&\multicolumn{1}{c}{Median Days on the Market}&\multicolumn{1}{c}{Sales to List Ratio}\\
\midrule
Data Quality $\times$ Digitization&      0.0133\sym{***}&                     &                     &                     &                     &                     &                     \\
            &  (0.000402)         &                     &                     &                     &                     &                     &                     \\
\addlinespace
Digitization $\times$ Entry&                     &       99.02         &     0.00446         &      0.0201         &      -135.6         &      -20.10         &    -0.00410         \\
            &                     &     (71.37)         &   (0.00447)         &    (0.0206)         &     (245.7)         &     (12.93)         &   (0.00400)         \\
\midrule
R-squared   &       0.953         &       0.927         &       0.829         &       0.975         &       0.973         &       0.665         &       0.664         \\
Adjusted R-squared&       0.949         &       0.920         &       0.812         &       0.972         &       0.970         &       0.619         &       0.618         \\
Observations&        3673         &        3674         &        3674         &        3670         &        3674         &        2648         &        2650         \\
County + Year FE&         Yes         &                     &         Yes         &         Yes         &         Yes         &         Yes         &         Yes         \\
Controls    &SocioEc + Housing         &SocioEc + Housing         &SocioEc + Housing         &SocioEc + Housing         &SocioEc + Housing         &SocioEc + Housing         &SocioEc + Housing         \\
F-Statistic &       123.1         &                     &                     &                     &                     &                     &                     \\
MP F-stat   &                     &      1003.9         &      1003.9         &      1003.7         &      1003.9         &       227.3         &       227.4         \\
MP 5\% Critical Value&                     &       37.42         &       37.42         &       37.42         &       37.42         &       37.42         &       37.42         \\
\bottomrule
\multicolumn{8}{l}{\footnotesize Standard errors in parentheses}\\
\multicolumn{8}{l}{\footnotesize \sym{*} \(p<0.10\), \sym{**} \(p<0.05\), \sym{***} \(p<0.01\)}\\
\end{tabular}
}
}}

\label{atab:county_falsification}
\end{center}
\end{table}

\begin{singlespace} 
\footnotesize
\noindent \textsc{Notes}: This table reports estimated effects of county-level digitization on housing market outcomes. Panel A presents OLS estimates for counties without algorithmic investor entry. Panel B reports 2SLS estimates based on Equation~\ref{eq:2sls_entry}, instrumenting algorithmic investor entry with pre-digitization measures of county data quality. The Montiel–Pflueger F-statistic (MP F-stat) tests for weak instruments under clustering, and the MP 5\% Critical Value indicates the threshold below which the maximum 2SLS bias may exceed 5\% of the OLS bias. All regressions include year and Census block group fixed effects and control for time-varying socioeconomic and housing characteristics from the American Community Survey. Observations are weighted by the number of housing transactions, and standard errors are clustered at the county level. Data are at the county–year level and drawn from Zillow, Redfin, Attom Data, county government records, and five-year ACS county-level extracts.
\end{singlespace}
\normalsize
\clearpage
\begin{figure}[ht!]
\begin{center}
\captionsetup{justification=centering}
\caption{\textsc{Figure \ref{afig:residprice}: Residualized Prices, by Seller Race}}
\makebox[\linewidth]{
\begin{tabular}{c}
\textsc{A. Before County Digitization}    \\
\includegraphics[scale=0.7]{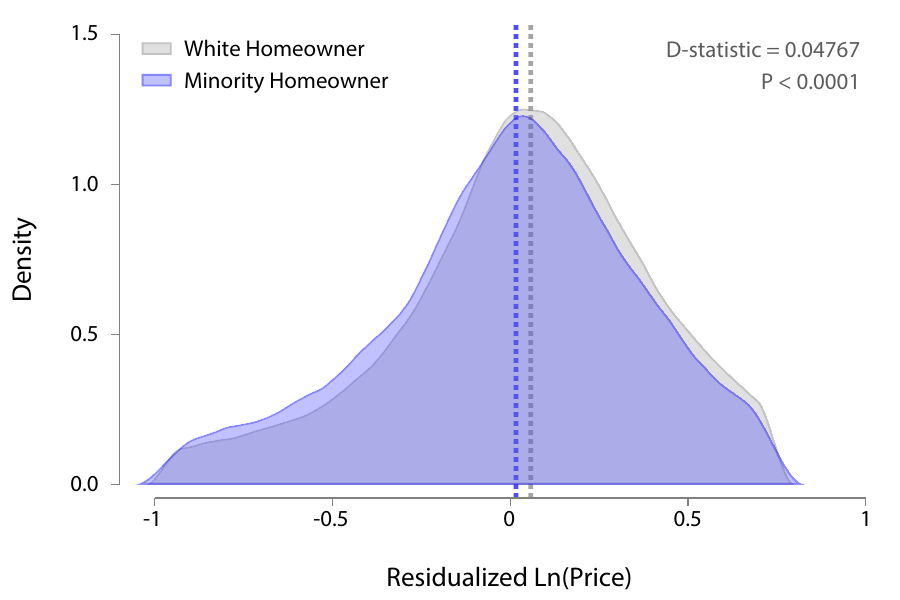}  \\
\textsc{B. After County Digitization} \\
\includegraphics[scale=0.7]{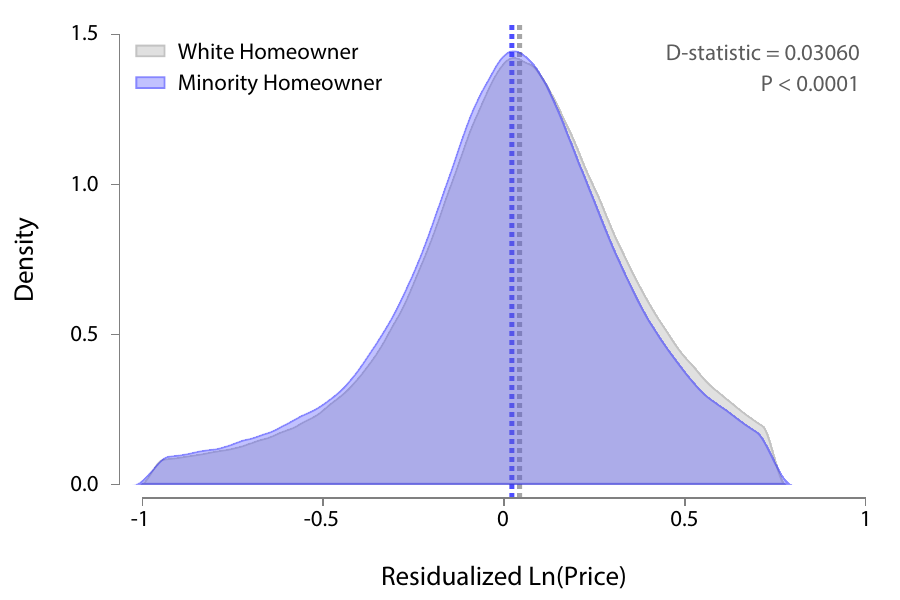} \\
\end{tabular}
} 
\label{afig:residprice}
\end{center}
\end{figure}

\begin{footnotesize} 
\begin{singlespace}
\noindent \textsc{Notes}: Panels A and B plot the distribution of residualized sale prices separately by seller race. Panel A shows transactions in county–years prior to digitization; Panel B shows transactions after digitization. Sale prices are residualized by regressing the natural log of sale price on observable housing characteristics (listed in Appendix~\ref{asec:housechars}) and block group–by–year fixed effects, following Equation~\ref{eq:racepenalty}. Distributions are plotted separately by seller race, and the Kolmogorov–Smirnov D-statistic is reported to quantify differences across distributions. Seller race is inferred using the \texttt{Ethnicolr} algorithm. Data are at the house–transaction level and drawn from Attom Data, county government records, and the U.S. Census.
\end{singlespace}
\end{footnotesize}

\clearpage
\begin{table}[ht!]
\begin{center}
                \caption{\textsc{Table \ref{atab:racepenalty_by_geo_bisg}: Race Penalty, by House Digitization with BISG Encoding }} 
                  \vspace{20pt}        
\scalebox{.8}{\makebox[\linewidth]{{
\def\sym#1{\ifmmode^{#1}\else\(^{#1}\)\fi}
\begin{tabular}{l*{3}{c}}
\toprule
                    &\multicolumn{1}{c}{(1)}         &\multicolumn{1}{c}{(2)}         &\multicolumn{1}{c}{(3)}         \\
\midrule
Minority Seller     &      -0.187\sym{***}&     -0.0871\sym{***}&     -0.0317\sym{***}\\
                    &    (0.0148)         &   (0.00831)         &    (0.0102)         \\
\addlinespace
Minority Seller $\times$ County Digitization&      0.0865\sym{***}&      0.0433\sym{***}&      0.0182\sym{*}  \\
                    &    (0.0143)         &   (0.00840)         &    (0.0103)         \\
\midrule
Dependent Variable Mean&       12.00         &       12.01         &       12.13         \\
R-squared           &       0.604         &       0.644         &       0.774         \\
Adjusted R-squared  &       0.592         &       0.618         &       0.697         \\
Observations        &     2141346         &     2123847         &     1488609         \\
Location            &Tract x Year         &Block Group x Year         &Block x Year         \\
Housing             &         Yes         &         Yes         &         Yes         \\
\bottomrule
\multicolumn{4}{l}{\footnotesize Standard errors in parentheses}\\
\multicolumn{4}{l}{\footnotesize \sym{*} \(p<0.10\), \sym{**} \(p<0.05\), \sym{***} \(p<0.01\)}\\
\end{tabular}
}
}}
  \label{atab:racepenalty_by_geo_bisg}
\end{center}
\end{table}

\begin{singlespace}
\footnotesize
\noindent \textsc{Notes}: This table reports estimated race penalty from Equation~\ref{eq:racepenalty} (\textit{Minority Seller}) and the marginal effect of county digitization on the race penalty (\textit{Minority Seller $\times$ County Digitization}). All regressions control for house characteristics, detailed in Appendix Section \ref{asec:housechars}, and include year-by-geography fixed effects. Standard errors are clustered at the relevant geographic level. \textit{Minority Seller} is an indicator for whether the seller is identified as Black or Hispanic (vs. White). Race is inferred using BISG. The data are at the house transaction level and come from Attom Data, the U.S. Census, and county governments. 
\end{singlespace}
\normalsize

\clearpage
\begin{table}[ht!]
\begin{center}
                \caption{\textsc{Table \ref{atab:racepenalty_coeff_tests}: Race Penalty Coefficient Tests}} 
                  \vspace{20pt}        
\scalebox{.7}{\makebox[\linewidth]{\begin{tabular}{lccccccc}
Test & Sample & Coefficient 1 & Coefficient 2 & Difference & Std. Error & P-value \\\\
\toprule
Pre-Period vs. Post-Period & All & -0.058 & -0.032 & -0.026 & 0.005 & 0.000 \\\\
Pre-Period vs. Post-Period (BISG) & All & -0.087 & -0.056 & -0.030 & 0.006 & 0.000 \\\\
Pre-Period vs. Post-Period (w/ AI Entry) & All & -0.058 & -0.024 & -0.034 & 0.005 & 0.000 \\\\
Pre-Period vs. Post-Period (w/o AI Entry) & All & -0.058 & -0.052 & -0.006 & 0.006 & 0.328 \\\\
Post-Period (w/ AI Entry) vs. Post-Period (w/o AI Entry) & All & -0.024 & -0.052 & 0.028 & 0.004 & 0.000 \\\\
Post-Period (w/ AI Entry) vs. Post-Period (w/o AI Entry) & Human Investors & -0.011 & -0.043 & 0.032 & 0.025 & 0.199 \\\\
Pre-Period vs. Post-Period (w/o AI Entry) & Human Investors & -0.098 & -0.043 & -0.055 & 0.044 & 0.210 \\\\
Post-Period (w/ AI Entry) vs. Post-Period (w/o AI Entry) & Owner Occupiers & -0.026 & -0.050 & 0.024 & 0.004 & 0.000 \\\\
Pre-Period vs. Post-Period (w/o AI Entry) & Owner Occupiers & -0.055 & -0.050 & -0.005 & 0.006 & 0.389 \\\\
\end{tabular}
}}
  \label{atab:racepenalty_coeff_tests}
\end{center}
\end{table}

\begin{singlespace}
\footnotesize
\noindent \textsc{Notes}: This table reports p-values from tests of equality in estimated race penalty coefficients across key subsamples: (i) before vs. after county digitization, (ii) by whether algorithmic investors entered the county post-digitization, and (iii) by buyer type. The corresponding coefficient estimates are shown in Figure~\ref{fig:racepenalty_subsampleanalysis}. Data are at the house–transaction level and drawn from Attom Data, county government records, and the U.S. Census.
\end{singlespace}
\normalsize

\clearpage
\begin{table}[ht!]
\begin{center}
                \caption{\textsc{Table \ref{atab:racepenalty_mechanism}: Indirect Effects of Digitization on the Race Penalty}} 
                  \vspace{20pt}        
\scalebox{.6}{\makebox[\linewidth]{{
\def\sym#1{\ifmmode^{#1}\else\(^{#1}\)\fi}
\begin{tabular}{l*{3}{c}}
\toprule
                    &\multicolumn{1}{c}{(1)}         &\multicolumn{1}{c}{(2)}         &\multicolumn{1}{c}{(3)}         \\
\midrule
Minority Seller     &     -0.0701\sym{***}&     -0.0665\sym{***}&     -0.0987\sym{**} \\
                    &   (0.00539)         &   (0.00545)         &    (0.0408)         \\
\addlinespace
County Digitized $\times$ Minority Seller&    -0.00401         &     -0.0104         &      0.0557         \\
                    &    (0.0113)         &    (0.0115)         &    (0.0563)         \\
\addlinespace
Minority Homeowner=1 $\times$ County Digitization=1 $\times$ Digitized House=1&      0.0321\sym{***}&      0.0335\sym{***}&      0.0305         \\
                    &    (0.0102)         &    (0.0105)         &    (0.0402)         \\
\midrule
Dependent Variable Mean&       11.90         &       11.93         &       11.44         \\
R-squared           &       0.521         &       0.526         &       0.691         \\
Adjusted R-squared  &       0.477         &       0.480         &       0.534         \\
Observations        &     2433151         &     2279310         &       99716         \\
Location x Year FE  &Block Group x Year         &Block Group x Year         &Block Group x Year         \\
Housing FE          &         Yes         &         Yes         &         Yes         \\
Sample              &No Algo. Inv.         &Owner-Occupiers         &Human Investors         \\
\bottomrule
\multicolumn{4}{l}{\footnotesize Standard errors in parentheses}\\
\multicolumn{4}{l}{\footnotesize \sym{*} \(p<0.10\), \sym{**} \(p<0.05\), \sym{***} \(p<0.01\)}\\
\end{tabular}
}
}}
  \label{atab:racepenalty_mechanism}
\end{center}
\end{table}

\begin{singlespace}
\footnotesize
\noindent \textsc{Notes}: This table reports estimated coefficients from Equation~\ref{eq:triple_diff_digit_racepenalty}. All regressions control for house characteristics and include year-by-block group fixed effects. Standard errors are clustered at the block group level. \textit{County Digitization} is an indicator equal to one if the county has adopted digitized property records. \textit{Digitized House} is an indicator for whether the specific house has been digitized. \textit{Minority Seller} is an indicator equal to one if the seller is identified as Black or Hispanic (vs.\ White), with race inferred using the \texttt{Ethnicolr} procedure. Column 1 includes purchases by human investors and owner-occupiers, excluding algorithmic investors. Column 2 includes only owner-occupier purchases, and Column 3 includes only human investor purchases. The data are at the house transaction level and are drawn from Attom Data, the U.S. Census, and county governments.
\end{singlespace}
\normalsize

\clearpage

\clearpage
\begin{figure}[ht!]
\begin{center}
\captionsetup{justification=centering}
\caption{\textsc{Figure \ref{afig:scrapedimages}: House Images}}
\makebox[\linewidth]{
\begin{tabular}{c}
\textsc{A.  Sample House Exterior Image}\\
\includegraphics[scale=0.4, angle=270]{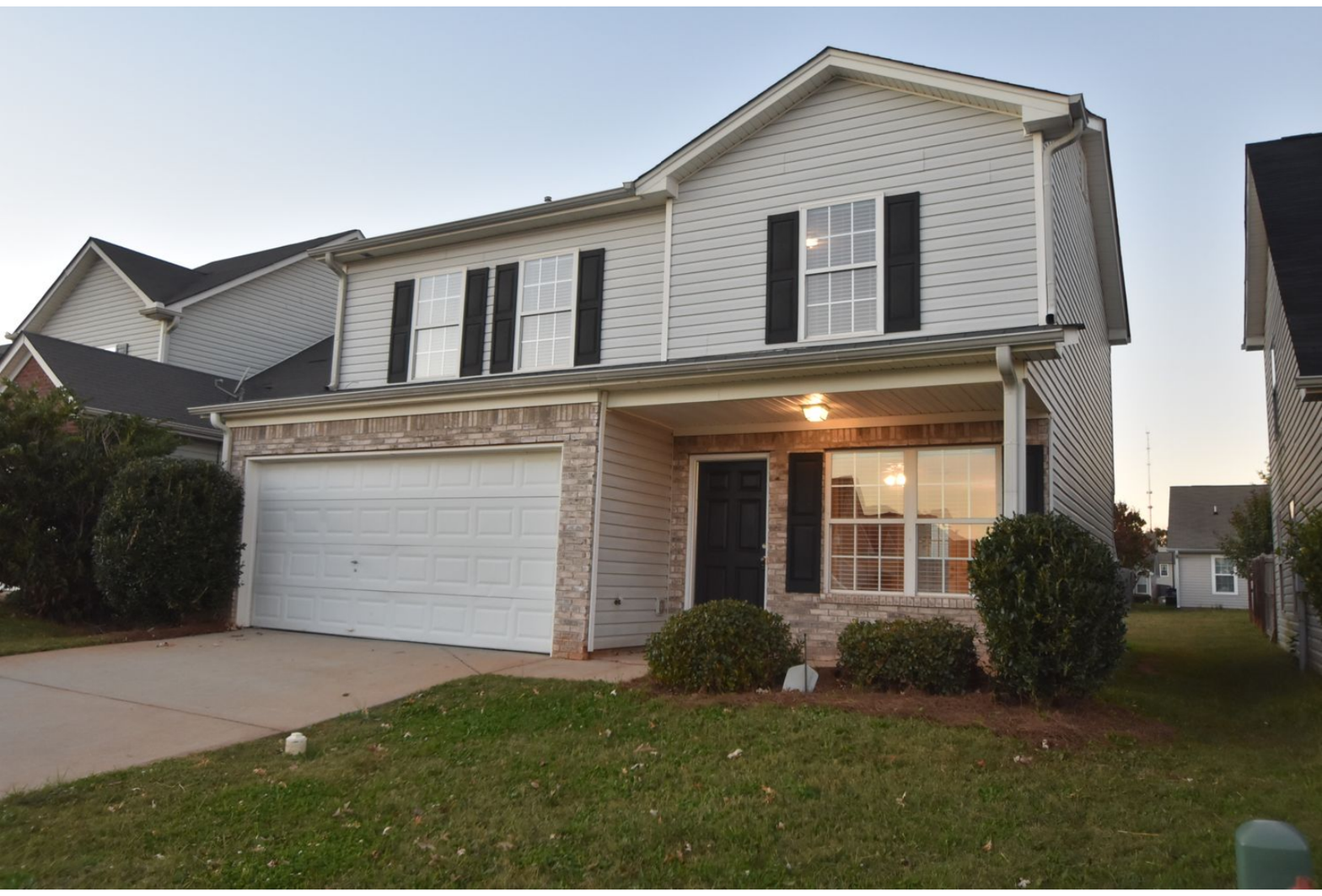}  \\
\textsc{B.  Sample House Interior Image}\\
\includegraphics[scale=0.4, angle=270]{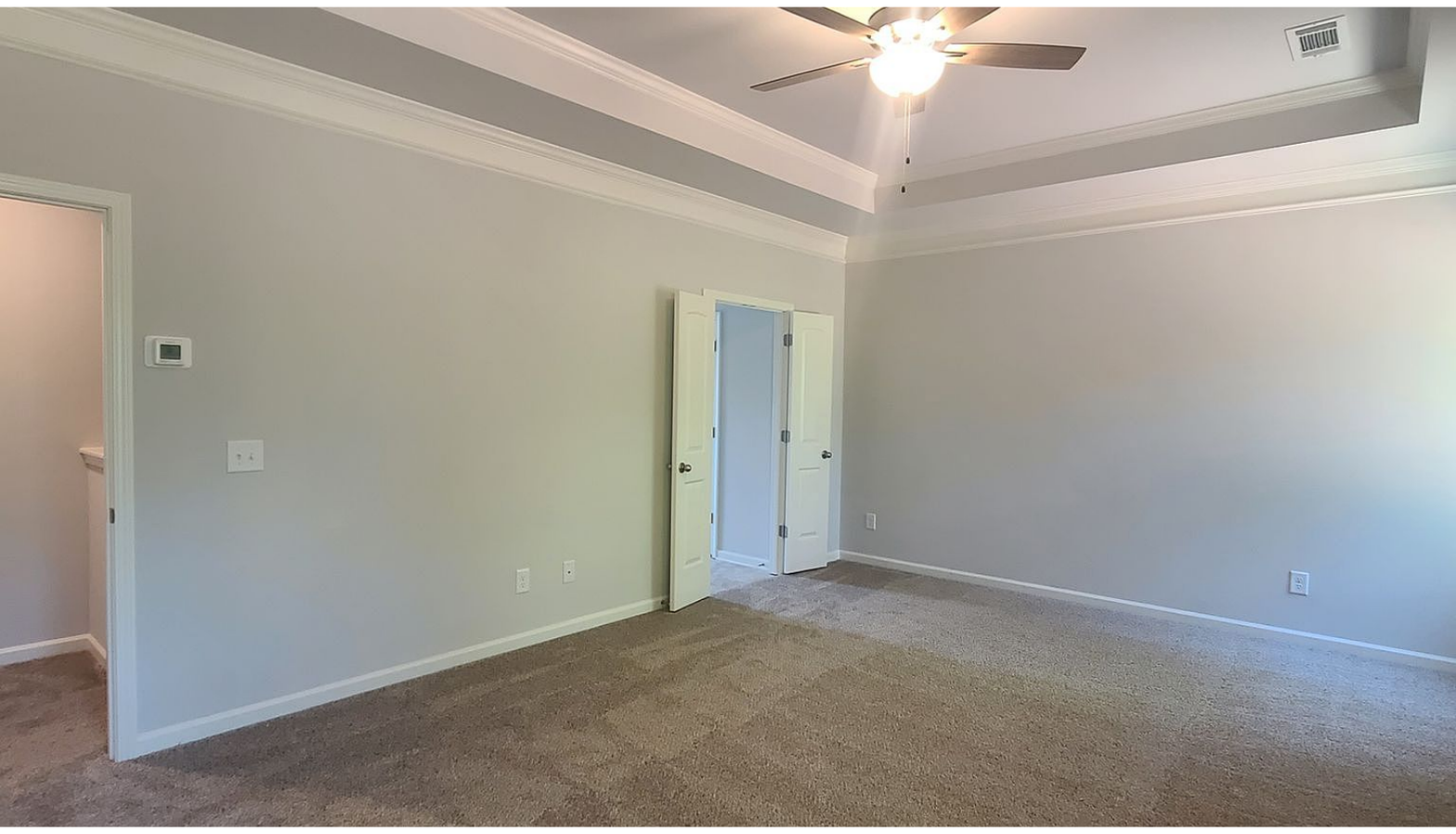} \\
\end{tabular}
}
\label{afig:scrapedimages}
\end{center}
\end{figure}

\begin{footnotesize} 
\begin{singlespace}
\noindent \textsc{Notes}: This figure shows an example of the exterior (Panel A) and interior (Panel B) images used to create embeddings with the deep learning model described in Appendix Section \ref{asec:images}. House images were collected from Zillow's website for houses listed for sale in the summer of 2023. 
\end{singlespace}
\end{footnotesize}

\clearpage
\begin{table}[ht!]
\begin{center}
                \caption{\textsc{Table \ref{atab:summary_stats_houseswithimages}: Summary Statistics of Houses Matched to Images and Full Sample}} 
                  \vspace{20pt}        
\scalebox{.8}{\makebox[\linewidth]{{
\def\sym#1{\ifmmode^{#1}\else\(^{#1}\)\fi}
\begin{tabular}{l*{2}{c}}
\toprule
                    &\multicolumn{1}{c}{(1)}&\multicolumn{1}{c}{(2)}\\
                    &\multicolumn{1}{c}{Full Sample}&\multicolumn{1}{c}{Houses with Images}\\
\midrule
Sale Price          &  207,248.27         &  176,105.82         \\
                    &(754,033.05)         &(475,621.82)         \\
\addlinespace
Bedrooms            &        2.11         &        2.58         \\
                    &      (3.22)         &      (1.54)         \\
\addlinespace
Bathrooms           &        2.14         &        2.19         \\
                    &      (2.37)         &      (0.97)         \\
\addlinespace
Partial Baths       &        0.27         &        0.31         \\
                    &      (0.48)         &      (0.48)         \\
\addlinespace
Stories             &        1.25         &        1.36         \\
                    &      (0.75)         &      (0.63)         \\
\addlinespace
Buildings           &        0.06         &        0.06         \\
                    &      (0.58)         &      (0.37)         \\
\addlinespace
Garage              &        0.56         &        0.63         \\
                    &      (0.50)         &      (0.48)         \\
\addlinespace
Fireplace           &        0.59         &        0.70         \\
                    &      (0.49)         &      (0.46)         \\
\addlinespace
Basement            &        0.17         &        0.13         \\
                    &      (0.37)         &      (0.34)         \\
\addlinespace
Parking Spaces      &        0.75         &        0.68         \\
                    &      (8.83)         &      (2.54)         \\
\addlinespace
House Age           &       31.13         &       32.61         \\
                    &     (25.76)         &     (27.48)         \\
\addlinespace
Age Since Remodel   &       24.44         &       26.08         \\
                    &     (21.08)         &     (23.66)         \\
\midrule
N                   &   6,727,758         &      35,499         \\
\bottomrule
\multicolumn{3}{l}{\footnotesize mean coefficients; sd in parentheses}\\
\multicolumn{3}{l}{\footnotesize \sym{*} \(p<0.05\), \sym{**} \(p<0.01\), \sym{***} \(p<0.001\)}\\
\end{tabular}
}
}}
  \label{atab:summary_stats_houseswithimages}
\end{center}
\end{table}

\begin{singlespace}
\footnotesize
\noindent \textsc{Notes}: This table presents statistics on house characteristics for the sample of houses transactions in Column 1 and those matched to interior and exterior images in Column 2. Means and standard deviations, are reported in parentheses. Data are at the house transaction level and are drawn from Attom Data. 
\end{singlespace}
\normalsize

\clearpage
\begin{table}[ht!]
\begin{center}
                \caption{\textsc{Table \ref{atab:racepenalty_images}: Race Penalty, with House Images}} 
                  \vspace{20pt}        
\scalebox{.8}{\makebox[\linewidth]{{
\def\sym#1{\ifmmode^{#1}\else\(^{#1}\)\fi}
\begin{tabular}{l*{3}{c}}
\toprule
                    &\multicolumn{1}{c}{With Available House Characteristics}&\multicolumn{1}{c}{With House Image Embeddings}&\multicolumn{1}{c}{With Interior Images}\\\cmidrule(lr){2-2}\cmidrule(lr){3-3}\cmidrule(lr){4-4}
                    &\multicolumn{1}{c}{(1)}         &\multicolumn{1}{c}{(2)}         &\multicolumn{1}{c}{(3)}         \\
\midrule
Minority Seller     &     -0.0588\sym{***}&     -0.0507\sym{***}&     -0.0516\sym{***}\\
                    &    (0.0169)         &    (0.0177)         &    (0.0177)         \\
\midrule
Dependent Variable Mean&       11.78         &       11.78         &       11.81         \\
R-squared           &       0.874         &       0.887         &       0.887         \\
Adjusted R-squared  &       0.795         &       0.802         &       0.802         \\
Observations        &      16,893         &      16,893         &      16,893         \\
Location FE         &Block Group x Year         &Block Group x Year         &Block Group x Year         \\
Year FE             &         Yes         &         Yes         &         Yes         \\
Housing Attrs       &          No         &         Yes         &         Yes         \\
\bottomrule
\multicolumn{4}{l}{\footnotesize Standard errors in parentheses}\\
\multicolumn{4}{l}{\footnotesize \sym{*} \(p<0.10\), \sym{**} \(p<0.05\), \sym{***} \(p<0.01\)}\\
\end{tabular}
}
}}
  \label{atab:racepenalty_images}
\end{center}
\end{table}

\begin{singlespace}
\footnotesize
\noindent \textsc{Notes}:  This table reports estimated race penalty coefficients following Equation~\ref{eq:racepenalty}. Each column reports the results of separate regressions. Column 1 include controls for house characteristics listed in Appendix Section \ref{asec:housechars} and geography-by-year fixed effects. Column 2 add vector embeddings of exterior property images, generated using a deep learning model, to account for visual features of the home’s appearance. Column 3 further add vector embeddings of interior images. The sample is limited to transactions prior to county digitization and includes only homes for which both exterior and interior images could be programmatically scraped from Zillow in the summer of 2023. All regressions include housing characteristics and year-by-geography fixed effects. Standard errors are clustered at the relevant geographic level. The data are at the house-transaction level and are drawn from Attom Data, the U.S. Census, and Zillow.
\end{singlespace}
\normalsize

\clearpage

\begin{table}[ht!]
\begin{center}
                \caption{\textsc{Table \ref{atab:assessimp}: Assessed Improvements by Geography, Race, and Investor Type}}
                  \vspace{20pt}        
\scalebox{0.75}{\makebox[\linewidth]{{
\def\sym#1{\ifmmode^{#1}\else\(^{#1}\)\fi}
\begin{tabular}{l*{6}{c}}
\toprule
                    &\multicolumn{2}{c}{Tract}                  &\multicolumn{2}{c}{Block Group}            &\multicolumn{2}{c}{Block}                  \\\cmidrule(lr){2-3}\cmidrule(lr){4-5}\cmidrule(lr){6-7}
                    &\multicolumn{1}{c}{(1)}         &\multicolumn{1}{c}{(2)}         &\multicolumn{1}{c}{(3)}         &\multicolumn{1}{c}{(4)}         &\multicolumn{1}{c}{(5)}         &\multicolumn{1}{c}{(6)}         \\
\midrule
Minority Seller     &       21.67         &     -5211.3\sym{***}&      -150.0         &     -4210.9\sym{***}&      -55.80         &     -2422.5\sym{***}\\
                    &     (561.4)         &     (309.6)         &     (498.8)         &     (297.9)         &     (567.8)         &     (299.7)         \\
\midrule
Dependent Variable Mean&    173476.0         &    192534.5         &    174029.5         &    192625.9         &    177105.3         &    208144.9         \\
R-squared           &       0.843         &       0.505         &       0.861         &       0.545         &       0.916         &       0.736         \\
Adjusted R-squared  &       0.820         &       0.496         &       0.832         &       0.521         &       0.875         &       0.649         \\
Observations        &       69693         &     3593735         &       64843         &     3590017         &       38363         &     2637760         \\
Location            &Tract x Year         &Tract x Year         &Block Group x Year         &Block Group x Year         &Block x Year         &Block x Year         \\
Housing             &         Yes         &         Yes         &         Yes         &         Yes         &         Yes         &         Yes         \\
Sample              &Algo. Investors         &Owner Occupiers         &Algo. Investors         &Owner Occupiers         &Algo. Investors         &Owner Occupiers         \\
\bottomrule
\multicolumn{7}{l}{\footnotesize Standard errors in parentheses}\\
\multicolumn{7}{l}{\footnotesize \sym{*} \(p<0.10\), \sym{**} \(p<0.05\), \sym{***} \(p<0.01\)}\\
\end{tabular}
}
}}

  \label{atab:assessimp}
\end{center}
\end{table}

\begin{singlespace}
\footnotesize
\noindent \textsc{Notes}: This table reports the assessed value of home improvements made by algorithmic investors, human investors, and owner-occupiers during their period of ownership, as recorded by county assessor offices. Columns 1--2 compare houses within the same Census tract; Columns 3--4 compare within the same block group; and Columns 5--6 compare houses within the same Census block. \textit{Minority Seller} is an indicator equal to $1$ if the previous homeowner was identified as Black or Hispanic, and $0$ if White, with race inferred using the \texttt{Ethnicolr} procedure. All specifications include assessment year and sale year by geography fixed effects and observable house characteristics. Standard errors are clustered at the geographic level, and data are at the house transaction level. All data comes from Attom Data. 
\end{singlespace}
\normalsize


\clearpage
\begin{figure}[ht!]
\begin{center}
\captionsetup{justification=centering}
\caption{\textsc{Figure \ref{afig:xgboost_resids}: House Predictability Out-of-Sample Accuracy and Residuals}}
\makebox[\linewidth]{
\begin{tabular}{c}
\textsc{A. Predicted versus Actual Ln(Sales Price)} \\
\includegraphics[scale=0.4]{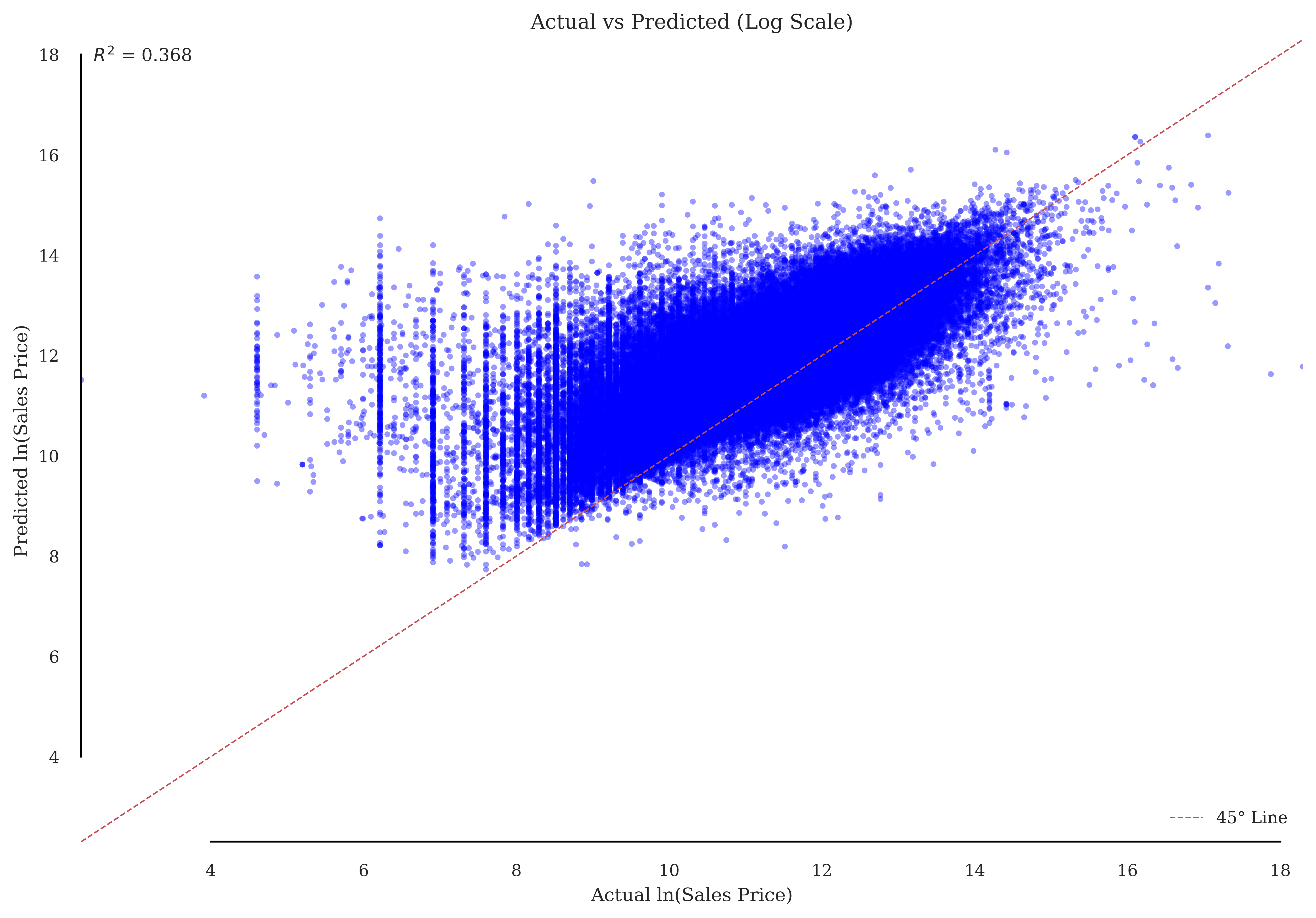} \\
\textsc{B. Sextiles of House Predictability} \\
\includegraphics[scale=0.3]{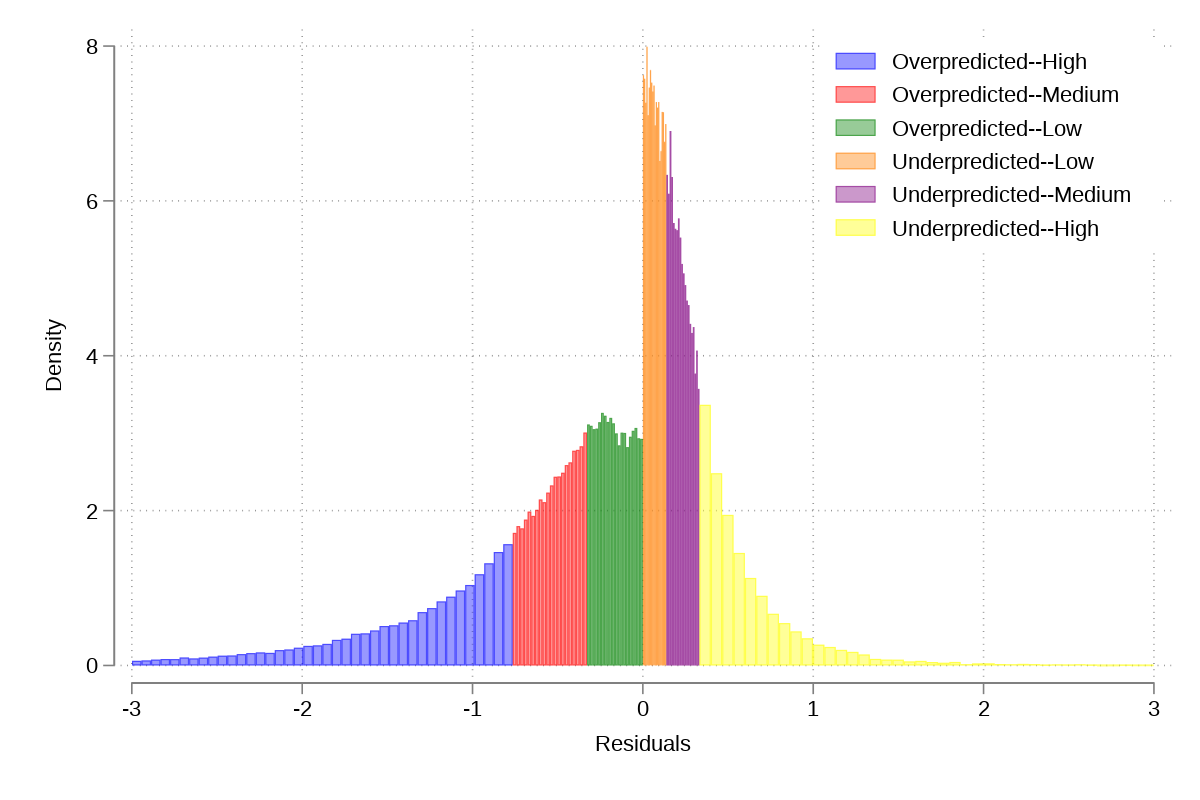} \\

\end{tabular}
}
\label{afig:xgboost_resids}
\end{center}
\end{figure}

\begin{footnotesize} 
\begin{singlespace}
\noindent \textsc{Notes}:  Panel A plots actual versus predicted log sale prices for pre-digitization transactions in the hold-out test set, using the extreme gradient boosting (XGBoost) model described in Section~\ref{asec:buildingpredictability}.
Panel B shows the distribution of baseline prediction residuals (predicted minus actual price) for these transactions, partitioned into sextiles separately for overpredicted and underpredicted homes. 
\end{singlespace}
\end{footnotesize}

\clearpage
\begin{table}[ht!]
\begin{center}
                \caption{\textsc{Table \ref{atab:specialization}: Heterogeneous Effects of Digitization by  Baseline Predictability}}
                  \vspace{20pt}        
\scalebox{.8}{\makebox[\linewidth]{{
\def\sym#1{\ifmmode^{#1}\else\(^{#1}\)\fi}
\begin{tabular}{l*{3}{c}}
\toprule
                    &\multicolumn{1}{c}{(1)}&\multicolumn{1}{c}{(2)}&\multicolumn{1}{c}{(3)}\\
                    &\multicolumn{1}{c}{Human Investment}&\multicolumn{1}{c}{Owner Occupier Purchase}&\multicolumn{1}{c}{Ln(Sales Price)}\\
\midrule
Digit x Under--High &      0.0352\sym{***}&     -0.0356\sym{***}&      0.0374         \\
                    &   (0.00982)         &   (0.00982)         &    (0.0257)         \\
\addlinespace
Digit x Under--Medium&      0.0115         &     -0.0118         &      0.0873\sym{***}\\
                    &   (0.00897)         &   (0.00897)         &    (0.0249)         \\
\addlinespace
Digit x Under--Low  &      0.0261\sym{***}&     -0.0264\sym{***}&      0.0591\sym{**} \\
                    &   (0.00861)         &   (0.00862)         &    (0.0238)         \\
\addlinespace
Digit x Over--Low   &      0.0200\sym{***}&     -0.0204\sym{***}&      0.0176         \\
                    &   (0.00738)         &   (0.00740)         &    (0.0224)         \\
\addlinespace
Digit x Over--Medium&      0.0126\sym{*}  &     -0.0127\sym{**} &      0.0252         \\
                    &   (0.00646)         &   (0.00646)         &    (0.0196)         \\
\addlinespace
Digit x Over--High  &     -0.0175\sym{**} &      0.0176\sym{**} &      0.0319         \\
                    &   (0.00815)         &   (0.00815)         &    (0.0249)         \\
\midrule
Dependent Variable Mean&      0.0680         &       0.932         &       11.57         \\
R-squared           &       0.258         &       0.258         &       0.722         \\
Adjusted R-squared  &      0.0810         &      0.0807         &       0.653         \\
Observations        &      195319         &      195319         &      174574         \\
Location FE         &Block Group x Year         &Block Group x Year         &Block Group x Year         \\
\bottomrule
\multicolumn{4}{l}{\footnotesize Standard errors in parentheses}\\
\multicolumn{4}{l}{\footnotesize \sym{*} \(p<0.10\), \sym{**} \(p<0.05\), \sym{***} \(p<0.01\)}\\
\end{tabular}
}
}}
  \label{atab:specialization}
\end{center}
\end{table}

\begin{singlespace}
\footnotesize
\noindent \textsc{Notes}: This table reports coefficients from regressions of (1) human investor purchase, (2) owner–occupier purchase, and (3) log sale price on the interaction between a house-level digitization indicator and a measure of baseline predictability. Predictability is defined as the out-of-sample prediction error, $y - \hat{y}$, from an XGBoost model trained on a random split of pre-digitization transactions to predict log sale price. Scores are assigned to homes in the pre-digitization validation sample, restricting the analysis to repeat-sale properties. Heterogeneous effects are estimated by sextile of prediction error: ``Under–High'' denotes the tercile with the largest positive residuals, and ``Over–High'' denotes the tercile with the largest negative residuals. Sextiles are constructed separately above and below zero based on the residual distribution. All regressions include block group–by–year fixed effects, with standard errors clustered at the block group level. The unit of observation is a house transaction; data are from Attom Data.
\end{singlespace}
\normalsize



\newpage
\clearpage
\section{Data and Model Appendix}\label{asec:datamodelsappendix}
\subsection{House Characteristics}  \label{asec:housechars}

Regressions that include ``house characteristics'' refer to a rich set of property details drawn from the assessor data to control for differences in physical features and structure. These include the number of bedrooms, bathrooms, half-bathrooms, and total rooms; indicators for the presence of a basement, garage, carport, fireplace, pool, and parking spaces; and specifics on building configuration, such as the number of stories; number of units and buildings; and whether the home is a split-level, townhouse, ranch, bungalow, or of contemporary, colonial French, cottage, Cape Cod or custom design. I also include square footage measures for the first and second floors, upper levels, finished and unfinished areas, decks, basements, and the total gross and building area. Additional controls capture construction material, roof type (e.g., flat or gable), foundation type, and lot size.

\subsection{Instrumenting for Algorithmic Investor Entry with Data Quality}\label{asec:instrumentingentry}

I instrument for the interaction term between digitization and algorithmic investor entry ($\text{TreatAlgoEntry}_{ct}$),
\begin{equation*}\text{TreatAlgoEntry}_{ct} = \text{Digitized}_{ct} \times \text{AlgoEntry}_c, \end{equation*} with the interaction between digitization and pre-period data quality ($Z_{ct}$):
\begin{equation*}
Z_{ct} = \text{Digitized}_{ct} \times \text{DataQuality}_c.
\end{equation*}
Data quality is defined as the comprehensiveness of the assessor characteristics collected by each county in the pre-digitization period. Counties differ in the specific house attributes their assessors record, reflecting local regulations that guide property tax assessments. I measure data quality as the average share of non-missing assessor fields across all transactions observed in a county before digitization. 

The first-stage regression is:
\begin{equation*}
\text{TreatAlgoEntry}_{ct} = \pi_0 + \pi_1 Z_{ct} + \gamma X_{ct} + \lambda_c + \delta_t + \varepsilon_{ct},
\end{equation*}
and the second-stage regression is:
\begin{equation}\label{eq:2sls_entry}
Y_{ct} = \beta_0 + \beta_1 \widehat{\text{TreatAlgoEntry}}_{ct} + \gamma X_{ct} + \lambda_c + \delta_t + \eta_{ct},
\end{equation}
where $Y_{ct}$ is the outcome of interest (e.g., human investor share, average sale price, sale-to-list ratio), $X_{ct}$ includes the standard set of time-varying county-level controls, $\lambda_c$ are county fixed effects, and $\delta_t$ are year fixed effects. Standard errors are clustered at the county level.


Panel B of Appendix Table~\ref{atab:county_falsification} reports the results. Data quality instrument is strong (first-stage F-statistic = 123 and Montiel-Pflueger F-statistic above 5\% maximal bias threshold) and predicts algorithmic investor entry. As shown in Column 1, a 1 percentage point increase in data quality raises the likelihood of algorithmic entry by 1.33 percentage points; equivalently, a one standard deviation increase in data quality increases entry by 22 percentage points. The 2SLS estimates in Columns 2–-7 are similar in magnitude to the OLS results in Panel A but less precise. Across outcomes, including human investor activity, average sale prices, and time on market, there is no evidence that digitization affects housing market dynamics once selection into algorithmic entry is taken into account.

\subsection{Inferring Race from Homeowner Name}  \label{asec:inferringrace}

\subsubsection{BISG: Bayesian Improved Surname Geocoding} \label{asec:bisg}
BISG combines information on individual surnames with the racial composition of neighborhoods to probabilistically impute race or ethnicity. Developed by the U.S. Census Bureau and RAND researchers, BISG applies Bayes’ Rule to update prior probabilities of race, drawn from the 2010 Census Surname List, using the racial distribution of the seller’s Census block group \citep{elliottUsingCensusBureau2009}. The Consumer Financial Protection Bureau (CFPB) has validated BISG for fair lending analysis \citep{consumer_financial_protection_bureau_using_2014}, finding that it correctly classifies race in approximately 79-84\% of cases. Reported AUC scores for Hispanic, White, Black, and Asian categories are 0.9446, 0.9430, 0.9540, and 0.9723, respectively. BISG outperforms methods that rely solely on surname or location, and its accuracy improves with more granular geographic identifiers (e.g., block group vs.\ ZIP code). The CFPB concluded that BISG is a reasonable and conservative proxy for race in settings without self-reported information.

\subsubsection{Ethnicolr: Machine Learning Approach} \label{asec:ethnicolr}
To complement BISG, I also implement the \texttt{ethnicolr} package, an open-source Python library developed by researchers at the University of Georgia and Florida State University for predicting race and ethnicity from names using deep learning methods \citep{ethnicolr2023}. \texttt{Ethnicolr} has been widely adopted in recent economics and social science research to impute race when survey or administrative data are unavailable \citep[e.g.,][]{box-couillard_racial_2024, blattner2021costly, koffi2024racial}. I use a model trained on a combination of U.S. Census data and voter registration records from Florida and North Carolina; it employs a Long Short-Term Memory (LSTM) neural network to model character sequences in full names. In out-of-sample validation, this model achieves classification accuracies of 92\%, 74\%, 86\%, and 63\% for White, Black, Hispanic, and Asian names, respectively, and significantly outperforms surname-only approaches.

Unlike BISG, which combines surname and geographic data via Bayes’ Rule, \texttt{ethnicolr} uses both first and last names and does not rely on geographic context. This makes it particularly useful when geographic identifiers are missing or when researchers seek to separate racial classification from neighborhood composition, as is relevant in this analysis. In practice,  BISG and \texttt{ethnicolr} produce highly similar race classifications in my data. I use \texttt{ethnicolr} as the primary specification, as it yields more conservative estimates of the race penalty.

\subsection{Effects of Competition versus Spillovers on the Race Penalty Specification Details}\label{asec:triplediff_racepenalty}
The estimating equation is:
\begin{align} 
\ln(p_{ibt}) =\;
& \beta_1 \cdot \text{Minority}_{it} + \mathbf{X}_{ibt}'\boldsymbol{\gamma} + \lambda_{bt} \notag \\
& + \sum_{d \in \{0,1\}} \big[ \; 
    \beta_2^{(d)} \cdot D_{ct} \cdot \mathds{1}\{D^{\text{house}}_{it} = d\} \notag \\
& \quad\quad\quad\; + \beta_3^{(d)} \cdot \text{Minority}_{it} \cdot D_{ct} \cdot \mathds{1}\{D^{\text{house}}_{it} = d\} \; \big] \notag \\
& + \varepsilon_{ibt}
\label{eq:triple_diff_digit_racepenalty}
\end{align}

\noindent where:
\begin{itemize}
\item $p_{ibt}$ is the sale price of house $i$ in Census block group $b$ and year $t$,
\item $\text{Minority}{it}$ is an indicator for whether the seller is a minority,
\item $D{ct}$ indicates whether county $c$ is digitized in year $t$,
\item $D^{\text{house}}_{it}$ indicates whether the specific house is digitized,
\item $\mathds{1}\{D^{\text{house}}_{it} = d\}$ denotes whether $d=1$ (digitized) or $d=0$ (not digitized),
\item $\lambda_{bt}$ are Census block group-by-year fixed effects,
\item $\mathbf{X}{ibt}$ is a vector of housing characteristics,
\end{itemize}

The coefficients of interest are $\beta_3^{(1)}$ and $\beta_3^{(0)}$:
\begin{itemize}
\item $\beta_3^{(1)}$ captures the effect of county digitization on the racial price gap for \textit{digitized} homes, which are more exposed to algorithmic bidding. A positive value implies algorithmic competition reduces the race penalty.
\item $\beta_3^{(0)}$ captures the effect of county digitization on the racial price gap for \textit{non-digitized} homes, which are harder to target but are possibly affected through spillover channels like appraisal anchoring or comp-based pricing. A positive value indicates spillovers reduce the race penalty.
\end{itemize}
This regression is estimated with OLS only on purchases by non-algorithmic buyers. 
Because the regression includes block group-by-year fixed effects, identification comes from comparisons between digitized and non-digitized homes within the same neighborhood and year. 

\subsection{Race Penalty Decomposition Details}\label{asec:oaxaca_decom}

I decompose the change in the average race penalty $\beta_M$, the residual price gap into two components: one capturing changes in who is buying the homes (composition effect) and the second capturing the change in the race penalty within each buyer group (within-buyer price effect). $s_{b}^{post}$ is the share of transactions bought by buyer type $b$ in the post period, and $\beta_{M,b}^{post}$ is the average race penalty by buyer type $b$ in the post period. Shares and race penalties in the pre-period are defined as  $s_{b}^{pre}$ and $\beta_{M,b}^{pre}$. Existing buyers are owner-occupiers and human investors, who overlap in the pre and post periods. Algorithmic investors only have positive market share in the post period, zero in the pre-period, and an undefined race penalty in the pre-period, so I separate out their contribution into a new group effect. 
\begin{align*} 
\Delta\beta_M &= \sum_b [s_{b}^{post}\beta_{M,b}^{post} - s_{b}^{pre}\beta_{M,b}^{pre}] \\
& = \sum_b [\underbrace{s_{b}^{post}(\beta_{M,b}^{post} - \beta_{M,b}^{pre})}_{\text{within-buyer effect}}  + \underbrace{\beta_{M,b}^{pre}(s_{b}^{post} - s_{b}^{pre})}_{\text{composition effect}}] \\
& = \sum_{b \in \text{existing buyers}} [\underbrace{s_{b}^{post}(\beta_{M,b}^{post} - \beta_{M,b}^{pre})}_{\text{within-buyer effect}}  + \underbrace{\beta_{M,b}^{pre}(s_{b}^{post} - s_{b}^{pre})}_{\text{composition effect}}] + \underbrace{s^{post}_{algo}\beta_{M,algo}^{post}}_{\text{new group effect}}
\end{align*}
 The composition effect captures how much of the change is due to shifts in the distribution of existing buyer types, and the within-buyer effect captures differences due to changes in the race penalty. 

The overall share driven by the composition effect is:
\begin{align*} 
 = \frac{\sum_{b \in \text{existing buyers}} [\beta_{M,b}^{pre}(s_{b}^{post} - s_{b}^{pre})]}{\Delta\beta_M}
\end{align*}
And overall share driven by the within-buyer effect is:
\begin{align*} 
 = \frac{\sum_{b \in \text{existing buyers}} [s_{b}^{post}(\beta_{M,b}^{post} - \beta_{M,b}^{pre})]}{\Delta\beta_M}
\end{align*}

And overall share driven by the new-group of buyers is:
\begin{align*} 
 = \frac{s^{post}_{algo}\beta_{M,algo}^{post}}{\Delta\beta_M}
\end{align*}
Mechanically, because the algorithmic investor race penalty is statistically indistinguishable from zero, its contribution to the overall change is small. 

\subsection{Building House Image Embeddings}\label{asec:images}
I collect property images primarily from Zillow, matching them to 35,046 transactions in my dataset. During the summer of 2023, I scraped Zillow listings to identify homes in the sample with active listings that included both interior and exterior photographs. Appendix Figure~\ref{afig:scrapedimages} displays representative examples of the images collected. Appendix Table~\ref{atab:summary_stats_houseswithimages} compares observable characteristics between homes with and without matched images. While homes with images are modestly lower in average sale price (\$175,000 vs. \$207,000), they are otherwise similar across structural attributes such as square footage and year built.

To construct structured representations of visual features, I apply a Vision Transformer (ViT-B/16) model pre-trained on ImageNet-21k, implemented via the \texttt{timm} library \citep{dosovitskiy2020, rw2019timm}. The model divides each image into patches and processes them using a transformer-based architecture to capture global spatial relationships. I extract the output of the final hidden layer (prior to classification) as a high-dimensional embedding that summarizes the visual content of each image. These embeddings capture aspects of the home's appearance—including roof condition, exterior symmetry, yard and driveway features, and signs of disrepair.

I include these image embeddings as covariates in regression models estimating the race gap in sale prices, alongside standard housing controls (e.g., lot size, number of bedrooms, year built) and census block group-by-year fixed effects. This approach allows me to assess how the estimated race penalty changes when incorporating high-dimensional information on property appearance.

\subsection{Estimating House Predictability}\label{asec:buildingpredictability}
I use the Extreme Gradient Boosting (XGBoost) algorithm to predict the transaction price \citep{Chen_2016}. Given the high-dimensionality of the data and significance of nonlinear relationships, non-parametric models outperform linear models when modeling houses. For example, even a slight increase in square footage could have a significant impact on price in a densely populated neighborhood, while the same would not be true in a rural area. Non-parametric models, such as tree-based algorithms, are able to capture nuanced, nonlinear relationships, particularly among the many variables that can influence house pricing like location, size, design, age, and local amenities.

The XGBoost algorithm operates on a gradient boosting framework in which new models are generated to correct the errors of pre-existing ones. In essence, the algorithm creates a robust overall model by combining multiple weak models. Doing so improves the accuracy of the prediction according to the regularized objective shown in Equation \ref{eq:xgboostobj}. $l$ is the differentiable convex loss function; $T$ is the number of leaves in each tree; and $w$ is the leaf weights \citep{Chen_2016}. Intuitively, this objective function balances training loss $l(\hat{y_i},y_i)$ with L1 regularization ($\gamma T$) and L2 regularization ($\lambda||w||^2 $) components, encouraging both simpler and more generalizable models. 

\begin{equation} \label{eq:xgboostobj}
   \mathcal{L}(\phi) = \sum_{i}l(\hat{y_i},y_i) + \sum_{k}\gamma T + \frac{1}{2}\lambda||w||^2 
\end{equation}

The model is built using pre-digitization data for each county to exclude any impacts from algorithmic investors. I randomly split the data into a training set and a 20 percent held-out test set. Using the training data set, I perform a random search through the hyperparameter space, then a grid search over a narrower parameter grid seeded by the random search, using 5-fold cross validation \citep{shen2022optimal, lavalle2004relationship}.

\end{appendix}

\end{document}